\documentclass[notitlepage,aps,superscriptaddress,showpacs,showkeys,nofootinbib,floatfix]{revtex4-1}

\usepackage{amsmath,amssymb,amscd,slashed,multirow,amsfonts,latexsym}
\usepackage{natbib}
\usepackage{mathrsfs}
\usepackage{accents}
\usepackage{graphicx}
\usepackage{epsf,epsfig,float,latexsym,amsthm,fancyhdr,rotating}
\usepackage{subcaption}
\usepackage{ragged2e}
\DeclareCaptionJustification{justified}{\justifying}
\usepackage{slashed}
\usepackage{esint}
\usepackage{nicefrac}
\usepackage{braket}
\usepackage{slashed}
\usepackage{color}
\usepackage{bbold}
\usepackage{multirow}
\usepackage[normalem]{ulem}
\usepackage{bm}

\newcommand{\nn}{\nonumber}

\usepackage[colorlinks,citecolor=blue,linktoc=all,linkcolor=cyan]{hyperref} 
\usepackage{tikz}

\definecolor{olive}{HTML}{668000}
\definecolor{lightolive}{HTML}{CCFF00}
\definecolor{darkolive}{HTML}{446600}
\definecolor{myorange}{HTML}{FF9900}
\definecolor{myblue}{HTML}{8080FF}
\definecolor{mygreen}{HTML}{008000}
\definecolor{purple1}{HTML}{BB99FF}
\definecolor{joker}{HTML}{990099}
\definecolor{fucsia}{HTML}{FF3399}
\definecolor{myred}{HTML}{FF0000}

\newcommand{\myemptypentagon}[1]{
\begin{tikzpicture}
\draw[#1] (1ex, 0.5257311121191336ex) -- (0.6180339887498949ex,1.0514622242382672ex) -- (0, 0.8506508083520399ex) -- (0, 0.2008114158862273ex) -- (0.6180339887498949ex, 0) -- (1ex, 0.5257311121191336ex);
\end{tikzpicture}
}

\newcommand{\myemptytriangle}[1]{
\begin{tikzpicture}
\draw[#1] (-0.5ex, 0) -- (0.5ex, 0) -- (0.ex, 0.866025403784ex) -- (-0.5 ex,0);
\end{tikzpicture}
}

\newcommand{\myemptysq}[1]{
\begin{tikzpicture}
\draw[#1] (-0.5ex, 0) -- (0.5ex, 0) --  (0.5ex, 1ex) -- (-0.5ex,1ex) -- (-0.5ex, 0);
\end{tikzpicture}
}

\def\onestatexp{\large{$\textcolor{black}\circ$}}
\def\onesysexp{\large{$\textcolor{red}\circ$}}
\def\twostatexp{{\rotatebox[origin=c]{90}{\large{\myemptypentagon{black}}}}}
\def\twosysexp{{\rotatebox[origin=c]{90}{\large{\myemptypentagon{red}}}}}
\def\threestatexp{{\rotatebox[origin=c]{0}{\large{\myemptysq{black}}}}}
\def\threesysexp{{\rotatebox[origin=c]{0}{\large{\myemptysq{red}}}}}
\def\fourstatexp{{\rotatebox[origin=c]{0}{\large{\myemptytriangle{black}}}}}
\def\foursysexp{{\rotatebox[origin=c]{0}{\large{\myemptytriangle{red}}}}}

\def\onestatth{\large{$\textcolor{black}\bullet$}}
\def\onesysth{\large{$\textcolor{red}\bullet$}}

\DeclareMathAlphabet{\mathantt}{OML}{antt}{l}{it}
\DeclareMathAlphabet{\mathpzc}{OT1}{pzc}{m}{n}

\def\beitem{\begin{itemize}}
\def\enitem{\end{itemize}}
\def\benum{\begin{enumerate}}
\def\ennum{\end{enumerate}}
\def\beq{\begin{equation}}
\def\enq{\end{equation}}
\def\bea{\begin{eqnarray}}
\def\eea{\end{eqnarray}}
\def\beqa{\begin{equation}\begin{array}{l}}
\def\eeqa{\end{array}\end{equation}}
\newcommand{\beql}[1]{\beq \label{#1}}
\newcommand{\beal}[1]{\bea \label{#1}}
\def\betab{\begin{table}}
\def\entab{\end{table}}

\def\eref#1{(\ref{eq:#1})}

\newcommand{\fr}[1]{FIG. \ref{#1}}
\newcommand{\frs}[1]{FIGs. \ref{#1}}
\newcommand{\eqr}[1]{Eq. (\ref{#1})}
 \newcommand{\eqrs}[1]{Eqs. (\ref{#1})}
 \newcommand{\secr}[1]{Sec. \ref{#1}}

\def\fref#1{\ref{fig:#1}}

\def\tref#1{Table \ref{#1}}
\def\tsref#1{Tables \ref{#1}}
\def\secref#1{Section \ref{sec:#1}}

\def\barr{\left(\begin{array}{c}}
\def\earr{\end{array}\right)}
\def\bmat{\left(\begin{array}{cc}}
\def\emat{\end{array}\right)}
\def\ie{i.e.~}
\newcommand\f{\frac}
\newcommand{\mc}{\mathcal}

\newcommand{\uN}{\mathcal{N}}

\newcommand{\uB}{\mathcal{B}}

 \def\astat{\alpha_{E1}}
 \def\bstat{\beta_{M1}}
 \def\chiboot{ \sum_{i=1}^{n} \left(\f{\uB_{ij}-T_{i}(\bm\psi_{j},\bm\theta)}{\sigma_{ij}}\right)^2}
  \def\chibootth{ \sum_{i=1}^{n} \left(\f{\mc M_{ij}-T_{i}(\bm\psi_{j},\bm\theta)}{\sigma_{ij}}\right)^2}
\def\chibootthnoerr{ \sum_{i=1}^{n} \left(\f{\mc M_{ij}-T_{i}(\bm\psi,\bm\theta)}{\sigma_{ij}}\right)^2}
\def\chibootmin{\sum_{i=1}^{n} \left(\f{\uB_{ij}-T_i(\bm\psi_{j},\hat{\bm\theta}_j)}{\sigma_{ij}}\right)^2}
\def\chibootminth{\sum_{i=1}^{n} \left(\f{\mc M_{ij}-T_i(\bm\psi_{j},\hat{\bm\theta}_j^\prime)}{\sigma_{ij}}\right)^2}
\def\chibootminthnoerr{\sum_{i=1}^{n} \left(\f{\mc M_{ij}-T_i(\bm\psi,\hat{\bm\theta}_j^\prime)}{\sigma_{ij}}\right)^2}

\def\sam{(1 + \delta_{k,j})( E_{i}+\gamma_{lj}\sigma_m) \quad \forall k=1, \ldots,n_s}
\def\samskew{(1 + \delta_{ij})( E_{i}+s_{ij})}
\def\samth{(1 + \delta_{ij})( T_{i}(\bm\psi, \hat{\bm\theta})+\gamma_{ij}\sigma_i)}
\def\chisingle{\sum_{i=1}^{n} \left(\f{E_i-T_i(\bm\psi, \hat{\bm\theta} )}{\sigma_i}\right)^2}
\def\myeps{\f{1}{\sigma_i}\left[T_i(\bm\psi, \hat{\bm\theta}) -\f{T_i(\bm\psi,\hat{\bm\theta}_j)}{1+\delta_{ij}}\right]}
\def\myeta{\f{1}{(1+\delta_{ij})\sigma_i}\left[T_i(\bm\psi,\hat{\bm\theta}_j) -T_i(\bm\psi_{j},\hat{\bm\theta}_j)\right]}
\def\myepsth{\f{1}{\sigma_i}\left[T_i(\bm\psi, \hat{\bm\theta}) -\f{T_i(\bm\psi,\hat{\bm\theta}_j^\prime)}{1+\delta_{ij}}\right]}
\def\myetath{\f{1}{(1+\delta_{ij})\sigma_i}\left[T_i(\bm\psi,\hat{\bm\theta}_j^\prime) -T_i(\bm\psi_{j},\hat{\bm\theta}_j^\prime)\right]}
\def\bwpdf{\f{I}{\pi\Gamma}\f{\Gamma^2}{(x-\mu)^2+\Gamma^2}}

\def\chimod{\sum_k \left\{\left[\sum_{i \in \text{ set k}}\left(\f{f_k E_i - T_i}{f_k\sigma_i}\right)^2\right]  +\left(\f{f_k - 1}{\sigma^{\text{sys}}_k}\right)^2\right\}}
\def\chimodone{\left[\sum_{\text{set 1}}\left(\f{f_1 E_i - T_i}{f_1\sigma_i}\right)^2\right] + \left[\sum_{\text{set 2,set 3}}\left(\f{ E_i - T_i}{\sigma_i}\right)^2\right] +\left(\f{f_1 - 1}{\sigma^{\text{sys}}_1}\right)^2}
\begin{document}

\title{A new Monte Carlo-based fitting method}
\author{P. Pedroni}
\email[Corresponding author; email: ]{paolo.pedroni@pv.infn.it}
\affiliation{Istituto Nazionale di Fisica Nucleare,
  Sezione di Pavia, 27100 Pavia, Italy}
\author{S. Sconfietti}
\affiliation{Dipartimento di Fisica,
Universit\`a degli Studi di Pavia, 27100 Pavia, Italy}
\affiliation{Istituto Nazionale di Fisica Nucleare,
Sezione di Pavia, 27100 Pavia, Italy}

	\begin{abstract}
          We present a new fitting technique based on the parametric bootstrap method, which relies on the idea to produce artificial measurements using the estimated probability distribution of the experimental data. In order to investigate the main properties of this technique, we develop a toy model and we analyze several fitting conditions with a comparison of our results to the ones obtained using both
the standard $\chi^2$ minimization procedure and a Bayesian approach.
Furthermore, we investigate the effect of the data systematic uncertainties both on the probability distribution of the fit parameters and on the shape of the expected goodness-of-fit distribution. Our conclusion is that, when systematic uncertainties are included in the analysis, only the bootstrap procedure is able to provide reliable confidence intervals and $p$-values, thus improving the results given by the standard $\chi^2$ minimization approach. Our technique is then applied to an actual physics process, the real Compton scattering off the proton, thus confirming both the portability and the validity of the bootstrap-based fit method.
	\end{abstract} 

\maketitle

	\section{A brief summary of a best-fit procedure}
	\label{sec:best_fit}
The main goal of a best-fit procedure is the estimate of some unknown parameters, which a given model depends on. The more commonly used algorithm is the so-called least squares method, which is based on the function:
	\beql{eq:chi2stand}
\chi_{stand}^2  (\bm\theta) = \sum_i{\left(\f{E_i - T_i(\bm\theta)}{\sigma_i}\right)^2}\ ,
	\enq
where $E_i$ are the experimental values, $\sigma_i$ are their corresponding statistical uncertainties in root mean square (rms) units and  $T_i$ are given by a theoretical model depending on the set of unknown parameters $\bm\theta$ to be evaluated from the data. The optimal parameter set $\hat{\bm\theta}$ is the one that minimizes $\chi_{stand}^2 $ and this solution can be written as:
	\beql{eq:chi2min}
\chi_{min}^2 = \sum_i{\left(\f{E_i - T_i(\hat{\bm\theta})}{\sigma_i}\right)^2}.
	\enq
Even though this procedure is commonly used in several scientific domains, its practical implementation often presents problems. One of them is the inclusion of the systematic uncertainties associated to the experimental data. If we consider the very simple case of a scaling factor parameter common to all data, the usual way to proceed is to modify \eqr{eq:chi2stand} as follows (see, for instance,~\cite{DAgostini1994}):
	\beq
 \label{eq:chi2sys}
 \chi_{mod}^2  (\bm\theta,f) = \sum_i{\left(\f{fE_i - T_i(\bm\theta)}{f\sigma_i}\right)^2}
 +\left(\f{f-1}{\sigma_{sys}}\right)^2.
	\enq 
        Here $f$ is a normalization factor to be treated as an additional fit parameter and $\sigma_{sys}$ is its estimated uncertainty (in rms units).
However this equation is strictly valid
 only in the case of Gaussian systematic uncertainties, since 
\beq
\chi_{mod}^2  (\bm\theta,f) =-2\ln{\mathcal{L}(\bm\theta,f)}\ ,
\enq

where the Likelihood function $\mathcal{L}(\bm\theta,f)$
is the product of the normal distributions with mean and standard deviations 
given by the experimental data multiplied by the normal distribution
modeling the common systematic scale uncertainty, \ie :

\beq
\mc{L}(\bm\theta,f) =  \prod_{i} \left[
\f{1}{\sigma^2_i\sqrt{2\pi}}
  e^{-\f{(fE_i - T_i(\bm\theta))^2}{2f\sigma^2_i}}
\right] \cdot \f{1}{\sigma^2_{sys}\sqrt{2\pi}} e^{-\f{(f - 1)^2}{2\sigma^2_{sys}}}.
%
\enq

Moreover,
especially with a large data base, this solution becomes unpractical since a different normalization parameter is needed for each subset and $\sigma_{sys}$ may as well change from point to point. Furthermore, when non-Gaussian and/or correlated uncertainties are present, as
in the previous case, the $\chi_{min}^{2}$ value does not generally follow the standard $\chi^{2}$-distribution, since it is not a sum of squared, indepedendent, standard Gaussian random variables\footnote{
One exception  is when both statistical and
correlated systematic Gaussian uncertainties are present. In this case \eqr{eq:chi2sys} can be replaced by the Mahalanobis distance, which can be shown to follow a $\chi^{2}$ distribution (see, for instance,~\cite{ref:maha}).}.
The evaluation of the goodness of fit then becomes quite difficult, since the $\chi^{2}$ test cannot be used.

The model $T$ may also not only depend on the parameter set $\bm\theta$, but also on some additional, non-fitted (nuisance) parameters $\bm\psi$ evaluated from experimental data, that can be written under the form:
	\beql{eq:thetaf}
\bm\psi = \bar{\bm\psi}_f \pm \bm\sigma_\psi.
	\enq
Here, $\bar{\bm\psi}$ and  $\bm\sigma_\psi$ are their estimated values and uncertainties (in rms units), respectively. In this case, another critical feature is to evaluate the effect of  $\bm\sigma_\psi$ on the final fit results. The total uncertainty on the fit parameters should be written as the sum of the pure contribution coming from the minimization itself and the uncertainty related to the effect of $\bm{\sigma_\psi}$ on the fit parameters. This last contribution can be evaluated according to the (linearly approximated) uncertainty propagation as
	\beq
\delta\hat{\bm\theta}_{extra,ab} \simeq\sum_{cd}  \left(\left.\f{\partial \bm\theta_a}{\partial \bm\psi_{c}}\right|_{\bm\theta_a = \hat{\bm\theta}_a}\right) \sigma_{\psi,cd} \left(\left.\f{\partial \bm\theta_b}{\partial \bm\psi_{d}}\right|_{\bm\theta_b = \hat{\bm\theta}_b}\right) ,
	\enq
where the indexes $a,b$ run over the components of $\bm\theta$, while $c,d$ on the components of $\bm\psi$. The quantity $\delta\hat{\bm\theta}_{extra,ab}$ thus includes both the covariances and the variances, obtained when $a\equiv b$. Furthermore, the terms in round brackets can be evaluated as
	\beq
\left.\f{\partial \bm\theta_x}{\partial \bm\psi_{y}}\right|_{\bm\theta_x = \hat{\bm\theta}_x} = \left[\left(\f{\partial T}{\partial \bm\theta_x}\right)^{-1} \f{\partial T}{\partial \bm\psi_{y}}\right]_{\bm\theta_x=\hat{\bm\theta}_x}.
	\enq
However, if the analytical structure of the model is complicated, the term $(\partial T)/(\partial \bm\psi_{y})$ could be hard to be obtained, even numerically, thus requiring the application of a different strategy.

Our new method is able to solve all these problems in a straightforward way and, even if we apply it within the least squares framework, it can, in principle, also be used with other minimization schemes, as the Maximum Likelihood (ML) approach.

The manuscript is organized as follows. In \secr{sec:outline} we give a general outline of our new method and we describe in detail its more relevant features by considering a general example of a fit of data with both statistical and systematic uncertainties. In \secref{toy} and \secref{skewgaus}
we perform an accurate check of the new method using
two different toy models
and simulated data. The results thus obtained under different fit conditions are also compared both to the ones coming form the standard $\chi^2$ fit procedure
and to the ones obtained 
  using, as an alternative approach, the Hierarchical Bayesian Model (HBM) described in~\cite{ref:bayes}.

In \secr{sec:compton} we apply our method to an actual physics process, the real Compton scattering off the proton. Here we briefly summarize the results that have already been published (see Ref.~\cite{Pasquini_2019}) and complement them with additional
information by giving an estimated of the experimental biases of the fitted data and by evaluating the expected goodness-of-fit distribution
both with the exclusion and the inclusion of the systematic uncertainties in the fit procedure. Finally, our conclusions are drawn in \secr{sec:concl}.

%
	\section{Outline of the new method}
	\label{sec:outline}

Our new method is based on the {\em parametric bootstrap} technique (see, for instance,~\cite{Davidson-Hinkley} and  references therein). It requires, for each point $E_i$, measured at a given set of known parameters $\bm x$, the knowledge of the probability density function $p({\bm x})$ of its evaluated uncertainty. The core idea is to assume each single $E_i$ to be the ML estimate of its true and unknown value $\mc E_i$.
In this case, the density  $p({\bm x},E_i)$ is taken as an approximation
of the true density $p({\bm x},{\mc E_i})$:
	\beq
p({\bm x},\mc E_i) \simeq p(\bm{x}, E_i) .
	\enq
Then a random bootstrap sample $E^b_1, E^b_2 \ldots  E^b_n$ is generated, for each  $E_i$, according to
$p(x,E_i)$.
Using this sample, an estimate of the
true model parameters $\bm\theta^b$ is obtained using  the standard minimization
tools (simplex, gradient, ...) applied to the function given in \eqr{eq:chi2stand}.

Repeating this bootstrap cycle a (very) large number $n_b$ of times, we get a sample
$\hat{\bm\theta}^b_1, \hat{\bm\theta}^b_2 \ldots \hat{\bm\theta}^b_{n_b}$ 
from which we are finally able to reconstruct 
the true probability distributions for every fit parameter. For instance, the sample mean and the sample standard deviation are given as:
	\beq
 \label{eq01:sample3}
 \hat{\bm\theta}^b  = \f{1}{n_b}  \sum^{n_b}_{i=1} {\hat{\bm\theta}^b_i}\quad ,  \quad \sigma_{\hat{\bm\theta}^b} = \left[\f{1}{n_b - 1}\sum_{i=1}^{n_b}\left(\hat{\bm\theta}_i^b - \hat{\bm\theta}^b\right)^2\right]^{1/2}.
	\enq 
	\subsection{A general example} 
As a general example, we consider 
the case of a database composed by different and independent subsets and
with a total of $n$ experimental points 
having both statistical and systematic uncertainties  independent of each other. The best estimate of the true value $\mc E_i$ of each experimental point
can then be written as:
	\beq
 \label{eq:errsys}
  E_i \pm \sigma_i^{\text{stat}} \pm \sigma_i^{\text{sys}}\ ,
	\enq 
        where $ \sigma_i^{\text{stat}}$ and
        $ \sigma_i^{\text{sys}}$ are the standard deviations of the statistical
        and systematic uncertainties, respectively.

        Now we suppose to have Gaussian-distributed statistical
        uncertaintes and, to be in the same conditions as 
        in \eqr{eq:chi2sys}, we also assume that all the points of each
        subset have the same scaling factor uncertainty $\Delta$.
        This parameter is different for each subset and represents
        the half width of a uniform distribution\footnote{These assumptions are just reasonable choices and they  can be easily changed to
          deal with every specific situation.}.
        In the first step of our procedure,
each artificial bootstrap ``measurement" is assumed to be Gaussian distributed around a given
experimental data point with a standard deviation given by its statistical
uncertainty  (see \eqr{eq:errsys}).
Then, all bootstrapped points of a given subset are shifted by the same random
quantity uniformly distributed within the estimated systematic uncertainty
interval.

If we define a {\em cycle} as when the number of bootstrapped points are equal to the total number of points 
in the considered experimental set, the bootstrap sampling
can be finally described 
for each subset $k$ as:

\beql{eq:sam}
\uB_{lj}= \sam ,
	\enq
        
where {$\uB_{lj}$} is a generic bootstrapped point with
the index {$l$} running over the number of data points
in each subset ($n_k$)  and 
the index $j$ indicates the $j^{th}$ bootstrap cycle.
The {$\gamma_{lj}$} 
parameters are sampled from the standard Gaussian distribution $\uN[0,1]$, while the {$\delta_{k,j}$} 
are random numbers uniformly distributed as
$\mathcal U[-\Delta_k,+\Delta_k]$,
being $\pm\Delta_k$ the percentage systematic uncertainty of each subset $k$
($k$ runs from 1 to the number of the different data subsets $n_{s}$).
If only statistical uncertainties have been taken into account, the systematic sources can be easily excluded from this procedure by just imposing
$\delta_{k,j}\equiv 0\ \ \forall k =1, \ldots, n_s $.

  After a complete cycle and once defined:
  \bea
\delta_{lj} &\equiv& \delta_{k,j}  \quad \forall l \in \text{ set k} \quad ; \quad
\forall k=1, \ldots,n_s \ ,\nn\\
\sigma_{lj} &\equiv& (1+\delta_{lj})\sigma_l \ , \label{eq:def}
  \eea

the minimization procedure is performed on the function:
	\beq
	\label{eq:xxx}
\chi_j^2 = \sum^n_{i=1} {\left(\f{\mc B_{ij} - T_i(\bm\theta)}{\sigma_{ij}}\right)^2},
	\enq

with the index $i$ running over the total number of data points $n$, 
and all the fit results are stored. 

The main advantages of the adopted technique are:
	\beitem 
\item[*] the straightforward inclusion of systematic uncertainties in the minimization procedure, as shown in \eqr{eq:sam}. 
This feature allows us to reduce the overall number of fit parameters with respect to the 
modified $\chi^2$ procedure, where a normalization factor for each data set is left as free parameters
(see \eqr{eq:chi2sys});
\item[*] any kind of uncertainty distribution of the experimental data can be easily implemented;
\item[*]  the probability
distributions of the fit parameters are not assumed {\em a priori},
but are directly evaluated from the distributions  assigned to the experimental data;
\item[*] the uncertainty on the fit parameters can be estimated also when the
  used mathematical minimization algorithm does not provide them as,
   for example, in the case of the simplex method.
	\enitem

When additional model parameters  $\bm\psi$ are present (see \eqr{eq:thetaf})
and their probability distribution $g(\bm\psi,\bm\sigma_\psi)$ is known, their uncertainties can be easily 
included in this algorithm by sampling at every cycle
an additional random variable $\bm\psi_{j}$ distributed as
$g(\bm\psi,\bm\sigma_\psi)$.
The minimization function of \eqr{eq:xxx} is accordingly generalized as:
\beq
\label{eq:chiboot}
\chi_{b,j}^2 = \chiboot , 
	\enq
and its minimum value can be written as:
 \beq
	\label{eq:chibmin}
 {\hat\chi^2_{b,j}} = \chibootmin .
	\enq

	\subsection{The meaning of \texorpdfstring{$\hat\chi_{b,j}^2$}. in parametric bootstrap}
	\label{sec:chi2dec}

 The value of ${\hat\chi^2_{b,j}}$ given in \eqr{eq:chibmin} 
cannot be treated as the  standard $\hat\chi^2$  value commonly
used to assess the goodness of a fit
in the standard procedure, \ie
	\beql{eq:hatchidef}
\hat \chi^2 = \chisingle ,
	\enq
due to the artificial statistical fluctuations inherent to
each bootstrapped sampling.

In the following, we will find the connection between  ${\hat\chi^2_{b,j}}$
and $\hat \chi^2 $.
After introducing the following definitions,
	\bea
\epsilon_{ij} &\equiv& \myeps,\nn\\
\eta_{ij} &\equiv& \myeta ,
	\eea
we can rewrite the $T_i(\bm\psi_{j},\hat{\bm\theta}_j)$ term of \eqr{eq:chibmin} as
	\beq
T_i(\bm\psi_{j},\hat{\bm\theta}_j) = (1+\delta_{ij})\left[T_i(\bm\psi, \hat{\bm\theta}) -\sigma_i(\epsilon_{ij}+\eta_{ij})\right].
	\enq
The $\epsilon_{ij}$ parameter,
once summed over $i$, quantifies the difference between the model evaluated at the global best
values of the fitting parameters $\hat{\bm\theta}$ and the model evaluated at the $j^{th}$ best values of $\bm\theta$ (\ie $\hat{\bm\theta}_j$), taking into account both the statistical and systematic uncertainties.
The $\eta_{ij}$ term is related to the effect that the uncertainties on the additional parameter set $\bm\psi$ have on the model evaluation of the generic observable $E_i$.
Thanks to the previous formalism, we can rewrite \eqr{eq:chibmin} as 
	\beql{eq:chi2exp}
{\hat\chi^2_{b,j}} = \hat \chi^2 + \sum_i \gamma_{ij}^2 + \sum_i \epsilon_{ij}^2 + \sum_i D_{ij} + \sum_i \Phi_{ij},
	\enq
where: 
	\bea \label{eq:chi2dec}
D_{ij} &\equiv& 2\left[\epsilon_{ij}\gamma_{ij} +\f{1}{\sigma_i}(\epsilon_{ij}+\gamma_{ij})(E_i -T_i(\bm\psi, \hat{\bm\theta}))\right],\nn\\
\Phi_{ij} &\equiv& \eta_{ij}^2 +  2\eta_{ij}\left[(\epsilon_{ij}+\gamma_{ij}) +\f{1}{\sigma_i}(E_i -T_i(\bm\psi, \hat{\bm\theta}))\right].
	\eea
Thanks to the decomposition given in \eqr{eq:chi2exp}, from the
${\hat\chi^2_{b,j}}$ parameter we can isolate and identify:
(i) the pure squared Gaussian term ($\sum_i \gamma_{ij}^2$); (ii) the main contribution related to the difference between the best evaluation of the model parameters obtained at the end of each bootstrap cycle and at the end of the full procedure 
          ($\sum_i \epsilon_{ij}^2$);
          (iii) the term containing the effect of the error due to the uncertainties on the additional model parameters ($\sum_i \Phi_{ij}$);
          and (iv) a parameter with the mixed contributions due to 
          the non-quadratic and $\eta$-independent terms ($\sum_i D_{ij}$).

Inverting the decomposition of \eqr{eq:chi2exp}, we can get the evaluation of $\hat\chi^2$ in the bootstrap framework.
Within the small numerical approximations introduced by the Monte-Carlo procedure, after each bootstrap cycle
such a value has to be identical to the one that can be directly computed from \eqr{eq:hatchidef}
at the very end of the bootstrap procedure.
This cross-check is crucial for the auto-consistency of the fitting method: if $\hat\chi^2 \neq {\hat\chi^2_{b,j}} -\left( \sum_i \gamma_{ij}^2 + \sum_i \epsilon_{ij}^2 + \sum_i D_{ij} + \sum_i \Phi_{ij}\right)$, there could be some mistakes in the sampling 
scheme or in the minimization procedure.

	\subsection{Evaluation of the expected goodness-of-fit distribution}
	\label{sec:chi2decth}

Once the analytical form of the minimization function and the decomposition of its minimum value have been established, it is
still necessary to determine a {\em goodness-of-fit distribution}, from which the associated $p$-value have to be computed. This
procedure is detailed below.

Within this framework, the expected distribution can be
evaluated assuming the model $T_{i}(\bm\psi, \hat{\bm\theta})$ to be correct
and by considering an ideal situation in which the experimental points
are {\em exactly} the values predicted by our model.

The sampling procedure outlined above (see \eqr{eq:sam})
can then be repeated replacing each experimental data with
$T_{i}(\bm\psi, \hat{\bm\theta})$. We thus obtain:
	\beql{eq:samth}
\mc M_{ij} = \samth.
	\enq
The minimization function can then be defined as:
	\beq
\chi_{th,j}^2 = \chibootth,
	\enq
and we denote its minimum value after the $j$-th cycle as:
	\beq
	\label{eq:chibminth}
 {\hat\chi^2_{th,j}} = \chibootminth.
 \enq
 The sampled parameters $\bm\psi_{j}$ are exactly the same as in \eqr{eq:chiboot}, while the fit values of the parameters at every bootstrap cycle are, in general, different from the ones obtained from the fit of the bootstrapped data: for this reason we use the symbol $\hat{\bm\theta}_j^\prime$ instead of $\hat{\bm\theta}_j$.
 According to this notation, we can apply the same decomposition as before,
 thus defining
	\bea
\epsilon_{ij}^\prime &\equiv& \myepsth,\nn\\
\eta_{ij}^\prime &\equiv& \myetath,\nn\\
D_{ij}^\prime &\equiv& 2\epsilon_{ij}^\prime\gamma_{ij},\nn\\
\Phi_{ij}^\prime &\equiv& {\eta_{ij}^\prime}^2 +  2\eta_{ij}^\prime\left(\epsilon_{ij}^\prime+\gamma_{ij} \right).
	\eea
The resulting decomposition for $ {\hat\chi^2_{th,j}}$ is 
	\beq\label{eq:chiteodecomp}
{{\hat\chi}^2_{th,j}} = \hat \chi_{th}^2 + \sum_i \gamma_{ij}^2 + \sum_i {\epsilon_{ij}^\prime}^2 + \sum_i D_{ij}^\prime + \sum_i \Phi_{ij}^\prime,
	\enq
 where $ \hat \chi_{th}^2$  is defined as in \eqr{eq:hatchidef}, replacing $E_i$ with $T_{i}(\bm\psi, \hat{\bm\theta})$.
This parameter
is identically zero by construction, but we explicitly leave this decomposition as a cross-check\footnote{We will discuss this point
later, in the comments related to \tref{tab:chi2}.} since, within the small numerical approximations introduced by
 the procedure itself,
 we should obtain:
\beq\label{eq:chiteozero}
 0 = {{\hat\chi}^2_{th,j}} -\left[ \sum_i \gamma_{ij}^2 + \sum_i {\epsilon_{ij}^\prime}^2 + \sum_i D_{ij}^\prime + \sum_i \Phi_{ij}^\prime\right] .
 \enq   
 It is interesting to notice that
in \eqr{eq:chiteodecomp}
the sensitivity of the theoretical model on the additional parameters set $\bm\psi$
is confined in the term $\Phi_{ij}^\prime$, which includes the dependence on
$\eta_{ij}^\prime$.
This allows us to define the unbiased theoretical distribution for the $\hat\chi^2$ value, which does not include the effect of the 
($\bm\psi-\bm\psi_{j}$) difference.
This feature corresponds to the definition of the following minimization function,
	\beq
\chi^2_{u,j} = \chibootthnoerr,
	\enq
with a minimum at
	\beq
	\label{eq:chibminth2}
 {\hat\chi^2_{u,j}}
 = \chibootminthnoerr = {{\hat\chi}^2_{th,j}} -\sum_i \Phi_{ij}^\prime.
 \enq
  This new parameter is 
  independent on any model or assumption about the probability distribution
  functions of the experimental data.
  The meaning of the different components of the ${{\hat\chi}^2_{th,j}}$
    parameter given in \eqr{eq:chiteodecomp}
    is the same as the one described for ${{\hat\chi}^2_{b,j}}$ at the end of
    \secr{sec:chi2dec}.

  When systematic uncertainties are not taken into account and the
  effect of the $\bm\psi_{j}$ 
  parameters can be neglected,
  the bootstrapped values of \eqr{eq:samth} 
  are 
  sampled from the Gaussian distribution $\mc N[T_i(\bm\psi, \hat{\bm\theta}),\sigma_i^2]$ and $ {\hat\chi^2_{th,j}}$
  (see \eqr{eq:chiteodecomp})
  is basically a sum of the squares of the independent standard Gaussian variables  $\gamma_{ij}$.
  The small additional corrections due to the $\sum_i {\epsilon_{ij}^\prime}^2$ and $\sum_i D_{ij}^\prime$ terms can introduce a tiny model-dependent distortion to this simple picture, as it will be discussed later.

  On the other hand, when systematic uncertainties are included in the fit procedure, these values are generated from the convolution $\mc U[-\Delta_k,\Delta_k] \ast \mc N[\ T_i(\bm\psi, \hat{\bm\theta}),\sigma_i^2]$.
  The terms $\sum_i {\epsilon_{ij}^\prime}^2$ and $\sum_i D_{ij}^\prime$ cannot thus be ignored, and an appreciable distortion is introduced to the
  standard $\chi^2$-distribution.\\
 The theoretical distribution reconstructed from the  $ {\hat\chi^2_{u,j}}$ values
 can then be written as:
  \beal{eq:chi2th}
\text{without systematics: } {\hat\chi^2_{u,j}} &=& \sum_i \gamma_{ij}^2 + \left.\sum_i {\epsilon_{ij}^\prime}^2\right|_{\delta_{ij}=0} + \left.\sum_i D_{ij}^\prime\right|_{\delta_{ij}=0} \sim \sum_i \gamma_{ij}^2, \nn\\
\text{with systematics: } {\hat\chi^2_{u,j}} &=& \sum_i \gamma_{ij}^2 + \left.\sum_i {\epsilon_{ij}^\prime}^2\right|_{\delta_{ij}\ne 0} + \left.\sum_i D_{ij}^\prime\right|_{\delta_{ij} \ne0}.
	\eea   
 For the sake of completeness, we list all the terms of \eqr{eq:chi2exp} and \eqr{eq:chiteodecomp},
 divided by the number of degrees of freedom $n_{dof}$, \ie
	\beal{eq:chi2note}
 \hat \chi^2_r &=& \f{1}{n_{dof}}\left\{ {\hat\chi^2_{b,j}} -\left[ \sum_i \gamma_{ij}^2 + \sum_i \epsilon_{ij}^2 + \sum_i D_{ij} + \sum_i \Phi_{ij}\right]\right\},\nn\\
 \hat \chi_{th,r}^2 &=& \f{1}{n_{dof}}\left\{ {\hat\chi^2_{th,j}} -\left[\sum_i \gamma_{ij}^2 + \sum_i {\epsilon_{ij}^\prime}^2 + \sum_i D_{ij}^\prime + \sum_i \Phi_{ij}^\prime\right]\right\},\nn\\
 \gamma^2_r &\equiv&  \f{1}{n_{dof}} \sum_i \gamma_{ij}^2 ,\quad  \epsilon_r^2 \equiv  \f{1}{n_{dof}}  \sum_i \epsilon_{ij}^2 , \quad D_r \equiv  \f{1}{ n_{dof}} \sum_i D_{ij} , \nn\\
  \Phi_r &\equiv&  \f{1}{ n_{dof}} \sum_i \Phi_{ij} , \quad  {\epsilon^\prime}^2_r \equiv  \f{1}{n_{dof}} \sum_i {\epsilon_{ij}^\prime}^2 ,\quad D^\prime_r \equiv  \f{1}{n_{dof}} \sum_i D_{ij}^\prime,\nn\\
  \Phi^\prime_r &\equiv&  \f{1}{ n_{dof}} \sum_i \Phi_{ij}^\prime, \quad \chi_b^2 = \gamma^2_r + \epsilon_r^2 + D_r^2 + \Phi_r, \quad \chi_{th}^2   = \gamma^2_r +{\epsilon^\prime}^2_r + {D^\prime_r}^2 \ ,
 \eea
 since they will be used later in the text.

 From a sufficiently  high number of bootstrap replicas, we can reconstruct the expected goodness-of-fit distribution $p({\chi^2_{th}})$. Once $p({\chi^2_{th}})$ is empirically evaluated,
 we are able to compute the $p$-value associated to the fit results
 using the two-sided $\chi^2$ test defined as:

 \[
 p\mbox{-value}= \left\{ 
 \begin{array}{ll}
  CDF(\mathbb X)  & \mbox{if  $CDF(\mathbb X) < 0.5$}, \\  & \\
   1-CDF(\mathbb X) & \mbox{if  $CDF(\mathbb X) \geq 0.5$} ,
 \end{array}\right.
 \]
where $\mathbb X$ is the value of the $\chi^2_r$ value obtained at the end of the fit and
 \beq
  CDF(\mathbb X) = \int_{-\infty}^{\mathbb X} p(\chi^2_{th}) d\chi^2_{th}.
 \enq
In the following, we will omit the $\mathbb X$ dependence in the cumulative distribution functions CDFs.\\ 
Using all the parameters defined in \eqr{eq:chi2note}, some important cross-checks about the validity of the overall procedure can be performed: 
	\beitem
  \item[*] the expected value of $\Phi_r$ should be small
 when the uncertainties of additional parameters set $\bm\psi$ do not give a relevant contribution to the final fit results.
 On the contrary, when the $\Phi_r$ term is not small, we can get some hints on how to deal with the $\bm\psi$ parameters:
 i) we should fit
 them or ii) we should reduce the uncertainties
   of the $\bm\psi$ terms with a more accurate evaluation.
\item[*] the probability distribution of $\gamma^2_r$ has to exactly follow the  $\chi^2$-distribution, otherwise the used pseudo-random number generator could be not good enough.
	\enitem 
%
	\section{A simplistic model to describe the new method}
	\label{sec:toy}
 	\subsection{Implementation}
  In order to investigate and check the features of the bootstrap-based fitting technique, we implemented a
   toy model using simulated random data from the 
   Breit-Wigner (BW) distribution, that can be written as:
	\beq\label{eq:bw}
 BW(x;I,\mu,\Gamma) = \bwpdf , 
	\enq
where $I$ is an overall scale factor, $\mu$ is the peak position and $\Gamma$
specifies the half-width at half-maximum.
   This function is chosen since
   it plays an important role in physics, being often used to model resonance phenomena, and it is also strongly non linear in the parameters space.
   
In order to generate our simulated data, we sample a variable $\xi$ from a uniform distribution $\mc U[0,1]$ and we use the well-known cumulative inversion method to obtain a value for $x$ which is distributed according to $BW(x; I ,\mu,\Gamma)$.
 The chosen values for the $\mu$, $\Gamma$, and  $I$ parameters of \eqr{eq:bw} are:
	\beql{eq:init}
\mu_0 = 0,\quad \Gamma_0 = 1, \quad I_0 = 250.
	\enq

Using this procedure, 30000 simulated events were generated and those falling
 within the $x$ interval $[-4, 4]$
were equally divided into 3 different subsets and grouped into the 100-bin histogram shown in \fr{fig:data_bw}. 

	\begin{figure}[h]
\includegraphics[scale=.8]{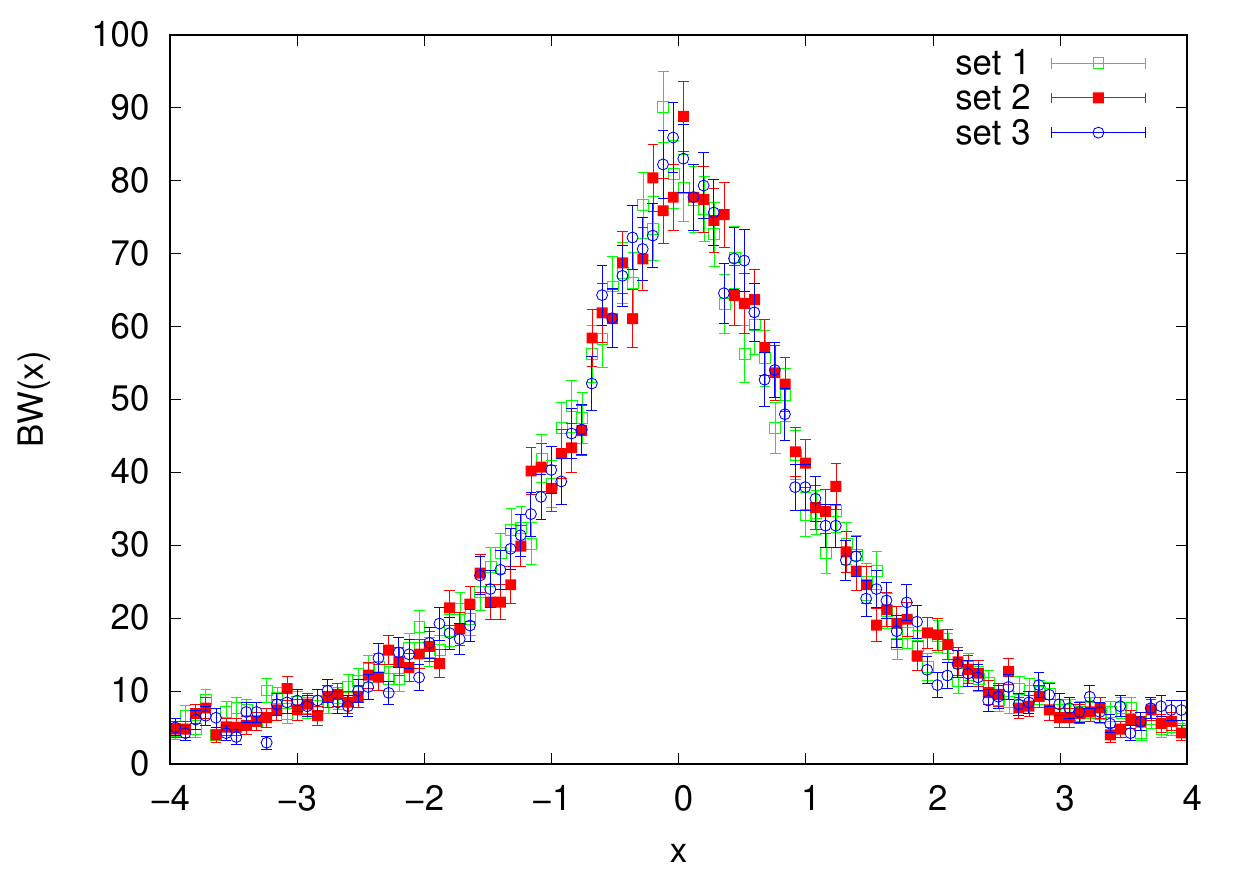}
\caption{Simulated data from the Breit-Wigner distribution. They are separated into the three subsets denoted by the different point
 styles.\label{fig:data_bw}}
	\end{figure} 
 If, for each subset, we denote by $B_i$ and $\sigma_{B,i}$ the content and the statistical uncertainty of the $i^{th}$ histogram bin,

 the bootstrapped data are given by \eqr{eq:sam}, where the experimental values $E_i$ and $\sigma_i$ are replaced by $B_i$ and $\sigma_{B,i}$, \ie
	\beql{eq:toy_sam01}
\uB_{ij} = (1+\delta_{ij}) \left[ B_i + \gamma_{ij}\sigma_{B,i}\right].
	\enq

 We are able now to implement our new fit method and to check it 
 in different conditions:
	\beitem
\item[*] 3 fit parameters ($I$, $\mu$ and $\Gamma$), labeled as Fit$_{3p}$;
\item[*] 2 fit parameters ($I$ and $\Gamma$), and one fixed parameter ($\mu = \mu_0$), labeled as  Fit$_{2p+1f}$;
\item[*] 2 fit parameters ($I$ and $\Gamma$), and one sampled parameter ($\mu \in \mc N[\mu_0,\sigma^2_{\mu_0}]$), labeled as  Fit$_{2p+1s}$. The $\sigma_{\mu_0}$ term is chosen as $k/100$, where $k=3,20$ in order to investigate how
  the size of the uncertainties on $\bm\psi$ affects the fit results.
	\enitem
We also assume that each subset is affected by 
 uniformly distributed  systematic scale 
 uncertainty with: 
 $\Delta_1 = 0.04$, $\Delta_2 = 0.06$ and $\Delta_3 = 0.03$.\\

 The number $N$ of bootstrap replicas to be generated is evaluated
   in the Fit$_{3p}$ case, and without the inclusion of systematic
    uncertainties,  with the goal to 
    to obtain a relative precision $e_r \leq 5\%$ (in rms units)
    on the central values of all fit parameters.
    Under these conditions, as shown in \fr{fig:err}, 
    the $\mu$ parameter has, for a given $N$,
    the highest  $e_r$ value and reaches the required precision at
    $N\simeq 10000$.

  For each condition, the fit is then
  performed with 10000 bootstrap replicas both without and with the inclusion of the systematic uncertainties.
  In the first case we simply set $\delta_{ij} = 0$ in \eqr{eq:sam} and \eqr{eq:toy_sam01},
  while in the second one we add the superscript $^\prime$ to
  every fit condition.

	\begin{figure}[h]
  \begin{subfigure}[b]{0.45\textwidth}
  \includegraphics[scale=.6]{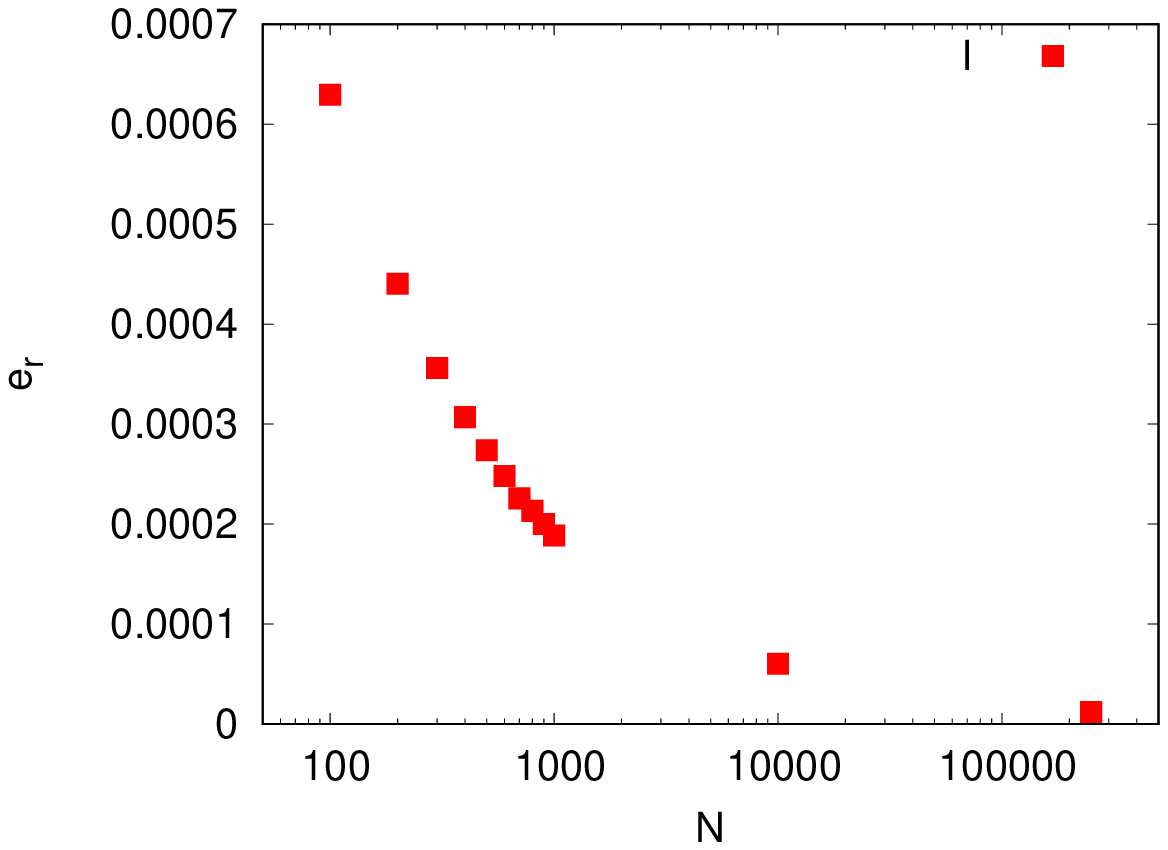}
  \subcaption{ \hspace{-0.8 truecm} }
  \label{fig:err_I}
  \end{subfigure}
  \begin{subfigure}[b]{0.45\textwidth}
  \includegraphics[scale=.6]{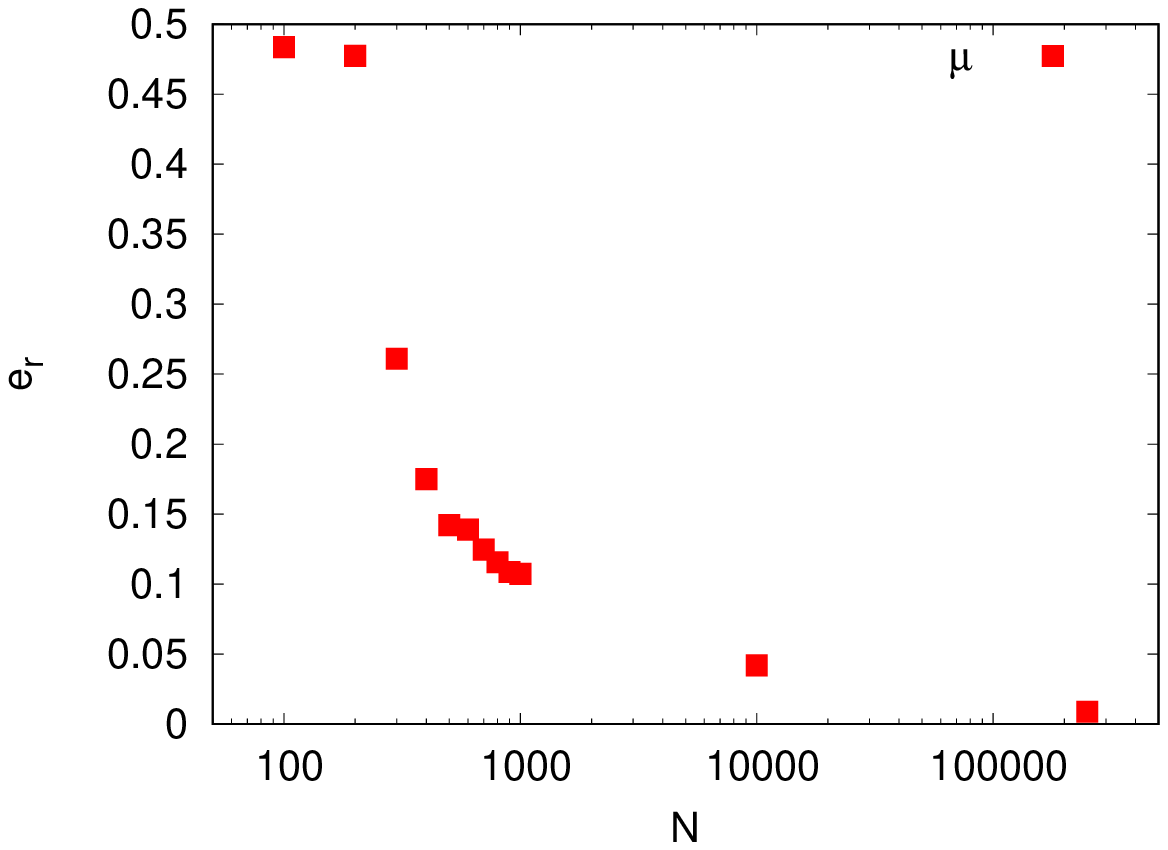}
  \subcaption{ \hspace{-1 truecm} 
  \label{fig:err_mu}
  }
  \end{subfigure}\\
  \begin{subfigure}[b]{0.45\textwidth}
  \includegraphics[scale=.6]{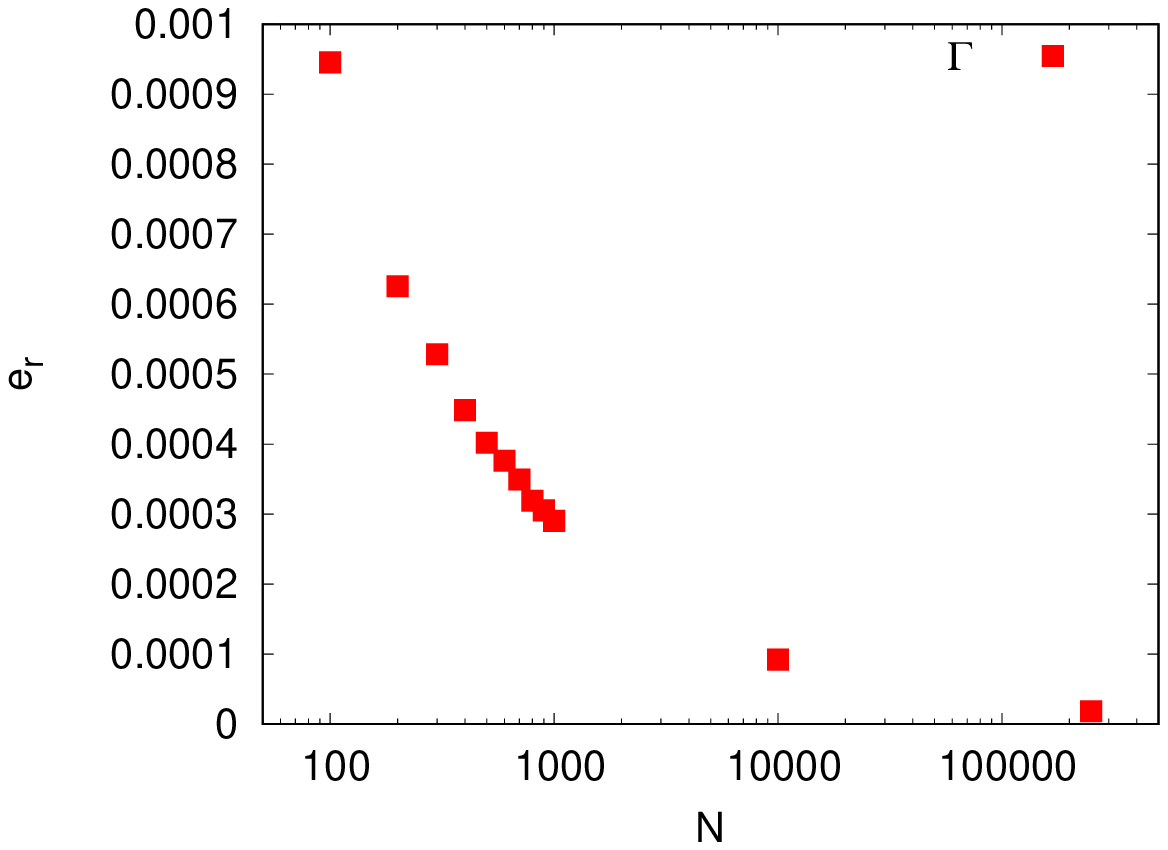}
  \subcaption{ \hspace{-0.8 truecm} }
  \label{fig:err_gamma}
  \end{subfigure}

   \caption{Relative error $e_r$ (in rms units) on the central values of the fit parameters   $I$ (a), $\mu$ (b) and $\Gamma$ (c)
      as a function of the number of bootstrap cycles $N$.
 \label{fig:err} }
	\end{figure}

	\subsection{Fit results}
	\label{sec:fitted}
 
 The final results of the fit performed under all the conditions described above
 are displayed in \tref{tab:res} and 
 the obtained distributions of the fit parameters are
 shown in \frs{fig:pdf01} and \fref{pdf02}.
 The expected values
 (labeled as $\mathbb E\left[\dots\right]$) and the distributions of the different components of  ${\hat\chi^2_{b,j}}$ 
 and ${\hat\chi^2_{th,j}}$ 
 (see \eqr{eq:chi2note} for notation) 
 are given in \tref{tab:chi2}
 and \frs{fig:chi2dec4} to \fref{chi2dec3}, respectively. Finally, the CDFs of the expected goodness-of-fit
 distributions are displayed in \frs{fig:cdf1}  and \fref{cdf3}, respectively.
 In the following a detailed discussion of all these results is given.

\begin{table}[]
\begin{tabular}{|c|c|c|c|c|c|c|}
\hline
\multicolumn{7}{|c|}{ DATA}  \\
\hline
\hline
 Fitting conditions &  $I$ & $\mu$ $(10^{-3})$  &  $\Gamma$ ($10^{-1}$) & $\hat\chi^2_r$ & $p$-value  & Symbol \\  
 \hline
  Fit$_{3p}$ & $ {247.0}^{+1.4}_{-1.5} $  & ${1.8}^{+7.1}_{-7.6} $  & $ {9.8} \pm 0.1$  & $ 0.98 $  & $ 45 \% $  & \onestatexp \\
  Fit$_{3p}^\prime$ & $ {247.0}^{+4.0}_{-4.2} $  & ${1.8}^{+7.1}_{-7.7} $  & $ {9.8} \pm 0.1 $  & $ 0.98 $  & $ 35 \% $  & \onesysexp  \\
  \hline
  Fit$_{2p+1f}$  & $ {247.0}^{+1.4}_{-1.5} $  & fixed  & $ {9.8} \pm 0.1 $  & $ 0.98 $  & $ 43\% $  & \twostatexp  \\
 Fit$_{2p+1f}^\prime$ & $ {247.0}^{+4.0}_{-4.2} $  & fixed  & $ {9.8} \pm 0.1 $  & $ 0.98$  & $ 34\% $  & \twosysexp \\
 \hline
 Fit$_{2p+1s}$ ($3\%$)  & $ {247.0}^{+1.4}_{-1.5} $  & sampled  & $ {9.8} \pm 0.1 $  & $ 0.98 $  & $ 43\% $  &  \threestatexp \\
 Fit$_{2p+1s}^\prime$ ($3\%$)& $ {247.0}^{+3.9}_{-4.3} $  & sampled  & $ 9.8 \pm 0.1 $  & $0.98  $  & $ 34 \% $  &  \threesysexp \\
\hline
Fit$_{2p+1s}$ ($20\%$)  & $ {243.0}^{+4.2}_{-5.0} $  & sampled  & $ {10.0}^{+0.2}_{-0.3} $  & $1.05 $  & $ 25\%  $  & \fourstatexp  \\
Fit$_{2p+1s}^\prime$ ($20\%$) & $ {242.0}^{+5.9}_{-6.2} $  & sampled  & $ {10.0}^{+0.2}_{-0.3} $  & $ 1.05$  & $19\% $  & \foursysexp  \\
\hline
 
\end{tabular}
\caption{Results from the fit applied to the simulated data in the different conditions described in the text. Each $p$-value is calculated from
  the expected goodness-of-fit distribution, which is reconstructed in the framework of the bootstrap technique.
The superscript $^\prime$ denotes the inclusion of systematic uncertainities   in the fit procedure and the
 different symbols refer to the point styles of \frs{fig:pdf01} and \fref{pdf02}.
}\label{tab:res}
\end{table}

	\begin{figure}[h]
\includegraphics[scale=1.2]{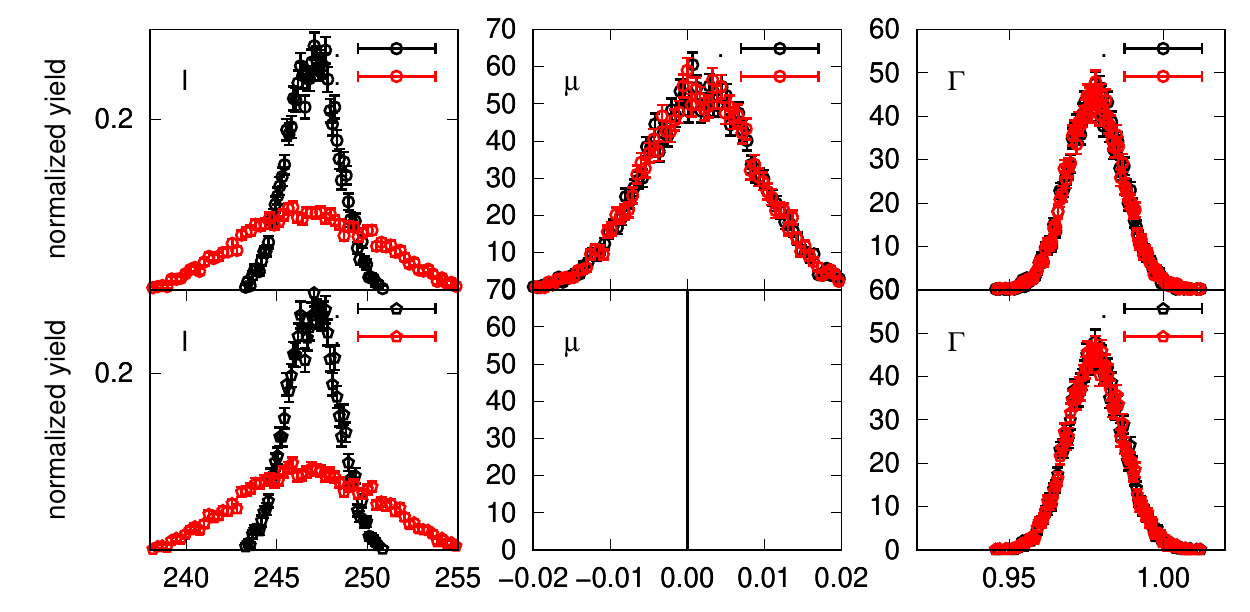}
\caption{Probability distributions for the fit parameters of the simulated data
for the Fit(')$_{3p}$ (upper panels) and Fit(')$_{2p+1f}$ (lower panels) configurations. 
From left to right: the scale factor $I$, the mean $\mu$ and the width $\Gamma$. The red (black) points indicate the inclusion
(exclusion) of the systematic uncertainties in the fit procedure.
In the  Fit$_{2p+1f}$ case, the mean value $\mu$ is not fitted  but kept fixed to zero. See~\tref{tab:res} for the meaning of the symbols. 
  \label{fig:pdf01}}
	\end{figure}
		\begin{figure}[h]
\includegraphics[scale=1.2]{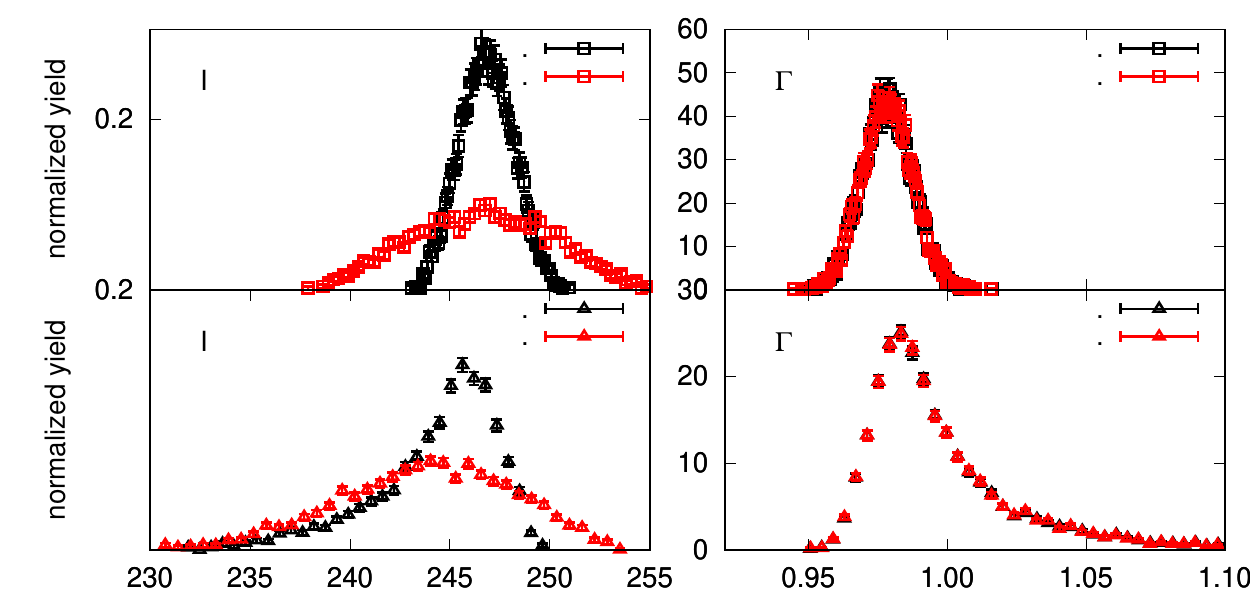}
\caption{Probability distributions for the fit parameters of the simulated data
  for the Fit(')$_{2p+1s}$($3\%$) (upper panels) and Fit(')$_{2p+1s}$($20\%$) (lower panels) configurations.
  From left to right: the scale factor $I$ and the width $\Gamma$.
  The red (black) points indicate the inclusion
(exclusion) of the systematic uncertainties in the fit procedure.
  The mean value $\mu$ is not shown here, being sampled from its known value $\mu_0$. See~\tref{tab:res} for the meaning of the symbols. 
  \label{fig:pdf02}}
	\end{figure}
	\subsubsection{Fit$_{3p}$}
  As a first consistency check, we compare the results obtained in this condition, shown in
  first line of \tref{tab:res},
 to the results from the standard fit procedure (see \eqrs{eq:chi2stand}
and \eref{chi2min}), which gives:
	\beql{eq:single}
I = 247.0 \pm 1.5 , \quad \mu = (1.7 \pm 7.4)\cdot 10^{-3}, \quad \Gamma = (9.8\pm0.1)\cdot 10^{-1}, \quad \hat\chi_r^2 = 0.98.
	\enq
All these values are in very good agreement with the numerical results of \tref{tab:res}. The only small difference
 is the asymmetry of the bootstrapped  1-$\sigma$ interval for $I$ and $\mu$.
 This feature is due to both the finite number of replicas (10000)
 and to the finite number of bins (100)  of the histograms used for the
 evaluation of the CDF for the goodness-of-fit distribution.
 
 Such a difference can be reduced by increasing the number of bootstrap cycles and of the
 classes used for the CDFs generation. As an example,
 when using 100000 replicas and 200-bin histograms, we obtain
 exactly the same confidence intervals as in the standard procedure. 

 This approximation can also be taken under control by examining
 the empirical probability distribution functions shown by the black open dots of
 \fr{fig:pdf01} (upper panels).
 Using a standard best-fit procedure, we checked that they follow a Gaussian distribution,
 in agreement with statistical expectations (see, for instance,~\cite{James:2006zz}).
 As an example, the probability distribution of the fit parameter $I$ obtained with
 100000 bootstrap replicas is shown in the left plot of \fr{fig:convolution} and
 compared to the best-fit Gaussian distribution.
 Using the output fit parameters, we obtain exactly the same 1-$\sigma$ range given in \eqr{eq:single}.
 
 The expected values and the distributions of the different 
 components of  ${\hat\chi^2_{b,j}}$ (see \eqr{eq:chi2exp}), are given in the first line
 of  \tref{tab:chi2} (upper part) and in the upper panel of \fr{fig:chi2dec4} (black curves),
 respectively.
 The statistical fluctuations of ${\hat\chi^2_{b,j}}$ are almost entirely due to the
 $\gamma^2_r$ term, since the contributions given by $\epsilon_r^2$ and $D_r$ are quite small and 
the values of $\mathbb E\left[\epsilon_r^2\right]$ and $\mathbb E\left[ D_r \right]$  are $O (10^{-2})$
 and almost negligible.  
 
 In the ${\hat\chi^2_{th,j}}$ case (see \eqr{eq:chiteodecomp} and \fr{fig:chi2dec1}),
 the term $\mathbb E\left[\hat\chi^2\right]$
 is numerically very close to zero (see \tref{tab:chi2}) and its distribution
 coincides with the reduced $\chi^2$-distribution, as expected from \eqr{eq:chi2th},  due to the quite small and almost opposite values of  $\mathbb E\left[\epsilon^\prime\right]$ and $\mathbb E\left[D^\prime_r\right]$. This result is shown
 by the black points in the left plot of \fr{fig:cdf1}.

 All these properties fully confirm the considerations outlined
 in \secref{chi2dec} and \secref{chi2decth}. 

	\subsubsection{Fit$_{3p}^\prime$ } 

 Due to the functional form of the BW distribution in \eqr{eq:bw}, a common
 scale uncertainty 
 only affects the uncertainty of $I$  and does not influence the estimate of the other
 parameters $\mu$ and $\Gamma$, as shown in the second line of  \tref{tab:res} 
 and in \fr{fig:pdf01} (top panels).
 
 We can compare these results to the ones obtained using  the $\chi^2_{mod}$ function of \eqr{eq:chi2sys}, leaving the normalization
 factors for each subset as additional free parameters: 
	\beql{eq:chimodbias}
 \chi^2_{mod} = \chimod ,
 \enq
 
 where $\sigma^{\text{sys}}_k = \Delta_k/\sqrt{3}, k=1,2,3$. In this case we obtain:
	\beql{eq:singl2}
I = 246.7 \pm 3.5 , \quad \mu = (1.7 \pm 7.5)\cdot 10^{-3}, \quad \Gamma = (9.8\pm0.1)\cdot 10^{-1}, \quad \hat\chi_r^2 = 0.99 .
	\enq
 These results are very similar to the ones obtained under the Fit$_{3p}^\prime$ configuration. The
 slightly smaller uncertainty on $I$ is
   mainly due to the fact that, as previously discussed,
   \eqr{eq:chimodbias} should only be applied
    in the case of Gaussian systematic uncertainties. 

This underestimation can be clearly seen if we  closely examine the probability distribution for the fit parameter $I$, the
only one that is significantly affected by the inclusion of the systematic uncertainties in the fit procedure.
This study is also interesting since 
 we cannot make any general assumption  about its functional form and our method
 empirically provides its shape.

 The obtained distribution with 100000 bootstrap cycles is shown in the central plot of \fr{fig:convolution}, where we can clearly
 notice a significant deviation from the pure Gaussian shape obtained 
 when systematic undertainties are neglected (left plot of of \fr{fig:convolution}).
 The simple Gaussian best-fit (solid blue line) gives for $I$
 the same confidence interval of \eqr{eq:singl2}, thus  underestimating
   the true rms value.
 Such a discrepancy becomes more relevant as the value of $\Delta_k$ is artificially
increased, as shown in right plot of \fr{fig:convolution}.
This behavior can be qualitatively explained by the fact that, given the 
 sampling defined in \eqr{eq:sam}, the distribution of the $I$ parameter results from the convolution
 of a uniform and a Gaussian function, with its shape depending  on the $\Delta_k/\sigma_i$ ratio.  

 Within the frequentist framework, the correct solution can be found 
 using the ML approach and by finding the parameter values that maximize: 

\beq\label{eq:like}
\mc{L}_{unif} =   \sum_k \left\{ \prod_{i \in \text{ set k}} \left[
\f{1}{\sigma^2_i\sqrt{2\pi}}
  e^{-\f{(f_kE_i - T_i(\bm\theta))^2}{2f_k\sigma^2_i}}
\right] \cdot \mc{U}[-\Delta_k,\Delta_k]
\right\}.
\enq

 As an alternative, the HBM model (see \cite{ref:bayes}) can also be used.
 In this framework,  $\mc{L}_{unif}$ can be rewitten as:

 \beq\label{eq:likebayes}
 \mc{L}_{unif} \equiv \mc{L}(E_i|\bm{\theta,f})\pi(\bm{f}) \ , 
 \enq
 
 where $\mc{L}(E_i|\bm{\theta,f})$ is the probability to obtain the experimental data for a given set of model parameter values and
 $\pi(\bm{f})$ is the prior distribution for the parameters $\bm{f}$.
 When a uniform non-informative prior distribution 
 is taken for the other fit parameters $\bm{\theta}$, we can write:

 \beq\label{eq:likebayes}
 p(\bm{\theta,f}|E_i) \propto  \mc{L}_{unif} \ , 
 \enq

 where
 $p(\bm{\theta,f}|E_i)$ is the posterior probability of a specific set of
 model parameter given the data.

 Under these conditions, the 68\% Bayesian credible intervals 
  for the fit parameters obtained from HBM and using a Markov chain
  Monte Carlo procedure\footnote{We run 12000 iterations and discard the first
  2000 ``burn-in'' draws.} are:
 
	\beql{eq:bayesint}
I = 246.6^{+4.0}_{-4.1} , \quad \mu = (1.7^{+6.6}_{-7.0})\cdot 10^{-3}, \quad \Gamma = (9.8\pm0.1)\cdot 10^{-1}. 
	\enq

        These values are in very good agreement with the ones given in
        \tref{tab:res} and also the distributions for all the fit parameters
        are the same as the ones shown in  \fr{fig:pdf01} and in the central plot of \fr{fig:convolution}.
        All these conclusions give an important consistency check on
        the validity
        of our fit procedure.

        Similarly to it, the HBM approach has also a very flexible implementation
        and can be used to easily model and reproduce 
        a wide variety of uncertainty distribution.
        However, our method takes into account the effect of           
the systematic uncertainties without the need of additional fit parameters.
This feature can give a sizable advantage
in all those cases where a significanty large number of these parameters
should be included in the fit procedure.

 Finally, If we examine the ${\hat\chi^2_{b,j}}$ decomposition (red curves in the upper panel of
 \fr{fig:chi2dec4})  we note that there are additional statistical fluctuations due to the
 increased dispersion of the $\epsilon_r^2$ distribution (see also \tref{tab:chi2}), even if 
 $\mathbb E\left[\hat\chi^²\right]$ has the same value as in the  Fit$_{3p}$ case. All this agrees to
 the results obtained with the $\chi^2_{mod}$ procedure.
 
 Due to the  correlations between the data caused by the systematic uncertainties,
 the expected goodness-of-fit distribution is now different from the reduced $\chi^2$-distribution.
 This feature can be clearly seen in a quantitative way in \fr{fig:chi2dec1} (red curves),
 where both the  ${\epsilon^\prime}^2_r$ 
 and the $D^\prime_r$ term now give a significant contribution
 to  ${\hat\chi^2_{th,j}}$, thus distorting
 the effect of the predominant $\gamma^2_r$ term.
 All these results are consistent with the considerations
 outlined in \ref{sec:chi2decth} (see, in particular, \eqr{eq:chi2th}).
 The size of this distortion depends both on the magnitude of the systematic uncertainties and 
 on the analytical structure of the model $T$.

 The CDF for the goodness-of-fit distribution is shown by the red dots of \fr{fig:cdf1}
 (left panel) and the resulting $p$-value (see \tref{tab:res}) is significantly different
 from the Fit$_{3p}$ case.
 All these considerations signal that this
 crucial fit parameter cannot be correctly evaluated within the
 $\chi^2_{mod}$ framework when the normalization uncertainties
follow a distribution that differs substantially from the Gaussian one.

	\begin{figure}[h]
\includegraphics[scale=1.0]{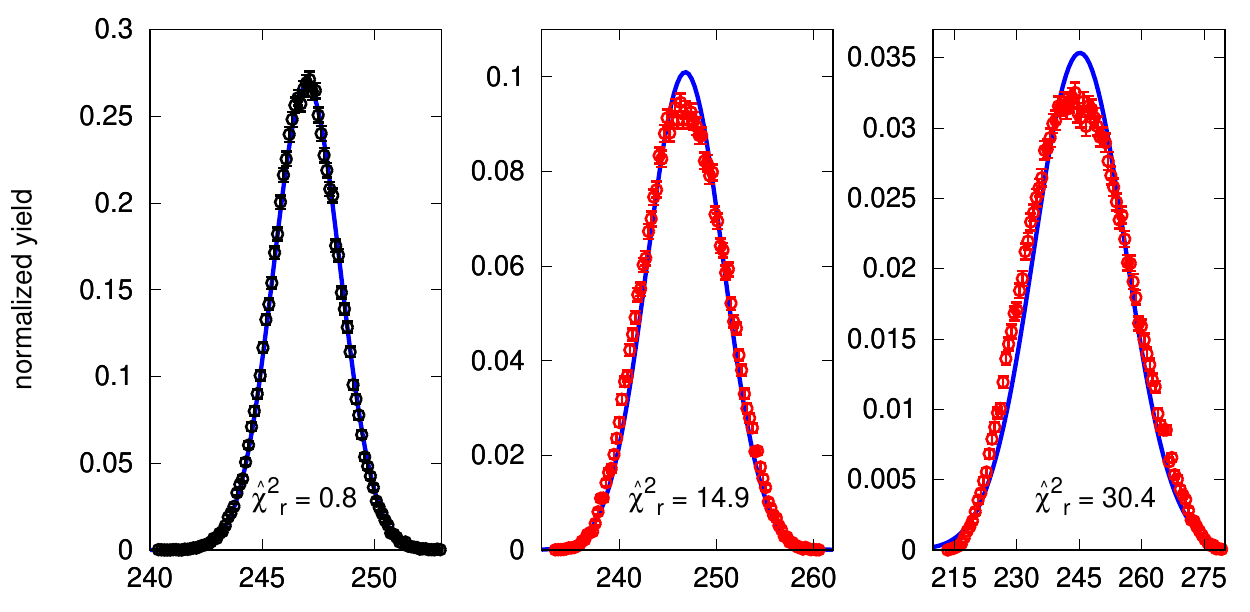}
\caption{Probability distributions for the fit parameter $I$, obtained with 100000 bootstrap cycles and 100 histogram bins under the Fit$_{3p}$ (left panel), Fit$_{3p}^\prime$ (central panel) and Fit$_{3p}^\prime$ with $3\cdot\Delta_k$ (right panel) configurations.
  These distributions are compared to the best-fit Gaussian curves (blue solid curves) and the corresponding  $\hat\chi^2_r$ values
  are also given at the bottom of each plot.\label{fig:convolution}}
	\end{figure}

	\subsubsection{Fit$_{2p+1f}$ }
 As expected, all the results obtained in this case, both with and without the inclusion of the
 systematic uncertainties, are basically the same as in the Fit$(^\prime)_{3p}$ conditions (see
 \tref{tab:res} and \tref{tab:chi2}). They will be used as a benchmark to compare the
 results of the Fit$(^\prime)_{2p+1s}$ conditions to quantitatively investigate the effects on the fit
 results given by the uncertainties related to additional model parameters.

	\subsubsection{Fit$_{2p+1s}$} 
The results obtained when $\sigma_{\mu_0} =3\%$ (fifth and sixth lines of
\tref{tab:res} and \tref{tab:chi2}, upper and lower parts of \fr{fig:chi2dec2} and  \fr{fig:cdf3}, left panel)
are very similar to the ones obtained in the Fit$(^\prime)_{2p+1f}$ conditions.
 The only relevant difference is the non-zero value of the  $\mathbb E\left[\Phi\right]$ parameter.
 However, due to the small uncertainty assigned to the  $\mu_0$ parameter, its effect is almost
 negligible.
 Also the probability distributions of the fit parameters $I$ and $\Gamma$ are still compatible with a Gaussian function. 
 
 On the contrary, the fit results significantly change when  
 $\sigma_{\mu_0}$ is large and equal to $20\%$, as shown in the seventh and eighth lines of
 \tref{tab:res} and \tref{tab:chi2}, upper and lower parts,  \fr{fig:chi2dec3} and  \fr{fig:cdf3}, right panel.

 The probability distributions of the fit parameters $I$ and $\Gamma$ are strongly asymmetric and significantly 
 different from all the previous cases. The $\Phi^\prime_r$ term gives a sizable contribution
 to ${\hat\chi^2_{th,j}}$ with a significant distortion of the goodness-of-fit distribution. 
 Also the effect on  the data systematic uncertainties is strongly reduced, as shown in \fr{fig:cdf3},
 due to the effect of the relevant uncertainty on $\mu_0$.
 As previously mentioned (see \secr{sec:chi2decth}), in this case it would be more meaningful to add  $\mu$ as an additional
 fit parameter.

\begin{table}[]
\begin{tabular}{|c|c|c|c|c|c|}
\hline
\multicolumn{6}{|c|}{ DATA}  \\
\hline
\hline
 Fitting conditions &  $\mathbb E\left[\hat\chi^2_r\right]$ & $\mathbb E\left[\gamma^2_r\right]$  &  $\mathbb E\left[\epsilon_r^2\right]$ & $\mathbb E\left[D_r\right]$ & $\mathbb E\left[\Phi_r\right]$   \\  
\hline  
Fit$_{3p}$ & $ 0.98 $  & $ 1.01 \pm 0.08 $  & $ (9.92 \pm 8.18)\cdot 10^{-3} $  & $ (-2.0\pm11.6)\cdot 10^{-2} $  & $ 0 $  \\
 Fit$_{3p}^\prime$ & $0.98 $  & $ 1.01 \pm 0.08 $  & $ (5.55 \pm 3.67)\cdot 10^{-2} $  & $ (-2.0\pm11.9)\cdot 10^{-2} $  & $ 0 $   \\
\hline
Fit$_{2p+1f}$ & $ 0.98 $  & $ 1.01 \pm 0.08 $  & $ (6.59\pm6.65)\cdot 10^{-3}$  & $ (-1.3\pm11.5)\cdot 10^{-2} $  & $ 0 $   \\
Fit$_{2p+1f}^\prime$ & $0.98 $  & $ 1.01 \pm 0.08 $  & $ (5.20 \pm 3.63)\cdot 10^{-2} $  & $ (-1.3\pm11.8)\cdot 10^{-2} $  & $ 0 $   \\
\hline
Fit$_{2p+1s}$ ($3\%$) & $ 0.98 $  & $ 1.01 \pm 0.08 $  & $ (6.76\pm 6.70)\cdot 10^{-3}$  & $ (-1.2\pm11.5)\cdot 10^{-2} $  & $ (5.32 \pm 7.99)\cdot 10^{-2} $   \\
Fit$_{2p+1s}^\prime$ ($3\%$)  & $ 0.98 $  & $ 1.01 \pm 0.08 $  & $ (5.15\pm 3.58)\cdot 10^{-2}$  & $ (-1.2\pm11.8)\cdot 10^{-2} $  & $ (5.32 \pm 7.99)\cdot 10^{-2} $    \\
\hline
Fit$_{2p+1s}$ ($20\%$) & $ 1.05 $  & $ 1.01 \pm 0.08 $  & $ (11.3\pm 32.4)\cdot 10^{-2}$  & $ (-1.5\pm20.4)\cdot 10^{-2} $  & $ 2.00 \pm 2.47 $    \\
Fit$_{2p+1s}^\prime$ ($20\%$) & $ 1.05 $  & $ 1.01 \pm 0.08 $  & $ (15.6\pm 32.2)\cdot 10^{-2}$  & $ (-1.8\pm 20.9)\cdot 10^{-2} $  & $ 2.00 \pm 2.47 $   \\
\hline
\hline
\multicolumn{6}{|c|}{ MODEL}  \\
\hline
\hline
 Fitting conditions &  $\mathbb E\left[\hat\chi_{th}^2\right]$ & $\mathbb E\left[\gamma^2_r\right]$  &  $\mathbb E\left[{\epsilon^\prime}^2_r\right]$ & $\mathbb E\left[D^\prime_r\right]$ & $\mathbb E\left[\Phi_r^\prime\right]$  \\  
\hline  
Fit$_{3p}$  & $\mc O ( 10^{-6} )$  & $ 1.01 \pm 0.08 $  & $ (9.96\pm 8.20)\cdot 10^{-3}$  & $ (-1.99\pm1.64)\cdot 10^{-2} $  & $ 0 $    \\
 Fit$_{3p}^\prime$& $ \mc O(10^{-4}) $  & $ 1.01 \pm 0.08 $  & $ (5.66\pm3.75)\cdot 10^{-2}$  & $ (-2.00\pm 3.02)\cdot 10^{-2} $  & $ 0 $    \\
\hline
Fit$_{2p+1f}$ & $ \mc O(10^{-7}) $  & $ 1.01 \pm 0.08 $  & $ (6.61\pm6.67)\cdot 10^{-3}$  & $ (-1.32\pm1.33)\cdot 10^{-2} $  & $ 0 $    \\
Fit$_{2p+1f}^\prime$& $ \mc O(10^{-4}) $  & $ 1.01 \pm 0.08 $  & $ (5.32\pm 3.71)\cdot 10^{-2}$  & $ (-1.34\pm2.86)\cdot 10^{-2} $  & $ 0 $    \\
\hline
Fit$_{2p+1s}$ ($3\%$) & $ \mc O(10^{-5}) $  & $ 1.01 \pm 0.08 $  & $ (6.78\pm 6.72)\cdot 10^{-3}$  & $ (-1.34\pm1.33)\cdot 10^{-2} $  & $ (5.46 \pm 8.16)\cdot 10^{-2} $    \\
Fit$_{2p+1s}^\prime$ ($3\%$) & $ \mc O(10^{-4})$  & $ 1.01 \pm 0.08 $  & $ (5.26\pm 3.67)\cdot 10^{-2}$  & $ (-1.39\pm 2.83)\cdot 10^{-2} $  & $ (5.46 \pm 8.15)\cdot 10^{-2} $   \\
\hline
Fit$_{2p+1s}$ ($20\%$) & $ \mc O(10^{-1}) $  & $ 1.01 \pm 0.08 $  & $ (11.0\pm 31.6)\cdot 10^{-2}$  & $ (-1.6\pm16.4)\cdot 10^{-2} $  & $ 1.97 \pm 2.43 $   \\
Fit$_{2p+1s}^\prime$ ($20\%$) & $ \mc O(10^{-1} )$  & $ 1.01 \pm 0.08 $  & $ (15.4\pm 31.4)\cdot 10^{-2}$  & $ (-1.9\pm17.0)\cdot 10^{-2} $  & $ 1.97 \pm 2.43 $  \\
\hline
\end{tabular}
\caption{The expected values, labeled as $\mathbb E\left[\dots\right]$
of the different components of  ${\hat\chi^2_{b,j}}$ 
and ${\hat\chi^2_{th,j}}$ are given in the upper and lower panels,
respectively. See \eqr{eq:chi2note} for notation.
 The different symbols refer to the point styles of \frs{fig:cdf1} and \fref{cdf3}.
 \label{tab:chi2}}
\end{table}

\begin{figure}[h]
\includegraphics[scale=.9]{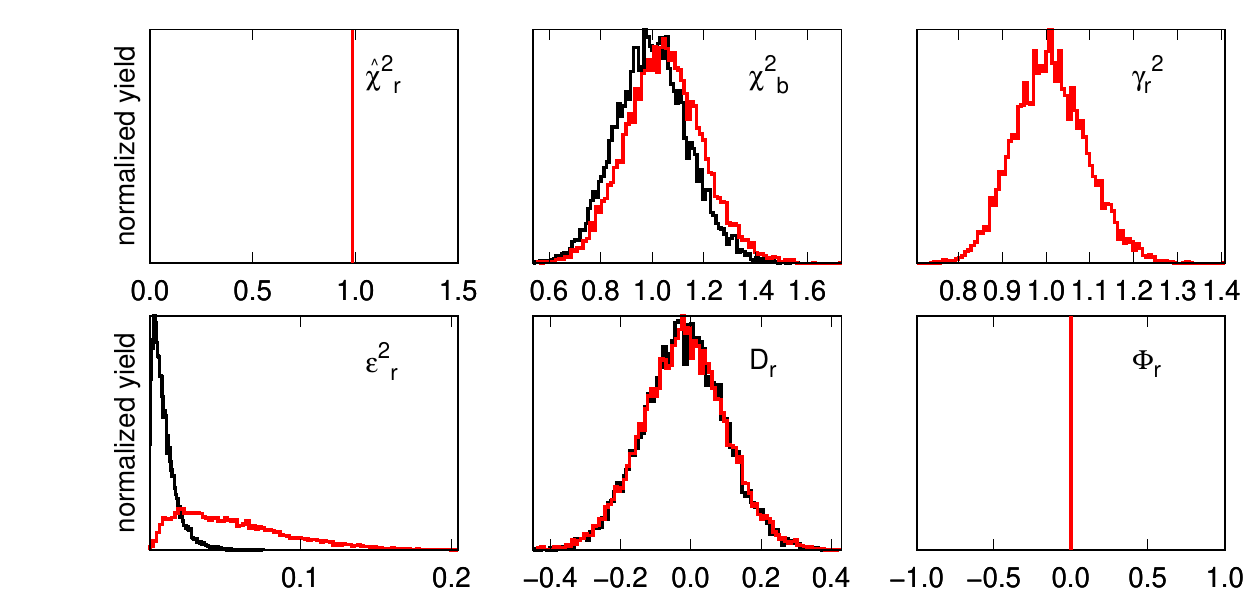}
\caption{Decomposition of the $\hat\chi^2_{b,j}$
  parameter in the Fit$_{3p}$ configuration when systematic uncertainties are excluded (black curves) and included (red curves).
  Upper panel (from left to right): $\hat\chi^2_r$, $\chi_b^2$ and $\gamma^2_r$ components. Lower panel (from left to right): $\epsilon^2_r$, $D_r$
  and $\Phi_r$ components.
  See text (and, in particular, \eqr{eq:chi2note}, \tsref{tab:res} and \ref{tab:chi2}) for the notation and the expected values of the different components. 
Black and red lines exactly overlap for the constant $\hat\chi^2_{r}$ 
and  $\Phi_r$ components.
  \label{fig:chi2dec4}}
	  \end{figure}

		\begin{figure}[h]
\includegraphics[scale=.9]{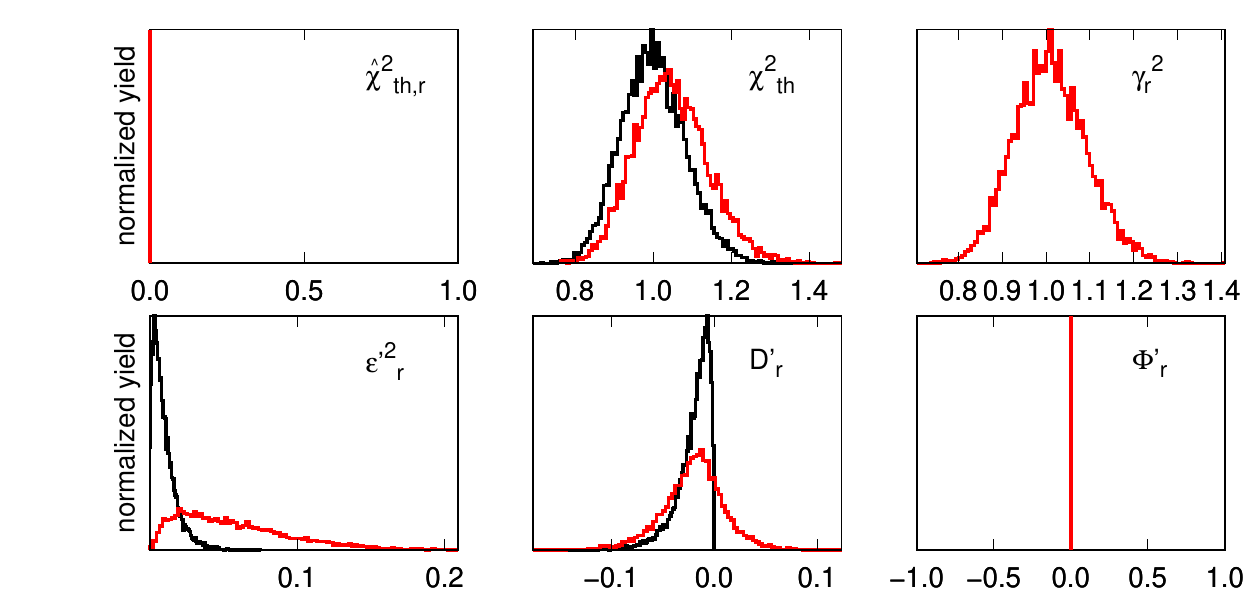}
\caption{Decomposition of the $\hat\chi^2_{th,j}$
  parameter  in the Fit$_{3p}$ configuration when systematic uncertainties are excluded (black curves) and included (red curves).
  Upper panel (from left to right): $\hat\chi^2_{th,r}$, $\chi^2_{th}$ and $\gamma^2_r$ components. Lower panel (from left to right): ${\epsilon^\prime}^2_r$,  $D^\prime_r$  and $\Phi^\prime_r$ components.
See text (and, in particular, \eqr{eq:chi2note}, \tsref{tab:res} and \ref{tab:chi2}) for the notation and the expected values of the different components. Black and red lines exactly overlap for the constant $\hat\chi^2_{th,r}$
  and  $\Phi^\prime_r$ components.
  \label{fig:chi2dec1}
}
	\end{figure}

			\begin{figure}[h]
\includegraphics[scale=.9]{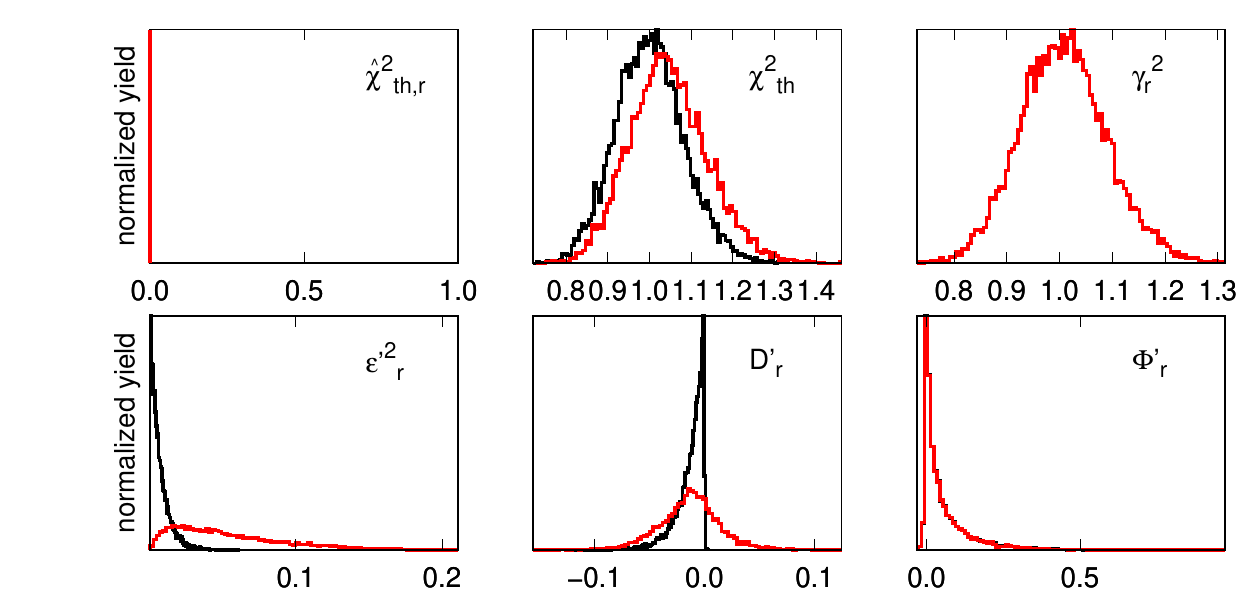}
\caption{Decomposition of the $\hat\chi^2_{th,j}$ parameter in the Fit$_{2p+1s}$ ($3\%$) configuration when systematic excluded (black curves) and included (red curves).
  Upper panel (from left to right): $\hat\chi^2_{th,r}$, $\chi^2_{th}$ and $\gamma^2_r$ components. Lower panel (from left to right): ${\epsilon^\prime}^2_r$,  $D^\prime_r$  and $\Phi^\prime$ components. See text for the notation.
  \label{fig:chi2dec2}}
	\end{figure}

		\begin{figure}[h]
\includegraphics[scale=.9]{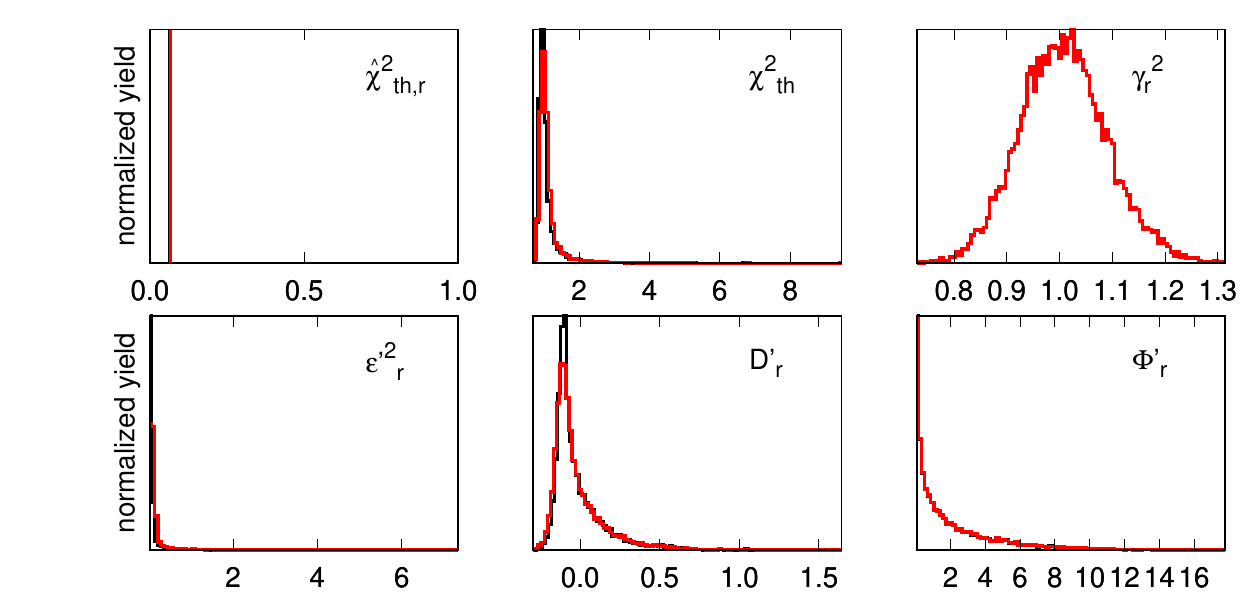}
\caption{Decomposition of the $\hat\chi^2_{th,j}$
  parameter for the  Fit$_{2p+1s}$ ($20\%$) configuration when systematic excluded (black curves) and included (red curves).
 Upper panel (from left to right): $\hat\chi^2_{th,r}$, $\chi^2_{th}$ and $\gamma^2_r$ components. Lower panel (from left to right): ${\epsilon^\prime}^2_r$,  $D^\prime_r$  and $\Phi^\prime$ components. See text for the notation.
  \label{fig:chi2dec3}}
	 \end{figure}

	\begin{figure}[h]
	  \begin{subfigure}[b]{0.45\textwidth}
\includegraphics[scale=.6]{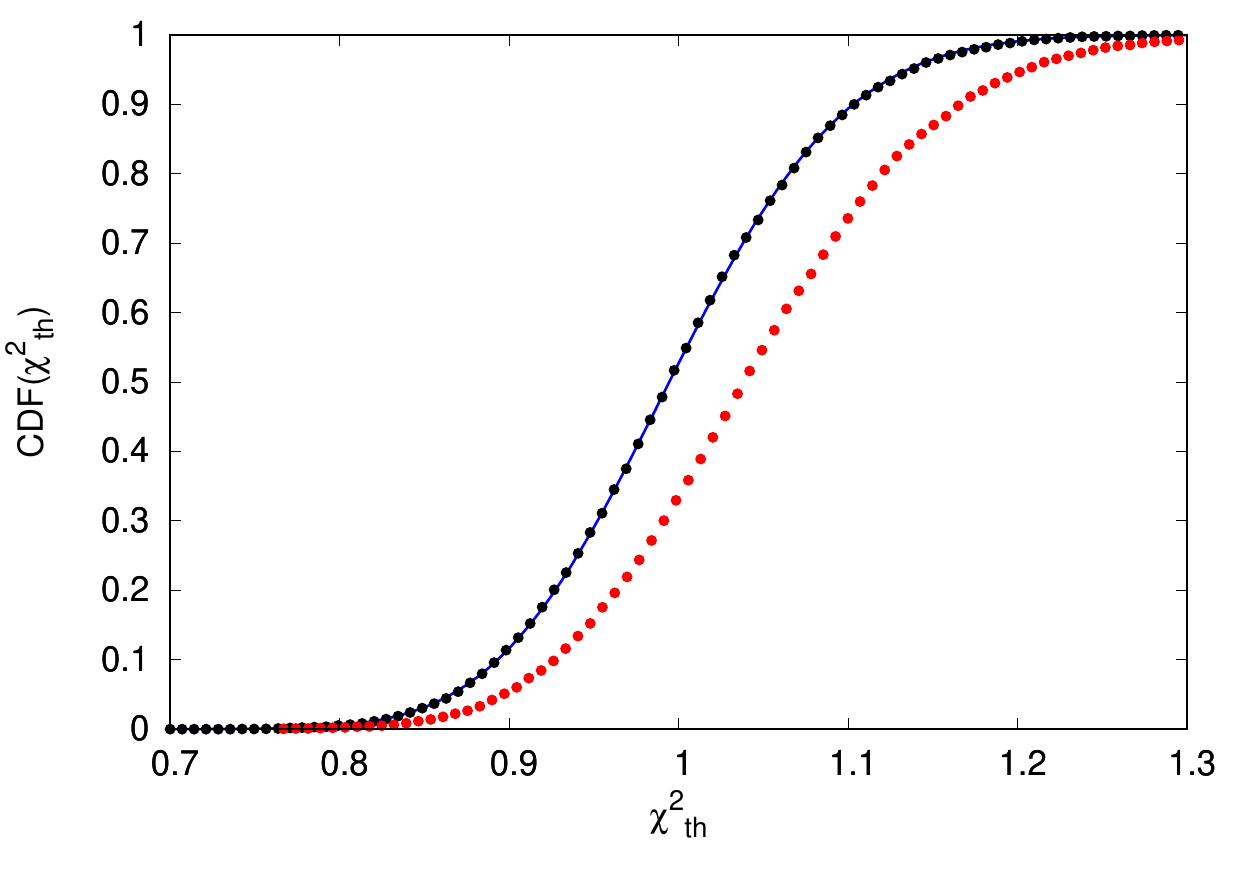}
  \end{subfigure} 
  \begin{subfigure}[b]{0.45\textwidth}
\includegraphics[scale=.6]{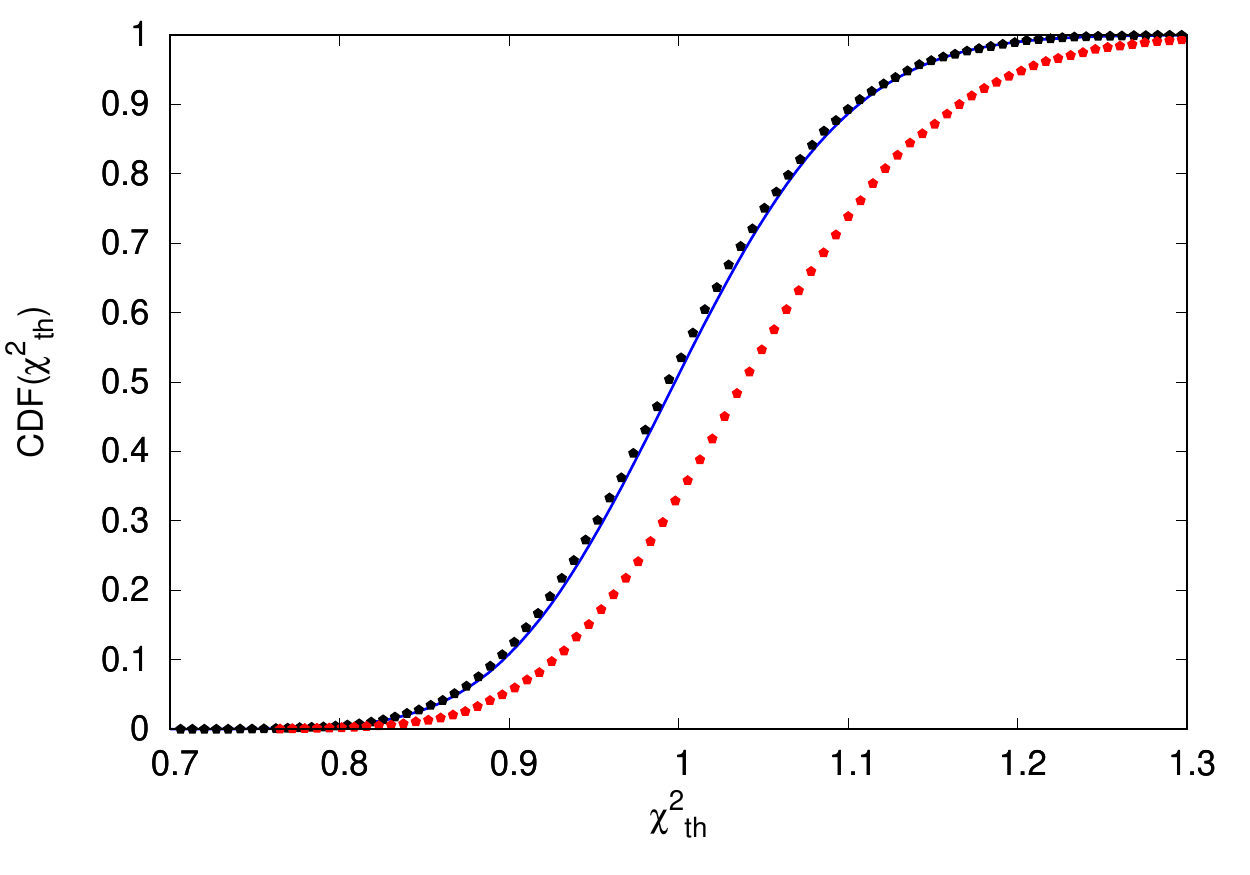}
  \end{subfigure} 
\caption{CDFs for the $\hat\chi^2_r$ parameter in the Fit$_{3p}$ (left panel) and Fit$_{2p+1f}$ configurations (right panel), when systematic uncertainties are included (red points) or discarded (black points). The solid blue line is  the CDF of the reduced $\chi^2$-distribution.\label{fig:cdf1}}
	\end{figure}
	
		\begin{figure}[h]
	  \begin{subfigure}[b]{0.45\textwidth}
\includegraphics[scale=.6]{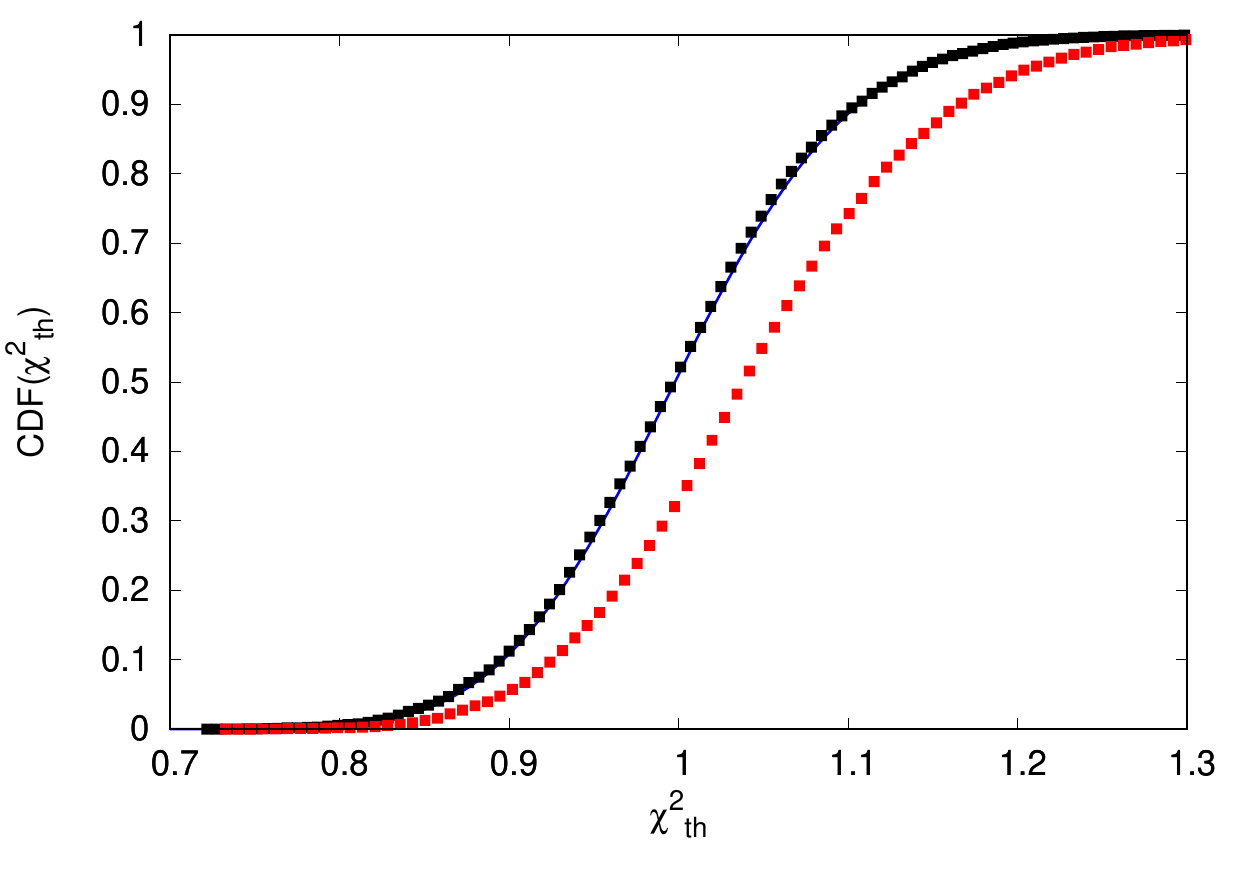}
  \end{subfigure} 
  \begin{subfigure}[b]{0.45\textwidth}
\includegraphics[scale=.6]{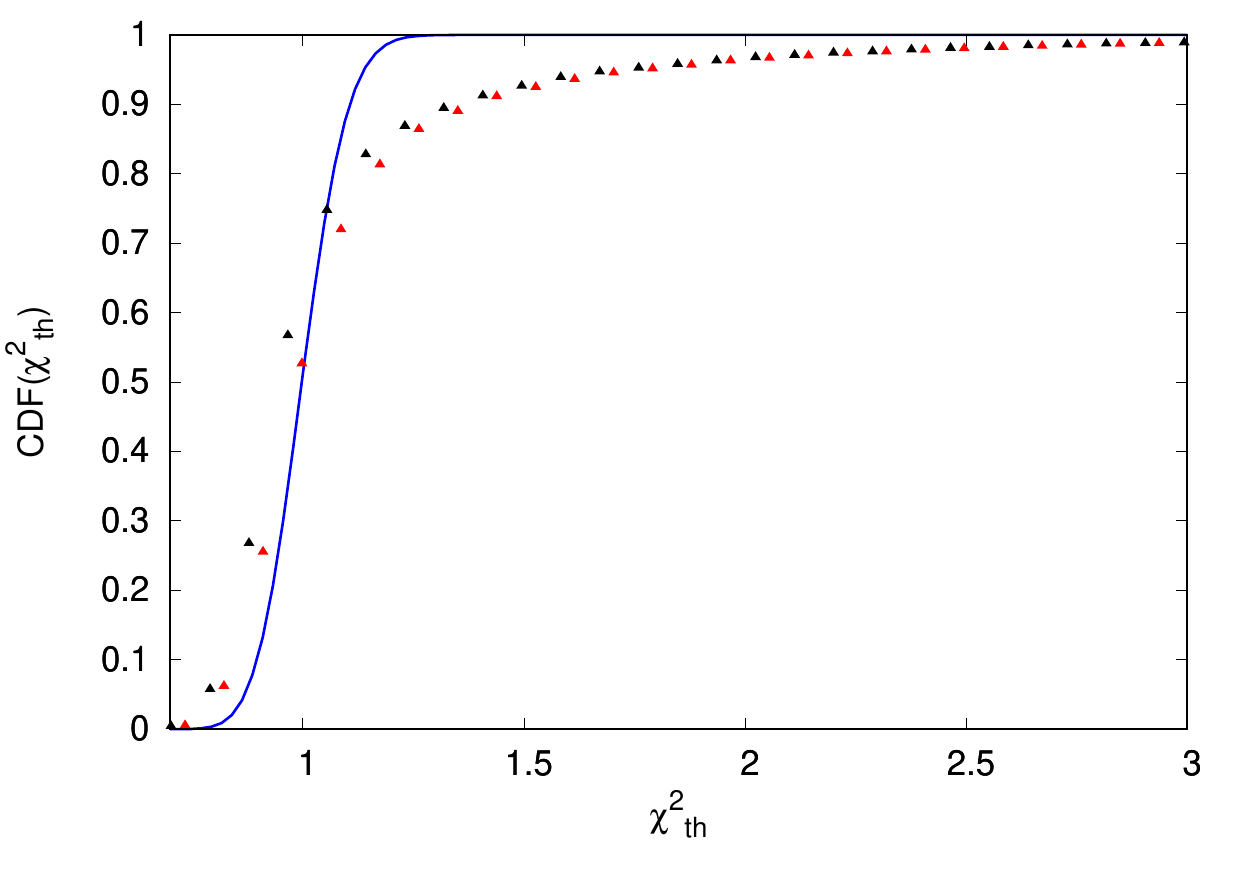}
  \end{subfigure} 
  \caption{CDFs for the $\hat\chi^2_r$ parameter in the Fit$_{2p+1s}$ ($3\%$) (left panel) and Fit$_{2p+1s}$ ($20\%$) (right panel) configurations, when systematic uncertainties are included (red points) or discarded (black points). The solid blue line is a the CDF of a reduced $\chi^2$ distribution. \label{fig:cdf3}}
	        \end{figure}

  We finally notice that in the HBM framework there is no need to
  choose between this and Fit(')$_{3p}$ condition. Any prior knowledge of $\mu$,
  such as the one incorporated in Fit(')$_{2p+1s}$, can
  be added to the Bayesian prior function $\pi$ defined in \eqr{eq:likebayes}
  and the final fit results will account for this additional information.

	\subsection{An additional complication: data with a systematic offset}
	\label{sec:profile}

 We showed in the previous sections how to deal with the systematic uncertainties associated to the
 experimental data. However, we implicitly assumed that the data themselves are not affected by any intrinsic offset.
 We now make a step further and outline a strategy that allow us to deal with a 
   priori unknown systematic offset of the data themselves.

 Within our toy model,
 we assume that each data set has an unknown multiplicative offset  $\delta_k^*$
 lying inside the estimated uncertainty interval  $[-\Delta_k,\Delta_k]$. 
 We then fix $\delta_1^* = 3\%$, $\delta_2^* = 4\%$ and $\delta_3^* = -2\%$
 and artificially rescale all the points of the $k^{th}$ subset ($k=1,2,3)$
 according to $E_i^* = (1+\delta^*)E_i$ and $\sigma_i^*= (1+\delta^*)\sigma_i$.

   If now we apply the bootstrap method in the Fit$_{3p}$ and Fit$_{3p}^\prime$ conditions, we obtain, respectively:
	\bea
\text{Fit$_{3p}$: } I &=& 250.7^{+1.4}_{-1.5}, \quad \mu = (1.9 \pm 7.4)\cdot 10^{-3}, \quad \Gamma =(9.8 \pm 0.1)\cdot 10^{-1}, \quad \hat \chi_r^2 = 1.07, \quad \text{$p$-value} = 19\% ,\nn \\
\text{Fit$_{3p}^\prime$: } I &=& 250.7 \pm 4.0, \quad \mu = (1.9^{+7.2}_{-7.6})\cdot 10^{-3}, \quad \Gamma =(9.8 \pm 0.1)\cdot 10^{-1}, \quad \hat \chi_r^2 = 1.07, \quad \text{$p$-value} = 30\%.
	\eea

Both the central values and the uncertainty intervals of the fit parameters are basically unchanged from
the previous results given in \tref{tab:res}, but the $\hat\chi_r^2$ value is now larger than before. Also the expected goodness-of-fit distribution, when systematic uncertainties are not taken into
account, is different form the reduced $\chi^2$-distribution. This effect is already visible in \fr{fig:bias} and, as expected, the discrepancy increases when $|\delta_k^*|$ increases.

	\begin{figure}[h]
\includegraphics[scale=.7]{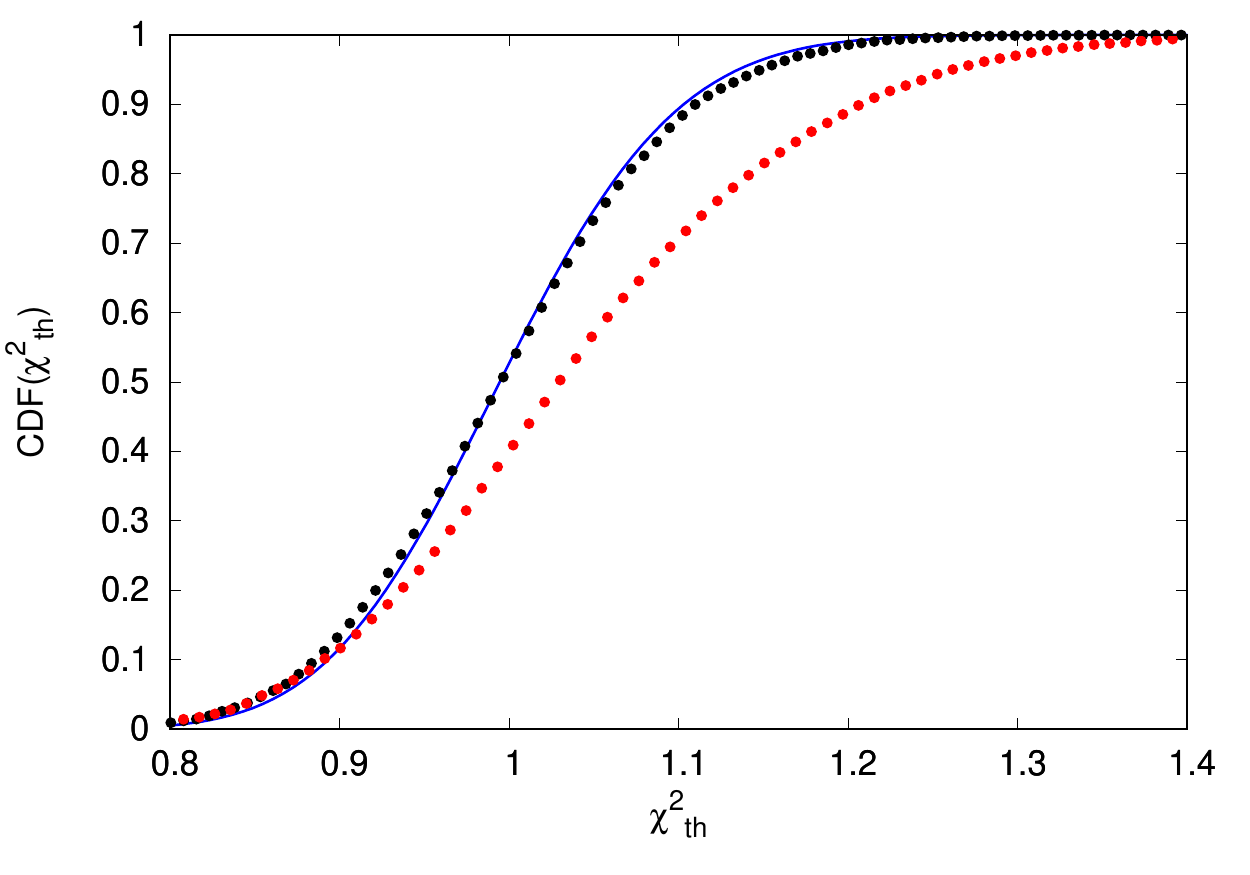}
\caption{CDFs for the $\hat\chi^2_r$ parameter in the Fit$_{3p}^\prime$  (red points) and Fit$_{3p}$ (black points) configurations. The solid blue line is a the CDF of a reduced $\chi^2$ distribution.\label{fig:bias}}
	\end{figure}

If the model $T$ correctly reproduces the data,
 we can use the bootstrap framework also to estimate the unknown data offset with the following procedure:

\benum 
 \item   
   apply the bootstrap fit in the Fit$_{3p}$ condition and allow the parameter $\delta_{ij}$ to span a range wider than $[-\Delta_k,\Delta_k]$ only for one chosen set\footnote{This wider interval allows to deal with the case of a systematic offset larger than the quoted systematic uncertainty interval.
     In our case, we fix $\delta_{ij} \in \mc U [-1/2,1/2]$.} and assuming that all other data sets do have any systematic offset. This choice is (at least partially) justified when several different and independent subsets have
 to be taken into account, since, in this case, the overall effect of the different biases should be 
 small due to compensation effects.
 
 \item
 the study of the behavior of the ${\hat\chi^2_{b,j}}$ parameter as a function of
 $\delta_{ij}$ allows one to find the  value $\tilde\delta$
 that gives the minimum value of $\mathbb E\left[\hat\chi^2_{b,j}(\delta_{ij})\right]$.
 Such a value can be taken as an estimate of the true, unknown data offset.
 
\item repeat the previous steps for each single subset to empirically evaluate all their different
  offsets.
  \ennum

The results of this strategy, where we choose to consider the offset only on set 1, are shown
in \fr{fig:prof_set1} and \tref{tab:prof}, from which we can see that $\tilde\delta = -3.1 \% \simeq -\delta_1^*$.
By generating a sufficiently high number of points, the
intrinsic error
 on $\tilde\delta$ can be made arbitrarily
small (in our case it is  of the order of $10^{-7}$).

This result means that if we want to force set 1 to be in good agreement with set 2 and set 3, we need to shift all its points back to their starting values, \ie rescaling them
by a factor $-\delta_1^*$.

This procedure is equivalent to use \eqr{eq:chimodbias} and
considering only the set 1 as affected by systematic uncertainties, \ie
	\beq
\chi^2_{mod,1} = \chimodone,
	\enq
 thus obtaining $1-f_1 = (2.60 \pm 1.16) \% \simeq \delta_1^*$, as expected.

	\begin{figure}[h]
  \begin{subfigure}[b]{0.45\textwidth}
  \includegraphics[scale=.6]{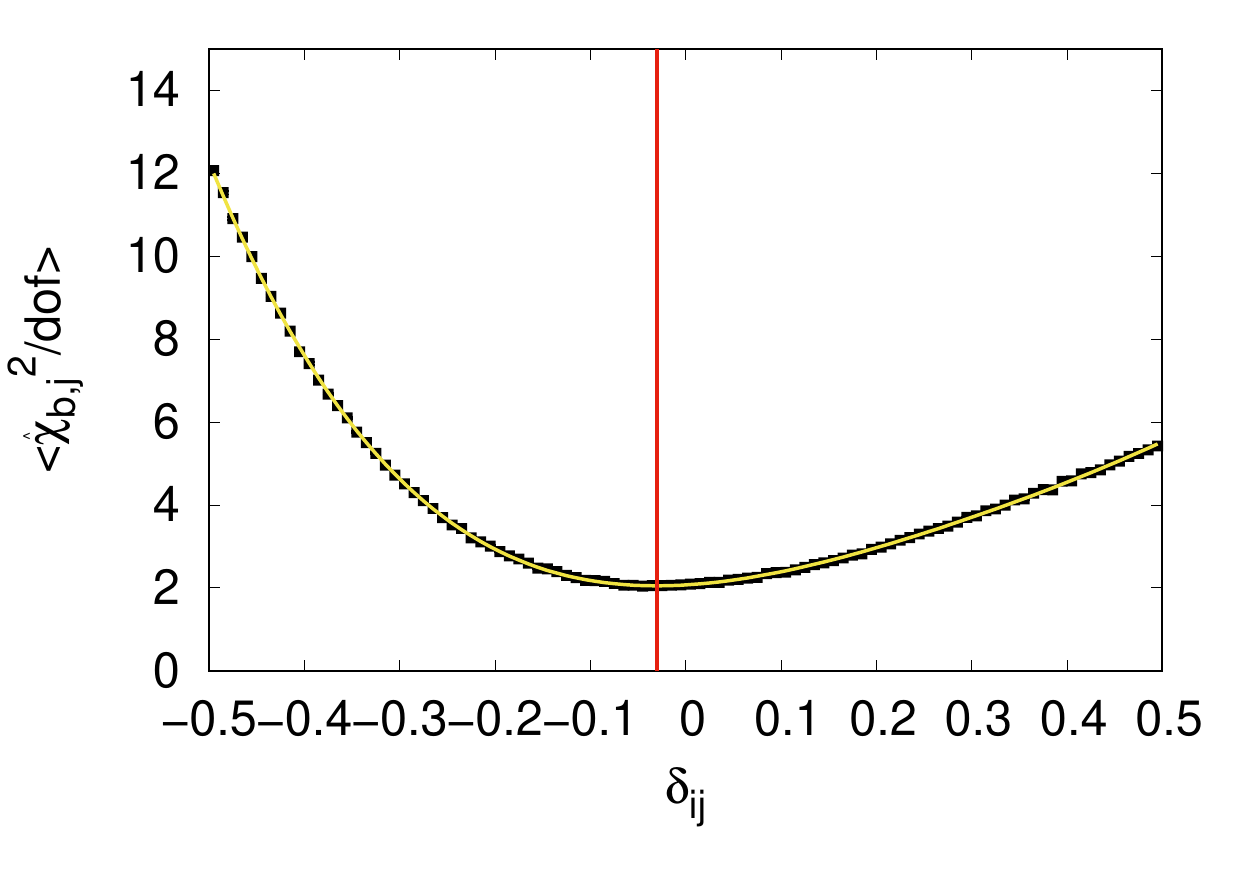}
  \subcaption{ \hspace{-0.8 truecm} }
  \label{fig:prof_set1}
  \end{subfigure}
  \begin{subfigure}[b]{0.45\textwidth}
  \includegraphics[scale=.6]{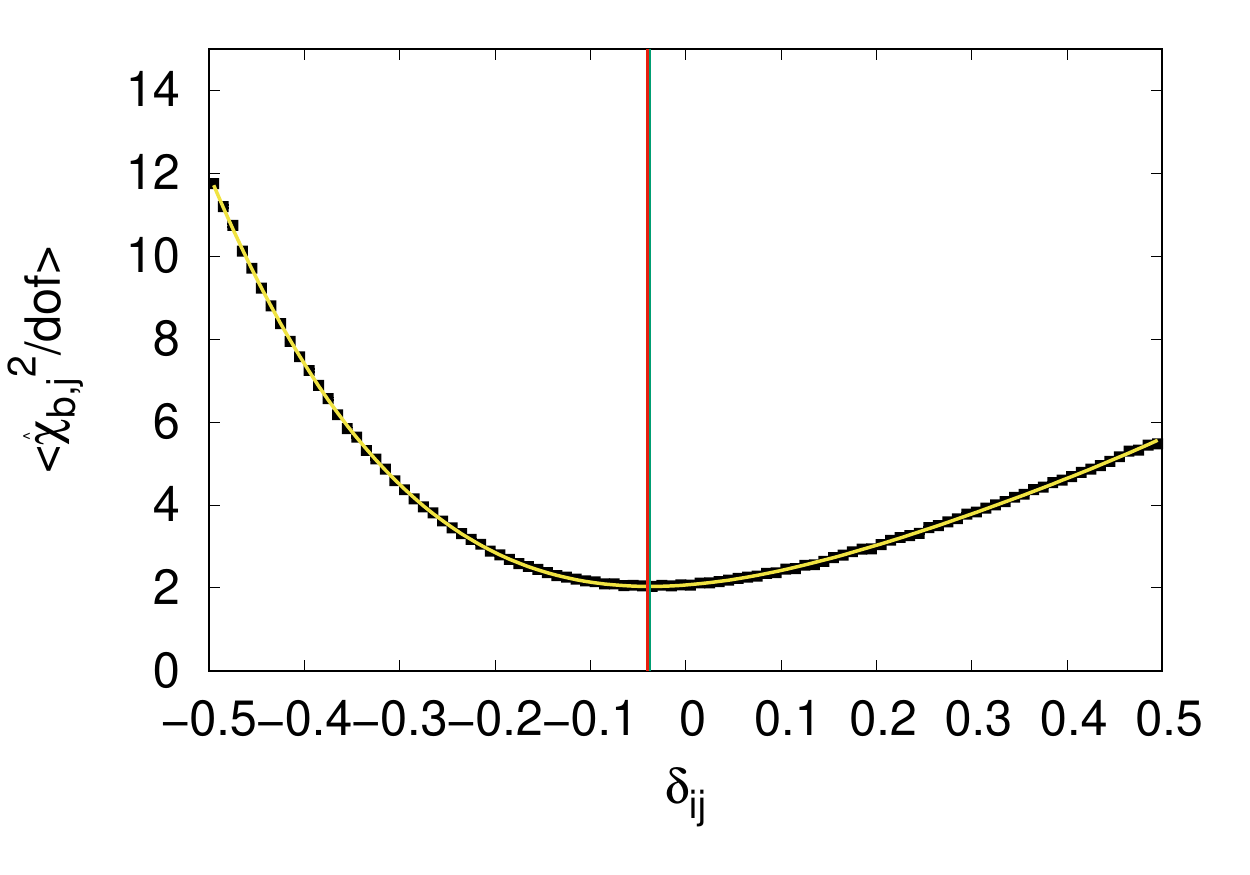}
  \subcaption{ \hspace{-1 truecm} 
  \label{fig:prof_set2}
  }
  \end{subfigure}\\
  \begin{subfigure}[b]{0.45\textwidth}
  \includegraphics[scale=.6]{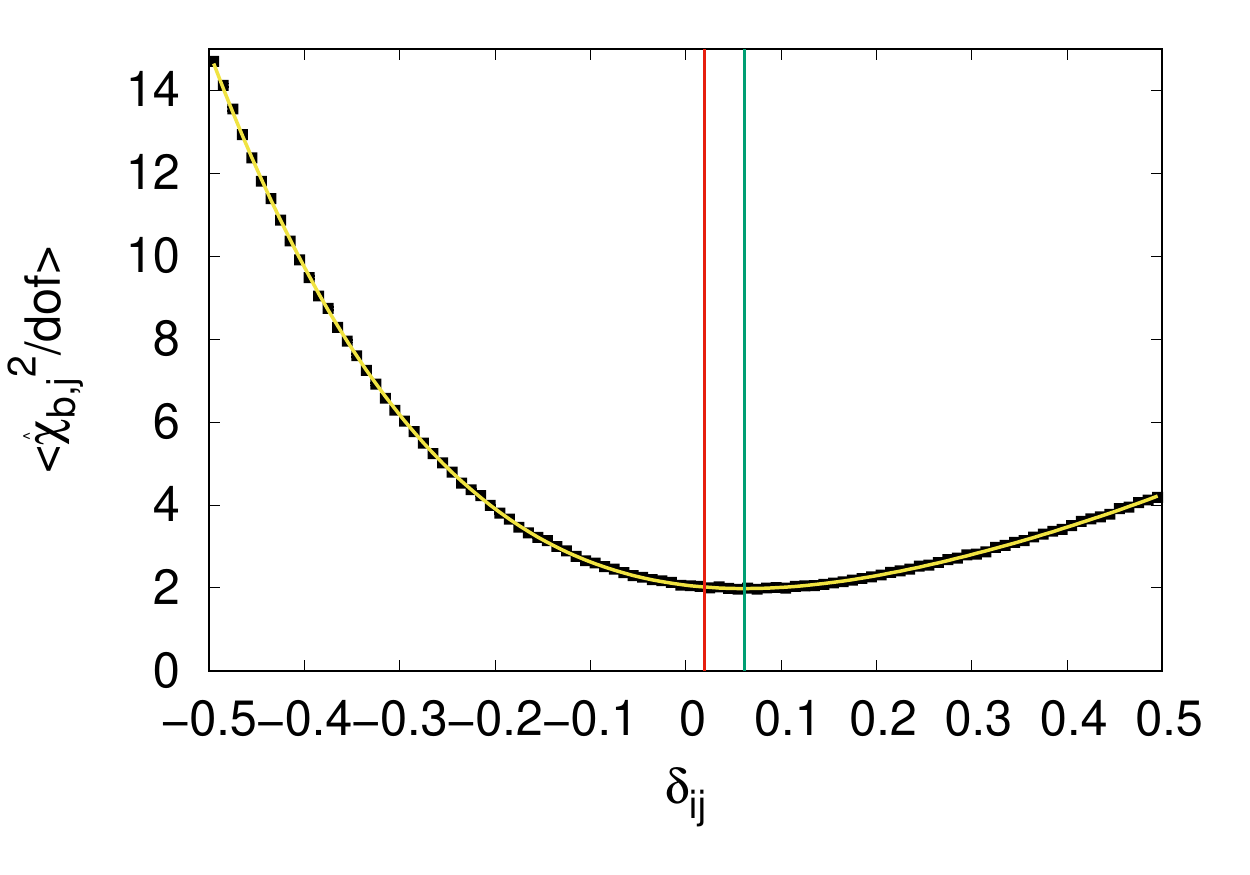}
  \subcaption{ \hspace{-0.8 truecm} }
  \label{fig:prof_set3}
  \end{subfigure}

  \centering
  \begin{subfigure}[b]{0.45\textwidth}

  \end{subfigure}
  \caption{The $\mathbb E\left[{\hat\chi^2_{b,j}} \right]$ value as a function of
  $\delta_{ij}$. The black points are from the bootstrap technique, fitted with a fourth-order polynomial fit (yellow line).
  In each plot, the
  $\tilde\delta$ value is represented by the vertical green line, compared to the known $-\delta^*$
  value (red line). The numerical results are given for subset 1 (a), 2 (b) and 3 (c).\label{fig:prof}}
	\end{figure}

 Applying this strategy to subsets 2 and 3, we obtain  the results shown in  \fr{fig:prof_set2} and  \fr{fig:prof_set3}
 and \tref{tab:prof}.
 They are compared to the results of the $\chi^2_{mod}$ approach  both when the $f_k$ parameters are fitted one by one and when they are all fitted simultaneously (see \eqr{eq:chimodbias}).
Even if the $\tilde\delta_k$ parameters
      are not directly fitted in our procedure,
   their uncertainty intervals
   can be assigned using the so-called MINOS method (\cite{James:2006zz}), \ie
   by finding  the values of $\tilde\delta_k$ that cause
   $\mathbb E\left[{\hat\chi^2_{b,j}} \right]$
   to vary by one unit. 
  The numerical values of all these intervals are found to be
  coincident with the uncertainties resulting from the $\chi^2_{mod}$ approach.

 From all these results, we can see that:

  \beitem
  \item[*]  the offsets of subsets 1 and 2 are well determined by this strategy, while there is a significant discrepancy between $\tilde\delta_3$ and $-\delta_3^*$.
  This disagreement is due to the fact that, as previously noted, when we estimate  $\delta_3^*$, we are implicitly assuming that the other two subsets are not affected by any systematics, while they are both rescaled by the positive parameters
 $\delta_1^*$ and $\delta_2^*$ respectively.
 On the other hand, when we try to estimate $\delta_1^*$ or $\delta_2^*$, the other two subsets are rescaled according to systematics of different signs, thus introducing a compensation that allows us to (almost) correctly determine their value.
 
 \item[*] the fit parameters obtained with the $\chi^2_{mod}$ procedure, \ie
	\beq
I = 247.7 \pm 3.6 , \quad \mu = (1.7 \pm 7.5)\cdot 10^{-3}, \quad \Gamma = 0.98\pm0.01, \quad \hat\chi_{mod,r}^2 = 1.01,
	\enq
 are, within their estimated uncertainties, in agreement with the Fit$_{3p}^\prime$ results shown in \tref{tab:res}. However, we are not able to give a reliable $p$-value, since, as previously mentioned, the correlations among the
 data of each subset induced by the systematic uncertainties give a goodness-of-fit distribution
 different from the reduced $\chi^2$-function. 
\enitem
 
	\betab
\begin{tabular}{|c|c|c|c|c|}
\hline
 set number & known sys. & bootstrap & $\chi_{mod,r}^2$ (one-by-one) &  $\chi_{mod,r}^2$ (simultaneous)  \\
\hline
  & $\delta_k^*$  ($\%$) & $-\tilde\delta_k$ ($\%$) & $1-f_k $ ($\%$) &  $1- f_k $ ($\%$) \\
\hline
\hline
1 & $3.0$  & $3.1$   & $2.2 \pm 1.1$ & $2.6 \pm1.5$\\   
2 & $4.0$  & $3.8$   & $3.2 \pm  1.1$ & $3.3 \pm 1.6$\\  
3  & $-2.0$  & $-6.2$   & $ -4.1\pm 1.0$ & $-2.1 \pm 1.4$ \\  
\hline
\end{tabular}
\caption{Estimated offset values for each subset: results from the bootstrap (third column) and from the $\chi_{mod}^2$ method
  when the normalization factors $f_k$ are fitted one-by one (fourth column) or simultaneously (last column).}\label{tab:prof}
\entab

The bootstrap-based estimates of the real systematic offsets can be included into the fitting procedure in several ways.
For instance, two alternative approaches are:
	\benum
\item rescale all the data points and their statistical uncertainties by a factor $(1+\tilde\delta_k)$ and perform a single minimization with the standard $\chi^2$ procedure. We then obtain:
\beq
I = 250.1 \pm 1.5 , \quad \mu = (1.4 \pm 7.5)\cdot 10^{-3}, \quad \Gamma =(9.8\pm0.1) \cdot 10^{-1}, \quad \hat\chi_{mod,r}^2 = 1.01, \quad \text{$p$-value } =  46\%;
\enq
\item rescale the experimental data as described in the previous point and then apply the bootstrap fitting technique,
  setting $\delta_{ij} = 0$ in \eqr{eq:sam}. We now get:
\beq
I = 250.1\pm1.5 , \quad \mu = (1.6 \pm 7.4)\cdot 10^{-3}, \quad \Gamma = (9.8\pm0.1) \cdot 10^{-1}, \quad \hat\chi_{mod,r}^2 = 1.01, \quad \text{$p$-value } =45 \%.	
\enq
\ennum
The results obtained in both the previous cases are very close, as expected, to the ones previously obtained in the Fit$_{3p}$ condition.
%
%

\section{A simplistic model with asymmetric statistical uncertainties}
\label{sec:skewgaus}

\subsection{Implementation}

To provide an additional check of our new technique, we also
apply our method to the case of asymmetric statistical uncertainties.
This situation, that could be for instance due to a non-uniform background
subtraction, is hardly treatable within the standard $\chi^2$ procedure.

To apply our method, we assume to have an experimental uncertainty distribution $f(x)$ described by the skew-Gaussian distribution 
(see, for instance, \cite{ref:azza1,ref:azza2}):

\beq
f(x) = \left({2}/{\omega}\right)  g(z)G(\lambda z) \quad ; \quad
G(\lambda z) =\int_{-\infty}^{\lambda z} g(t)\, dt \ ,
\enq

where:
\beq
z = \f{x-\xi}{\omega} \quad ; \quad g(z) \sim \mc N[0, 1]\ .
\enq

This function generalizes the usual Gaussian distribution
to accomodate  a certain amount of skewness. 
It is specified by 3 real-valued parameters: location($\xi$), scale ($\omega>0$)
and shape ($\lambda$) with the usual Gaussian distribution corresponding
to ($\lambda$=0).
For each experimental point $E_i$,
the $\lambda$, $\xi$ and $\omega$ parameters were choosen to be;

\beq\label{eq:skewpar}
\lambda=-3 \quad ; \quad
\xi = E_i -\omega \delta \sqrt{\frac{2}{\pi}} \quad , \quad 
\omega = \sigma_i  \sqrt{1-\frac{2d}{\pi}}\ ,
\enq

where $ d = {\lambda}/{\sqrt{1+\lambda^2}}$. In this way
we obtain\footnote{As discussed in~\cite{ref:dago2, ref:dago},
  the best way to present experimental results with asymmetric uncertainties
  is to give mean value and standard deviation.}:

\beq
\mathbb E\left[ x \right] = E_i \quad ; \quad
\text{Var}\left[ x \right] = \left( \sigma_i^{\text{stat}}\right)^2 \ .
\enq

As an example, the skew-Gaussian distribution having zero mean,
unit variance and
$\lambda=-3$ is shown in \fr{fig:skewgaus} and compared to $\mc N[0, 1]$.
\begin{figure}[h]
\includegraphics[scale=.9]{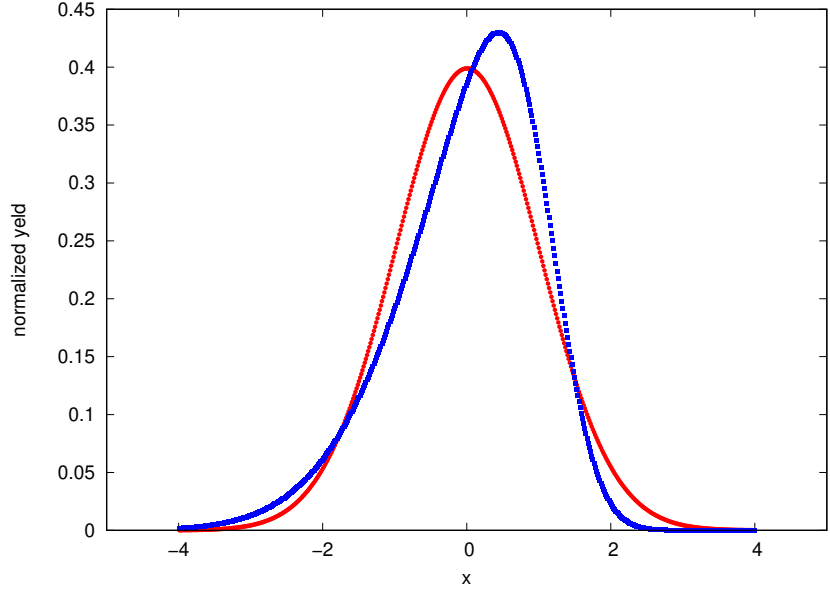}
\caption{The probability density function of the  skew-Gaussian distribution having zero mean, unit variance and $\lambda=-3$ (blue line) is compared to $\mc N[0, 1]$ (red line).
\label{fig:skewgaus}
}
	  \end{figure}

This particular functional form was chosen since, even if it is asymmetric,
we have:
\beq\label{eq:symmskew}
\left( \f{x-\xi}{\omega} \right)^2 \sim \chi^2 \ ,
\enq
independent of the value of $\lambda$.
To a good approximation, the same relation 
also holds for $[({x-E_i})/{\sigma_i}]^2$ because, in our case
$E_i \simeq \xi $ and $\sigma_i \simeq \omega$ (see \eqr{eq:skewpar}).
This property then allows us to cross-check and validate the results
we will obtain in this case with the ones found in the previous section.

\subsection{A generalised bootstrap formalism for \texorpdfstring{$\hat\chi_{b,j}^2$}. }

  In the previous case, the bootstrap sampling can be written as
  (see \eqrs{eq:sam} and \eref{def}):

\beql{eq:samskew}
\uB_{ij}= \samskew, 
\enq

where $s_{ij}$ is distributed according to a skew-Gaussian function having
mean zero and variance $(\sigma^{exp}_i)^2$. Using this notation, we can
basically adopt almost the same decomposition shown in \secr{sec:chi2dec}.
After introducing:

\beq
\rho_{ij}\equiv \f{s_{ij}}{\sigma_i}\ , 
\enq
 we only need to rewrite \eqr{eq:chi2dec} as:
	\bea
\tilde D_{ij} &\equiv& 2\left[\epsilon_{ij}\rho_{ij} +\f{1}{\sigma_i}(\epsilon_{ij}+\rho_{ij})(E_i -T_i(\bm\psi, \hat{\bm\theta}))\right],\nn\\
\tilde\Phi_{ij} &\equiv& \eta_{ij}^2 +  2\eta_{ij}\left[(\epsilon_{ij}+\rho_{ij}) +\f{1}{\sigma_i}(E_i -T_i(\bm\psi, \hat{\bm\theta}))\right] ,
\eea

where all the other parameters are defined as previously.
After this small modification, the decomposition of $\hat\chi_{b,j}^2$ can be written in the same way as in \eqr{eq:chi2exp}, \ie
	\beql{eq:chi2skew}
{\hat\chi^2_{b,j}} = \hat \chi^2 + \sum_i \rho_{ij}^2 + \sum_i \epsilon_{ij}^2 + \sum_i \tilde D_{ij} + \sum_i \tilde\Phi_{ij}. 
\enq
A very similar decomposition also holds for ${\hat\chi^2_{th,j}}$:

\beq
    {\hat\chi^2_{th,j}} = \hat \chi^2_{th} + \sum_i \rho_{ij}^{\prime 2} + \sum_i
    \epsilon_{ij}^{\prime 2} + \sum_i \tilde D'_{ij} + \sum_i \tilde\Phi'_{ij}. 
\enq  
  
thus allowing to rewrite all the components of \eqr{eq:chi2note} in the case of an asymmetric statistical error, with an obvious meaning
for the $\tilde\Phi^\prime_r$, $\tilde D^\prime_r$ and $\rho_r^\prime$ parameters.

In this way, our bootstrap formalism can then be easily adapted to deal with any
uncertainty distribution.

\subsection{Results}

The results of the fit performed under  the Fit(') $_{3p}$ condition
are displayed in \tref{tab:resskew} and the correponding distributions
for ${\hat\chi^2_{th,j}}$ and the  the CDFs of the expected goodness-of-fit
distributions are shown in \frs{fig:skew01} and and \fref{skew02},
respectively.
As expected, due to the symmetry properties of the skew-Gaussian function
(see \eqr{eq:symmskew} and comments to it), all these results
results are basically coincident with the ones shown in
\tref{tab:res}, \frs{fig:chi2dec1} and \fref{cdf1} (left plot).

\begin{table}[]
\begin{tabular}{|c|c|c|c|c|c|}
\hline
\multicolumn{6}{|c|}{ DATA}  \\
\hline
\hline
 Fitting conditions &  $I$ & $\mu$ $(10^{-3})$  &  $\Gamma$ ($10^{-1}$) & $\hat\chi^2_r$ & $p$-value  \\  
 \hline
  Fit$_{3p}$ & $ {247.0}^{+1.4}_{-1.5} $  & ${1.8}^{+7.1}_{-7.6} $  & $ {9.8} \pm 0.1$  & $ 0.98 $  & $ 45 \% $ \\
  Fit$_{3p}^\prime$ & $ {247.0}^{+4.0}_{-4.2} $  & ${1.8}^{+7.1}_{-7.7} $  & $ {9.8} \pm 0.1 $  & $ 0.98 $  & $ 35 \% $    \\
  \hline
\end{tabular}
\caption{Results from the fit with skew-Gaussian statistical uncertainties
 in the Fit$_{3p}$ and Fit$_{3p}^\prime$  configurations.}\label{tab:resskew}
\end{table}

\begin{figure}[h]
\includegraphics[scale=0.9]{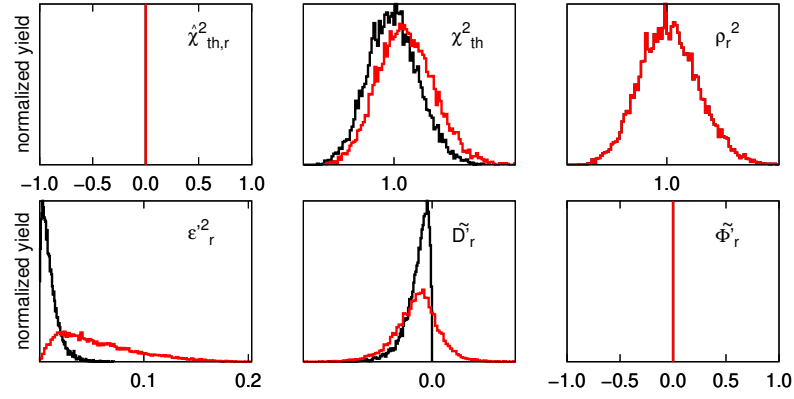}
\caption{
  Decomposition of the $\hat\chi^2_{th,j}$
  parameter with skew-Gaussian statistical uncertainties and in the Fit$_{3p}$ configuration when systematic uncertainties are excluded (black curves) and included (red curves).
  Upper panel (from left to right): $\hat\chi^2_{th,r}$, $\chi^2_{th}$ and $\rho^2_r$ components. Lower panel (from left to right): ${\epsilon^\prime}^2_r$,  $\tilde D^\prime_r$  and $\tilde \Phi^\prime_r$ components.
  See text
  for the notation. 
Black and red lines exactly overlap for the constant $\hat\chi^2_{th,r}$
  and  $\tilde\Phi^\prime_r$ components.
\label{fig:skew01}
}
	\end{figure}

	\begin{figure}[h]
\includegraphics[scale=.7]{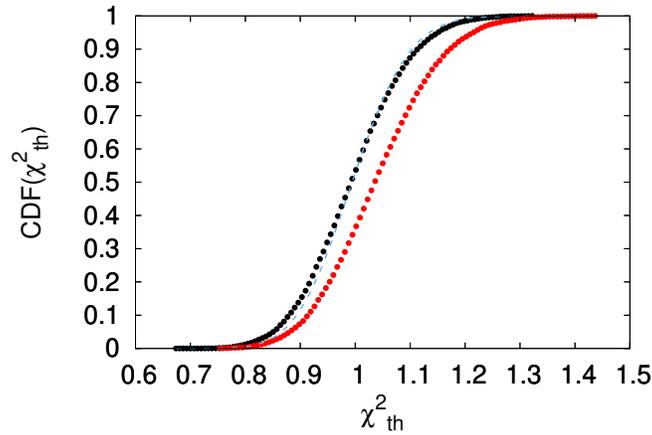}
\caption{CDFs for the $\hat\chi^2_r$ parameter in the Fit$_{3p}^\prime$  (red points) and Fit$_{3p}$ (black points) configurations with skew-Gaussian statistical uncertainties. The solid blue line is a the CDF of a reduced $\chi^2$ distribution.\label{fig:skew02}}
	\end{figure}

This very good agreement gives us confidence in the capability of our method
to also correcty deal with asymmetric distributions.


        
	\section{An application of the method: fit of real Compton scattering data}
	\label{sec:compton}

In this section we show an application of the bootstrap-based fitting method described in this work
to an actual physics case: the extraction of the proton dipole scalar polarizabilities
from the real Compton scattering (RCS) data, using fixed-$t$ subtracted dispersion relations~\cite{Pasquini:2017ehj, Pasquini_2019}. In the RCS process, a real photon scatters off a proton, whose internal structure is probed 
when the photon energy is at least a few tens of MeV.
The RCS differential cross section can be expressed, once the scattering angle and energy are fixed, in terms of 6 
parameters, defined as 
the dipole scalar electric ($\astat$) and magnetic ($\bstat$) polarizabilities,
 and 4 vector spin-dependent polarizabilities
($\vec\gamma_s$).
For a detailed description of RCS and the dispersion relation framework, the reader is addressed to Refs.~\cite{Pasquini:2018wbl,Drechsel:1999rf,Pasquini:2007hf,Drechsel:2002ar} and references therein.
For the purposes of this work, it is sufficient to recall that the RCS differential cross section $d\sigma/d\Omega$ can be written, once the photon scattering energy ($E_\gamma$) and angle ($\bm\theta_{\text{lab}}$) are fixed, as function of these 6 parameters, \ie $d\sigma/d\Omega(\astat,\bstat,\vec\gamma_s)$.
The experimental set used for the fit is made of 150 data collected at $E_\gamma \leq 150$~MeV, and
  divided into 13 independent subsets as shown in \tref{tab:data}.

	\betab[h]
\begin{tabular}{|c|c|c|c|c|c|}
\hline
&&&&&\\
set label & Ref. &first author&points number& $\bm\theta_{\text{lab}}$ ($^\circ$) & $E_\gamma$ (MeV)\\
\hline
1 & \cite{Oxley:1958zz} & Oxley & 4 & $70-150$ &$\simeq 60$  \\
2 & \cite{Hyman:1959zz} & Hyman&12 & $50,90$ & $55-95$\\
3 & \cite{GOLDANSKY1960473} &Goldansky&5 &$75-150$ & $55-80$\\
4 & \cite{Bernardini:1960wya} & Bernardini &2 & $\simeq 135$ & $\simeq 140$  \\
5 & \cite{Pugh:1957zz} & Pugh&16 & $50-135$ & $40-120$  \\
6 & \cite{Baranov:1974ec,Baranov:1975ju} &Baranov &3 & $90,150$ & $80-110$  \\
7 & \cite{Baranov:1974ec,Baranov:1975ju} &Baranov &4 & $90,150$ & $80-110$  \\
8 & \cite{Federspiel:1991yd} & Federspiel&16 & $60,135$ & $30-90$ \\
9 & \cite{Zieger:1992jq} & Zieger&2 & $180$ & $100,130$ \\
10 & \cite{Hallin:1993ft} & Hallin&13 & $45-135$ & $130-150$ \\
11 & \cite{MacGibbon:1995in} & MacGibbon&8 & $90,135$ & $95-145$ \\
12 & \cite{MacGibbon:1995in} & MacGibbon&10 & $90,135$ & $95-145$ \\
13 & \cite{OlmosdeLeon:2001zn}& Olmos de Leon &55 & $60-155$ & $60-150$ \\
\hline
\end{tabular}
\caption{Angular and energy coverage of the available experimental data on unpolarized cross section for proton RCS
at $E_\gamma \leq 150$~MeV.\label{tab:data}}
\entab

In Ref.~\cite{Pasquini_2019}
the bootstrap-based technique outlined in this work has already been applied
to extract $\astat$ and $\bstat$
from the fit of the RCS data listed above.
In one of the cases analyzed in Ref.~\cite{Pasquini_2019}, only the difference  $(\astat-\bstat)$ between the electric and magnetic polarizabilities was left
as free parameter. The values of ($\astat+\bstat$)  and of the 
remaining 4 spin-dependent polarizabilities $\vec\gamma_s$ were taken from the existing
experimental estimates and the corresponding uncertainties were propagated
into the fit procedure according to \eqr{eq:chiboot}. This analysis gives as final result (see~\cite{Pasquini_2019}):
	\beq
\astat = (12.03^{+0.48}_{-0.54})\times 10^{-4} \text{fm}^3 , \quad \bstat = (1.77^{+0.52}_{-0.54})\times 10^{-4} \text{fm}^3, \quad
\text{$p$-value} = 12\% .
	\enq
As an additional information about these fit outcomes, we show the estimate of the offset for each subset in the bootstrap framework (\secr{sec:com_bias}) and the reconstructed goodness-of-fit distribution obtained  in the fit conditions described above both with and without the inclusion of the systematic uncertainties (\secr{sec:com_chi2}).

	\subsection{Evaluation of the experimental bias}
	\label{sec:com_bias}

 In this work, we adopt the same conditions as in Ref.~\cite{Pasquini_2019}
 and we apply the strategy discussed in \secr{sec:profile} to evaluate the offset values
 of the different subsets.
 The results of this analysis are shown in \fr{fig:realsysall1} and \fr{fig:realsysall2}.

\begin{figure}
  \centering
  \begin{subfigure}[b]{0.45\textwidth}
 \includegraphics[scale=0.6]{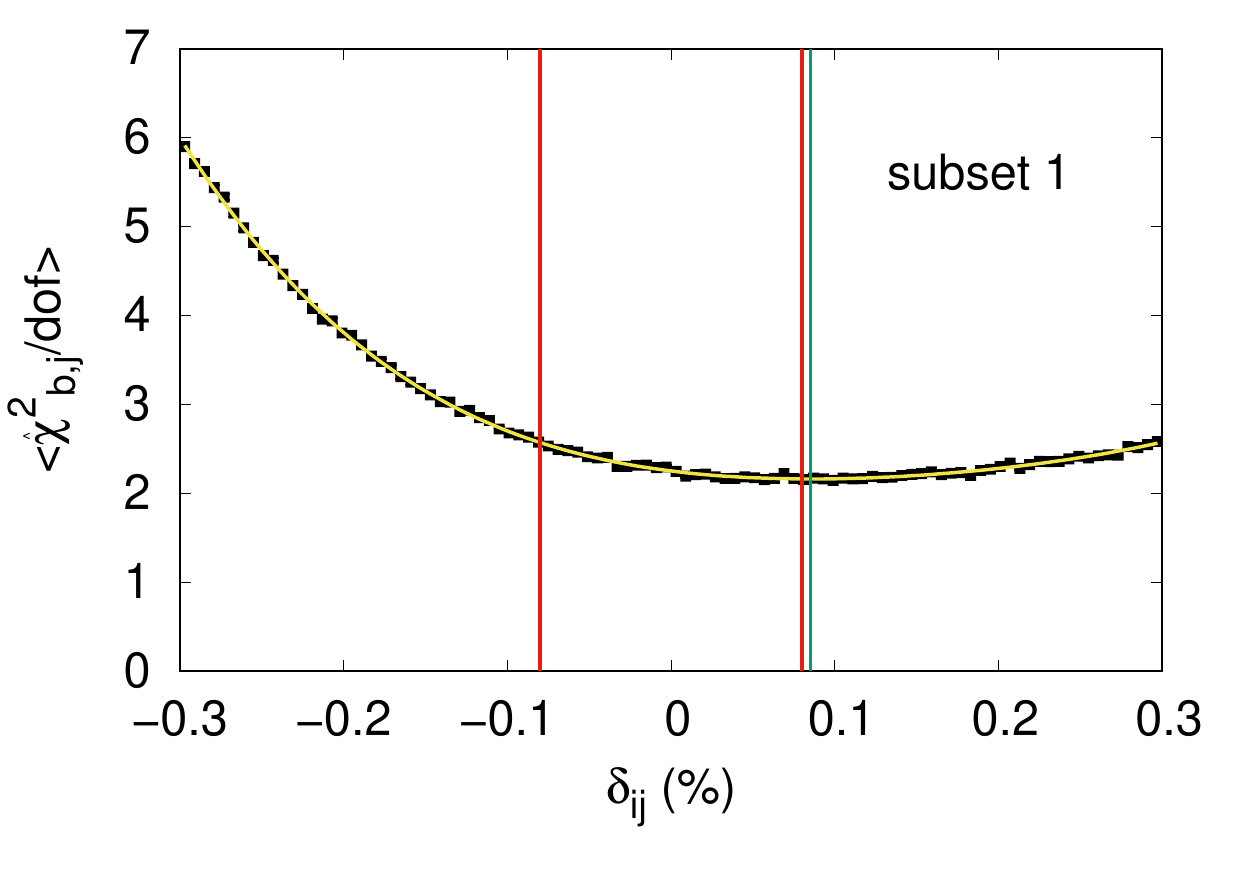}
\label{fig:realsys1}
  \end{subfigure}
  ~ 
  \begin{subfigure}[b]{0.45\textwidth}
 \includegraphics[scale=0.6]{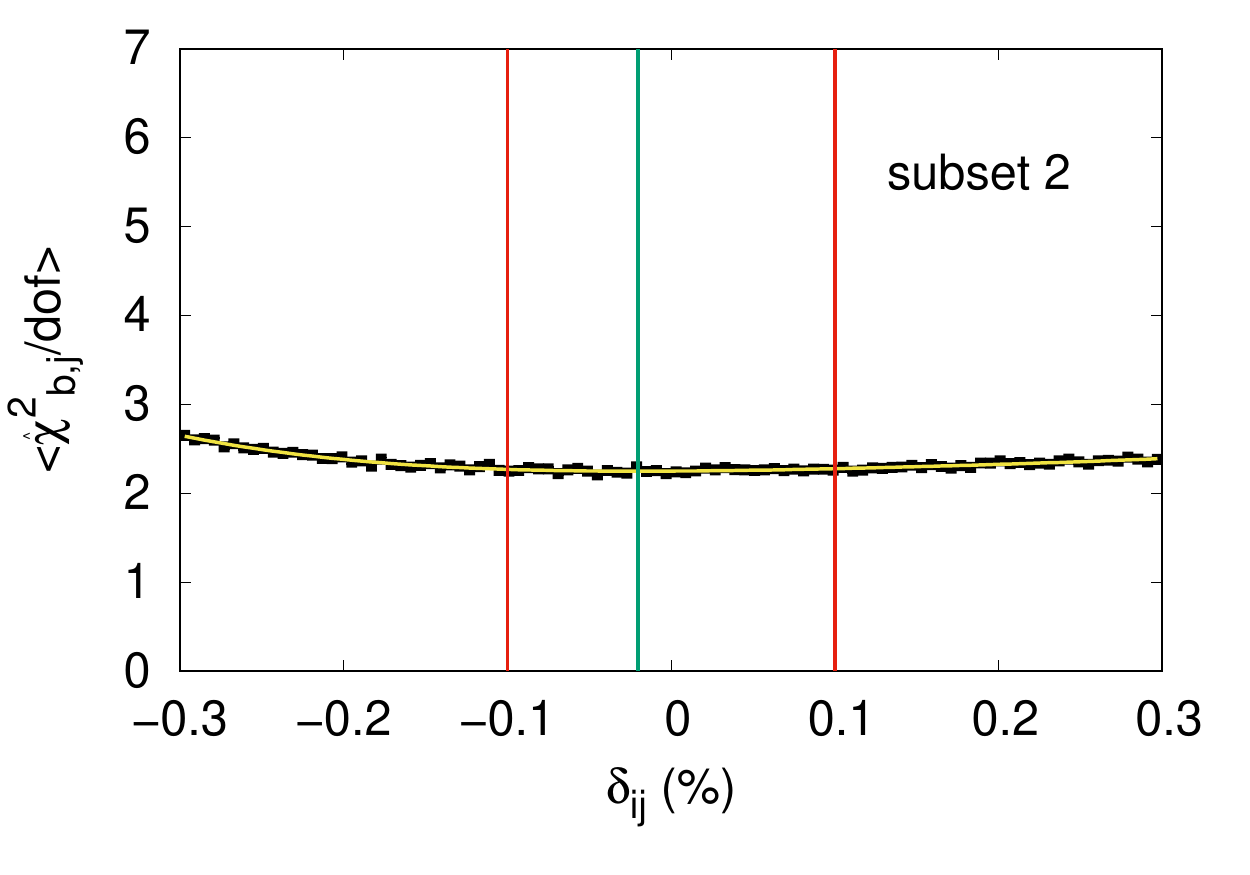}
\label{fig:realsys2}
  \end{subfigure}\\
 \begin{subfigure}[b]{0.45\textwidth}
 \includegraphics[scale=0.6]{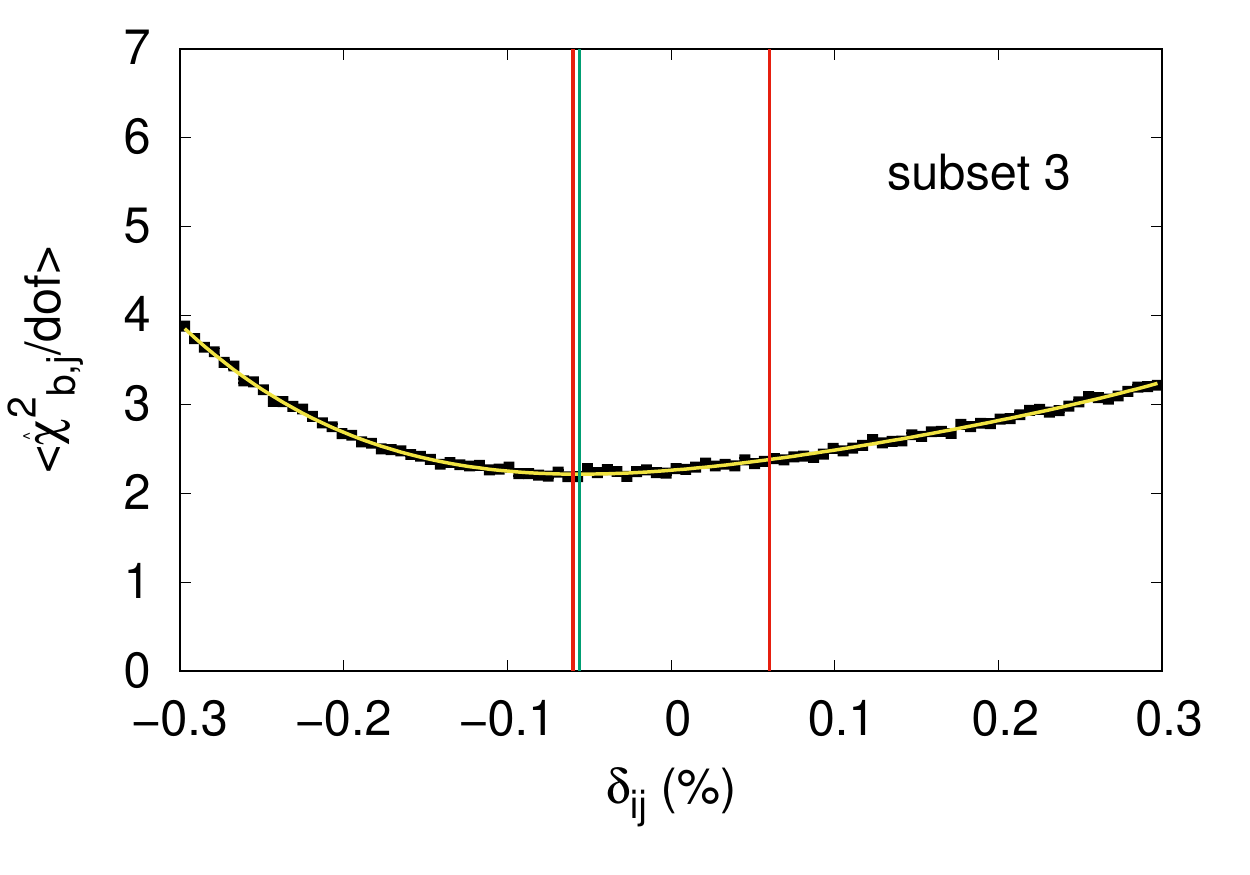}
\label{fig:realsys3}
  \end{subfigure}
  ~ 
  \begin{subfigure}[b]{0.45\textwidth}
 \includegraphics[scale=0.6]{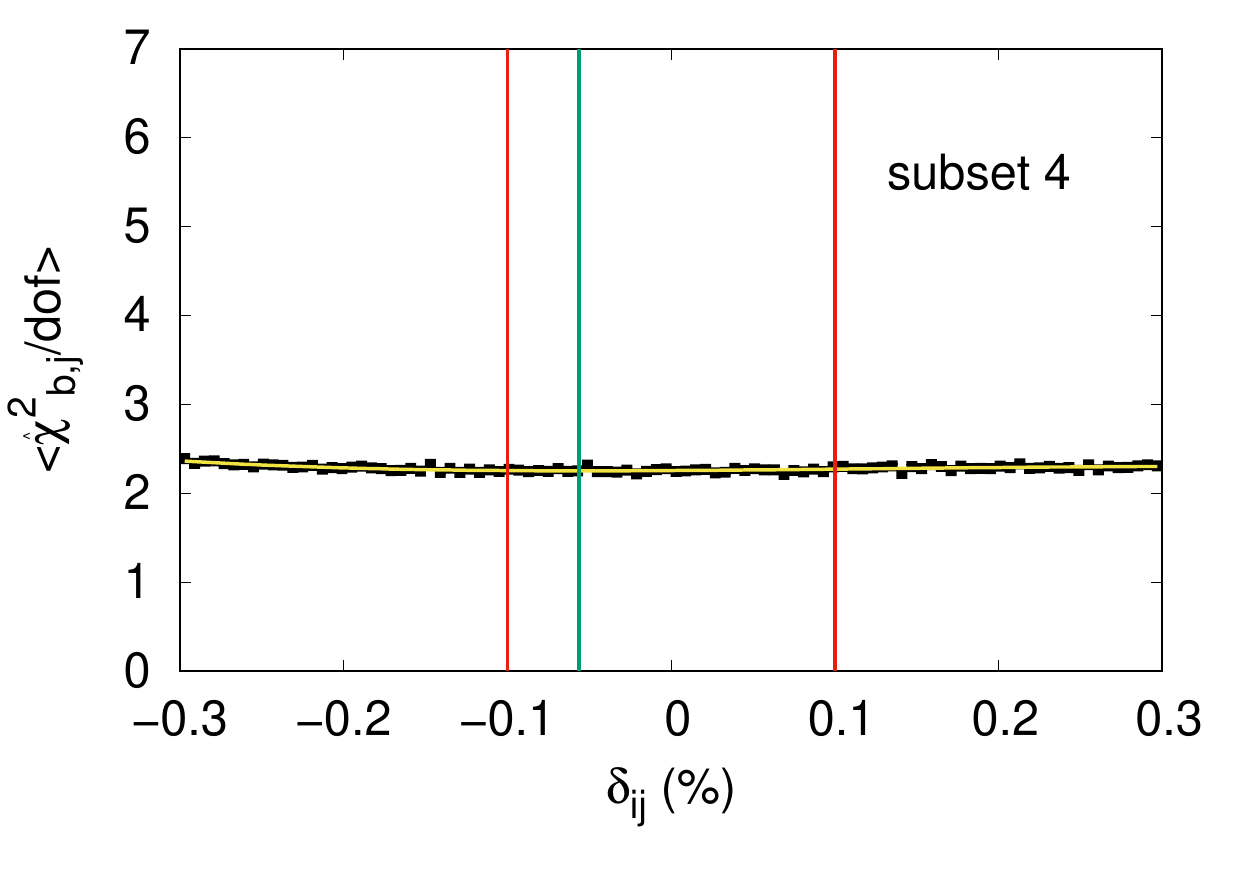}
\label{fig:realsys4}
  \end{subfigure}\\
 \begin{subfigure}[b]{0.45\textwidth}
 \includegraphics[scale=0.6]{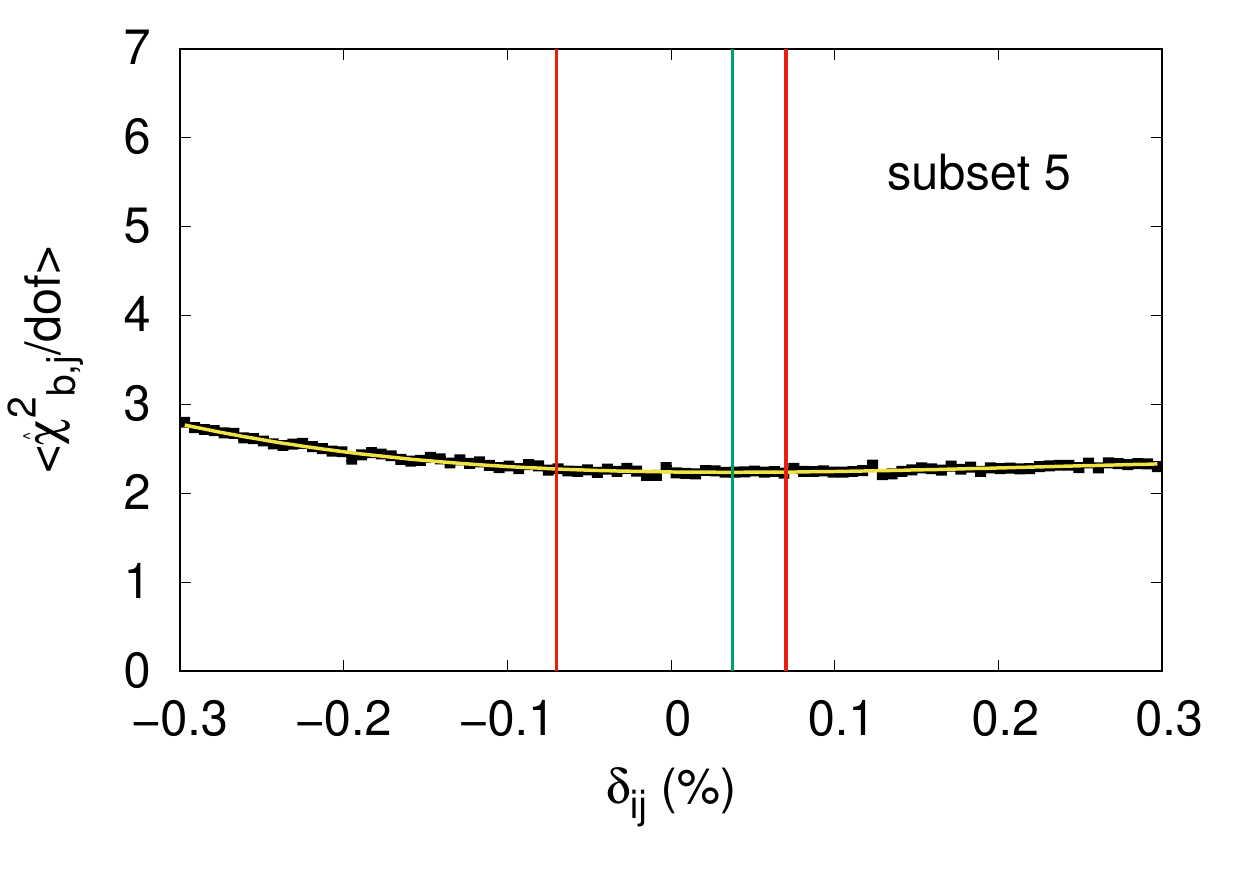}
\label{fig:realsys5}
  \end{subfigure}
  ~ 
  \begin{subfigure}[b]{0.45\textwidth}
 \includegraphics[scale=0.6]{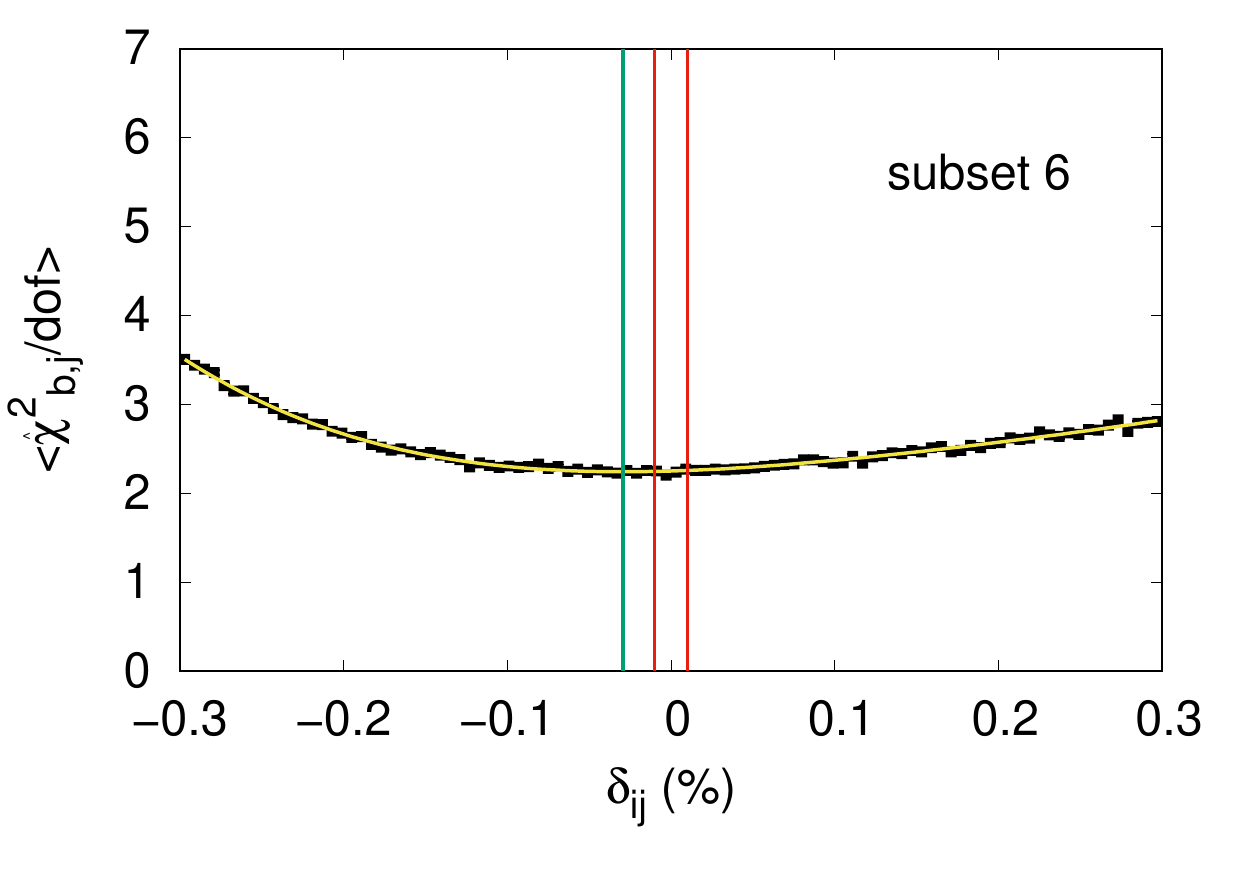}
\label{fig:realsys6}
  \end{subfigure}  
  \caption{Estimate of $-{\delta}^*$ for data subsets from 1 to 6. In each plot, the black points are the results of the preliminary bootstrap cycle, the yellow curve  gives the result of the quartic polynomial fit, the red curves are set at $\pm \Delta_k$ and the green line is set at $-{\delta}^*$.}\label{fig:realsysall1}
\end{figure}

\begin{figure}
  \centering
  \begin{subfigure}[b]{0.45\textwidth}
 \includegraphics[scale=0.6]{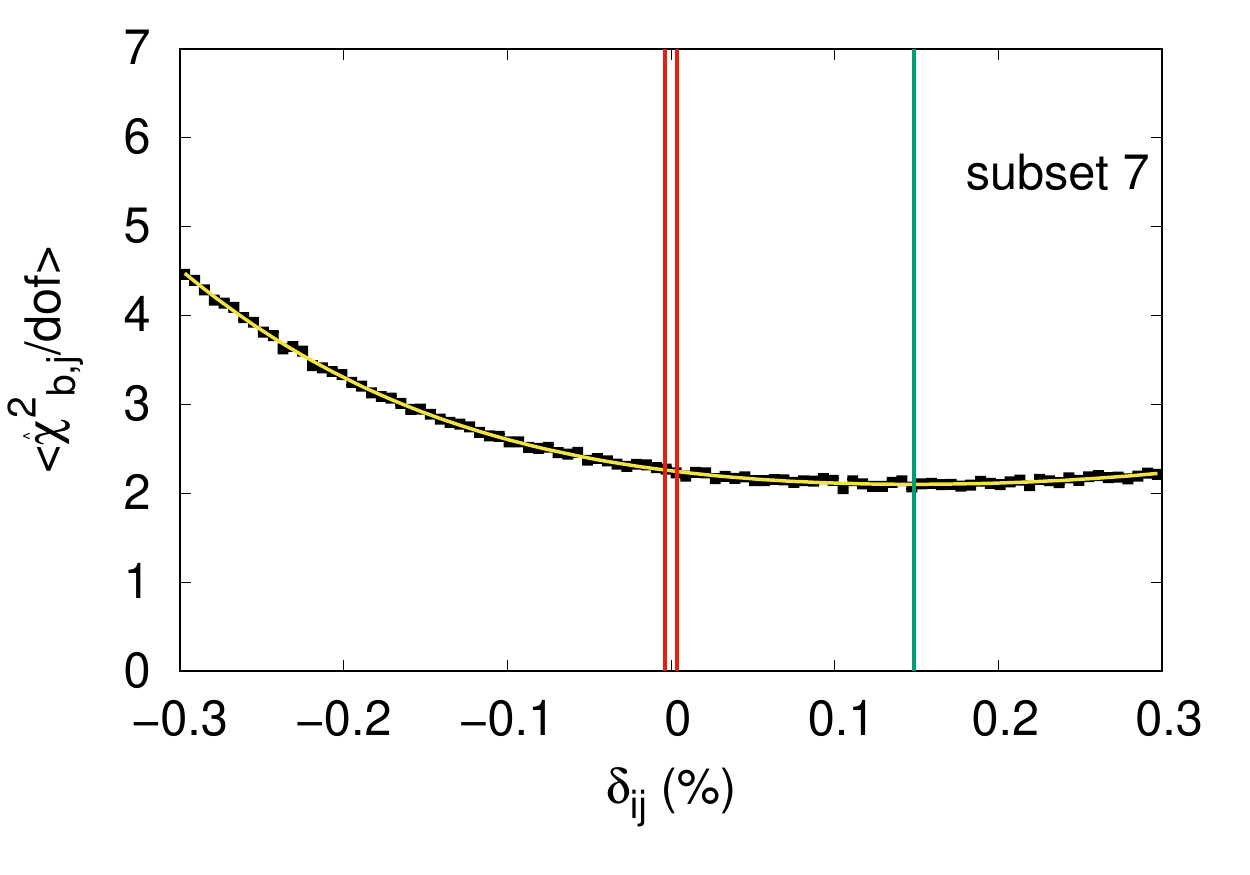}
\label{fig:realsys7}
  \end{subfigure}
  ~ 
  \begin{subfigure}[b]{0.45\textwidth}
 \includegraphics[scale=0.6]{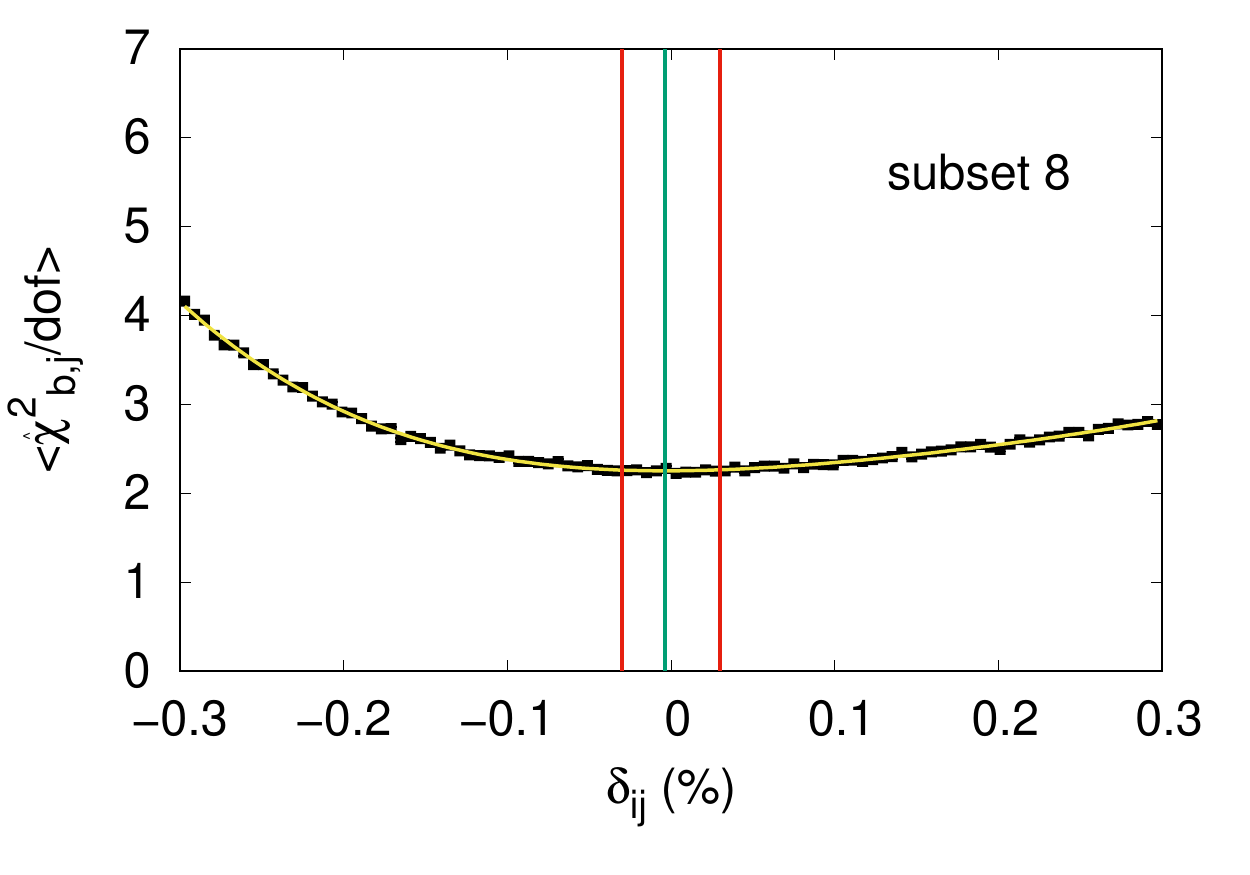}
\label{fig:realsys8}
  \end{subfigure}\\
 \begin{subfigure}[b]{0.45\textwidth}
 \includegraphics[scale=0.6]{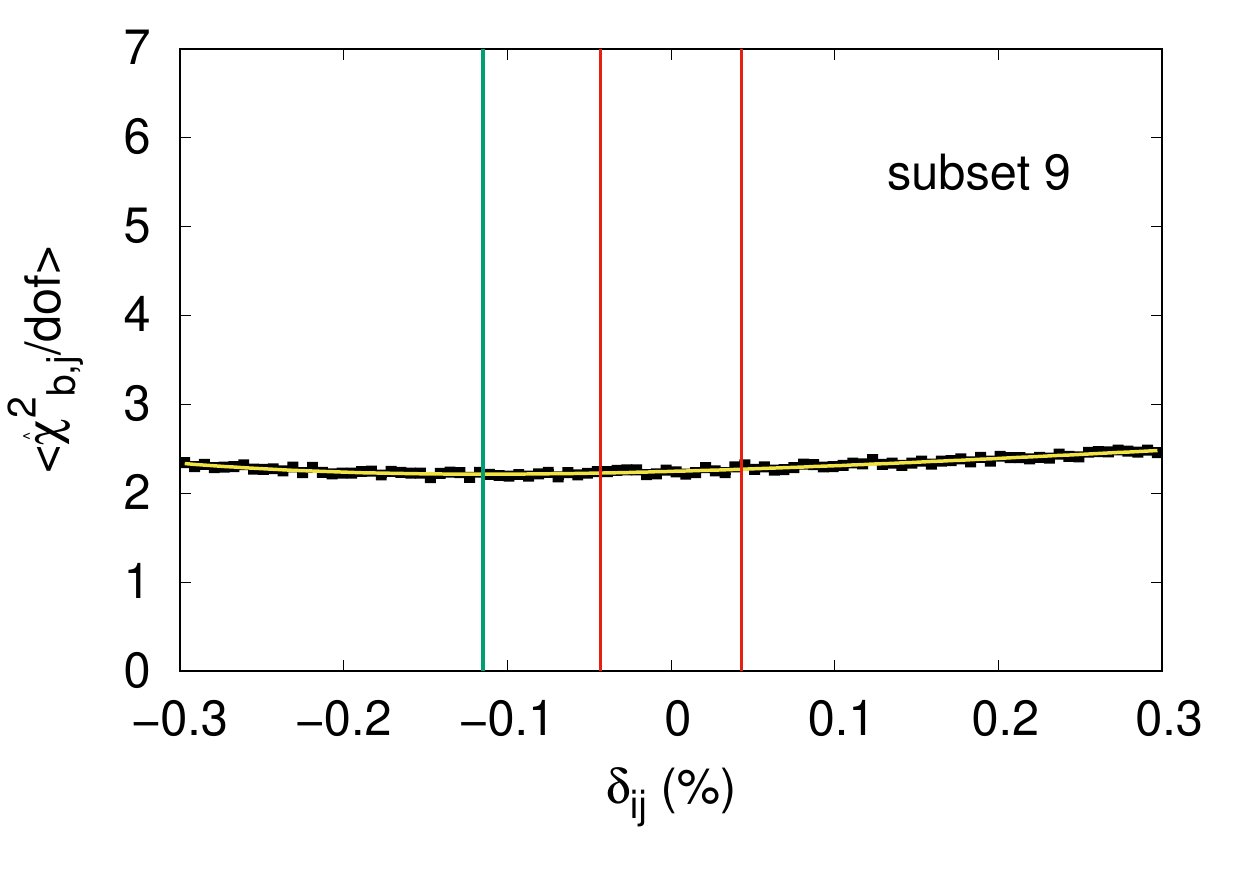}
\label{fig:realsys9}
  \end{subfigure}
  ~ 
  \begin{subfigure}[b]{0.45\textwidth}
 \includegraphics[scale=0.6]{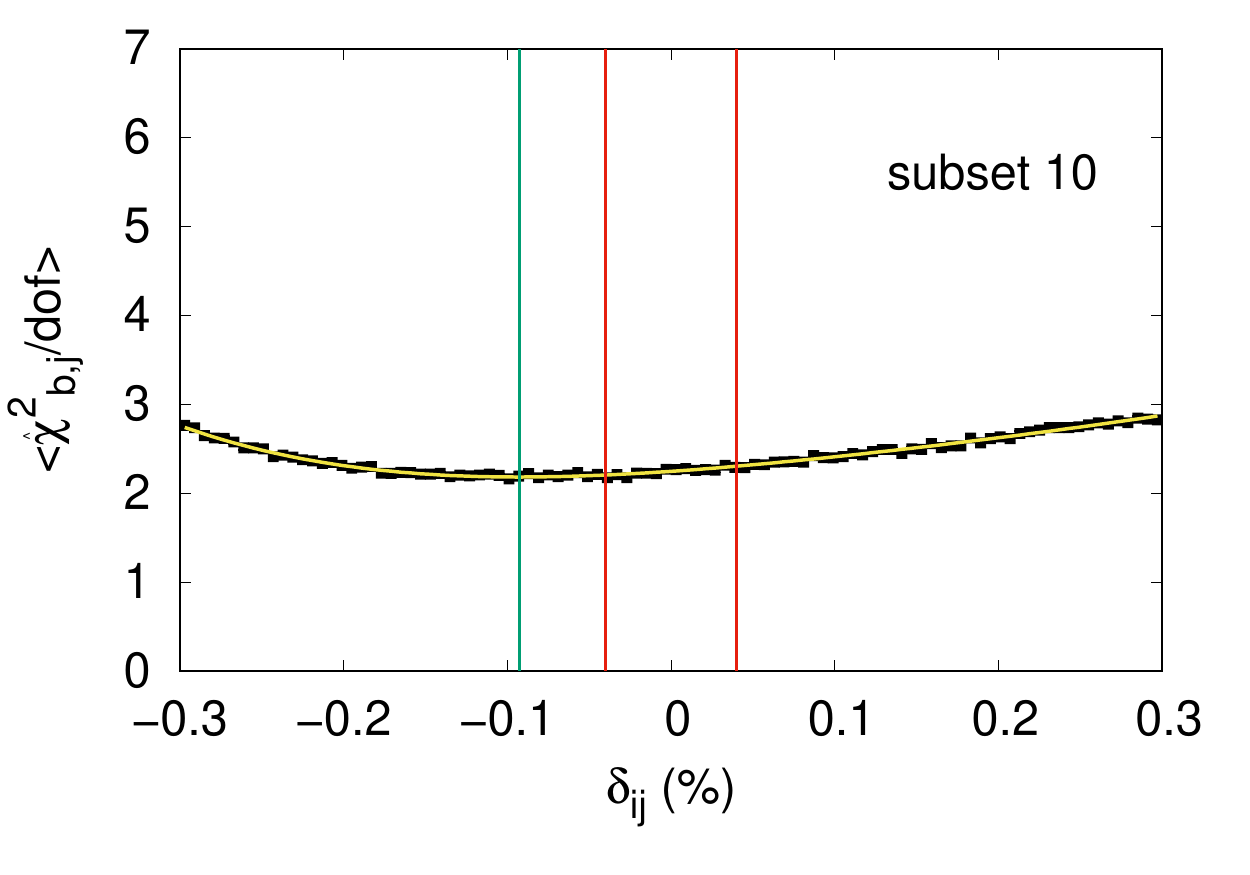}
\label{fig:realsys10}
  \end{subfigure}\\
 \begin{subfigure}[b]{0.45\textwidth}
 \includegraphics[scale=0.6]{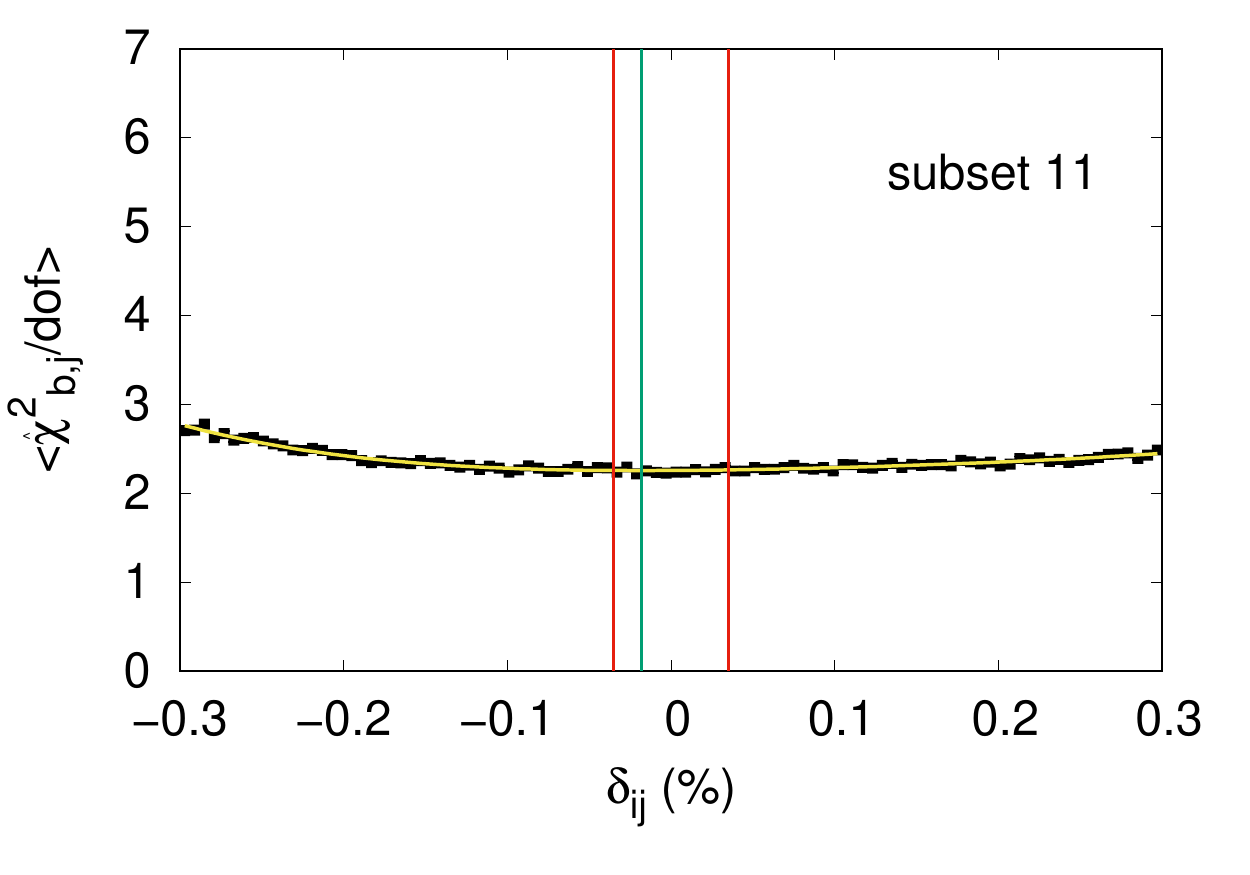}
\label{fig:realsys11}
  \end{subfigure}
  ~ 
  \begin{subfigure}[b]{0.45\textwidth}
 \includegraphics[scale=0.6]{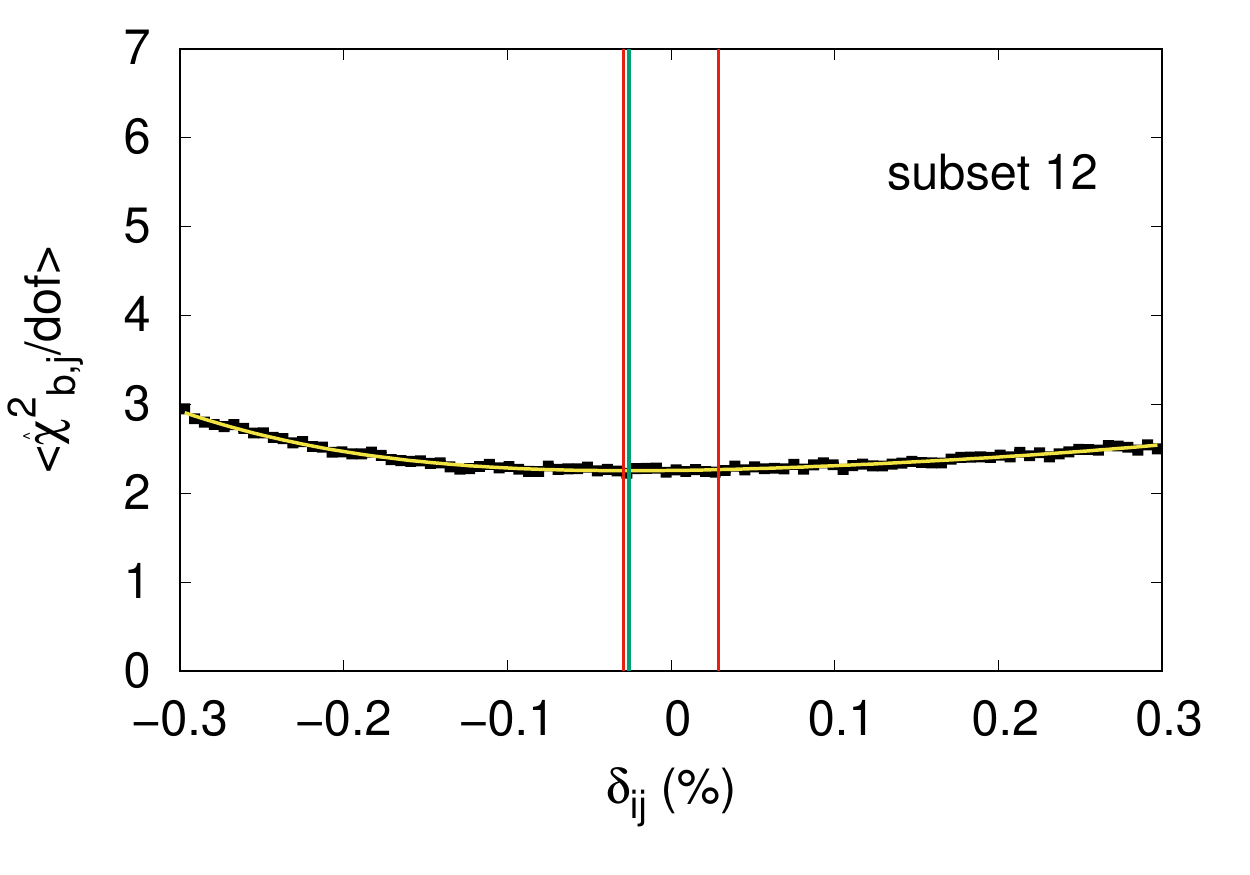}
\label{fig:realsys12}
  \end{subfigure} \\
 \begin{subfigure}[b]{0.45\textwidth}
 \includegraphics[scale=0.6]{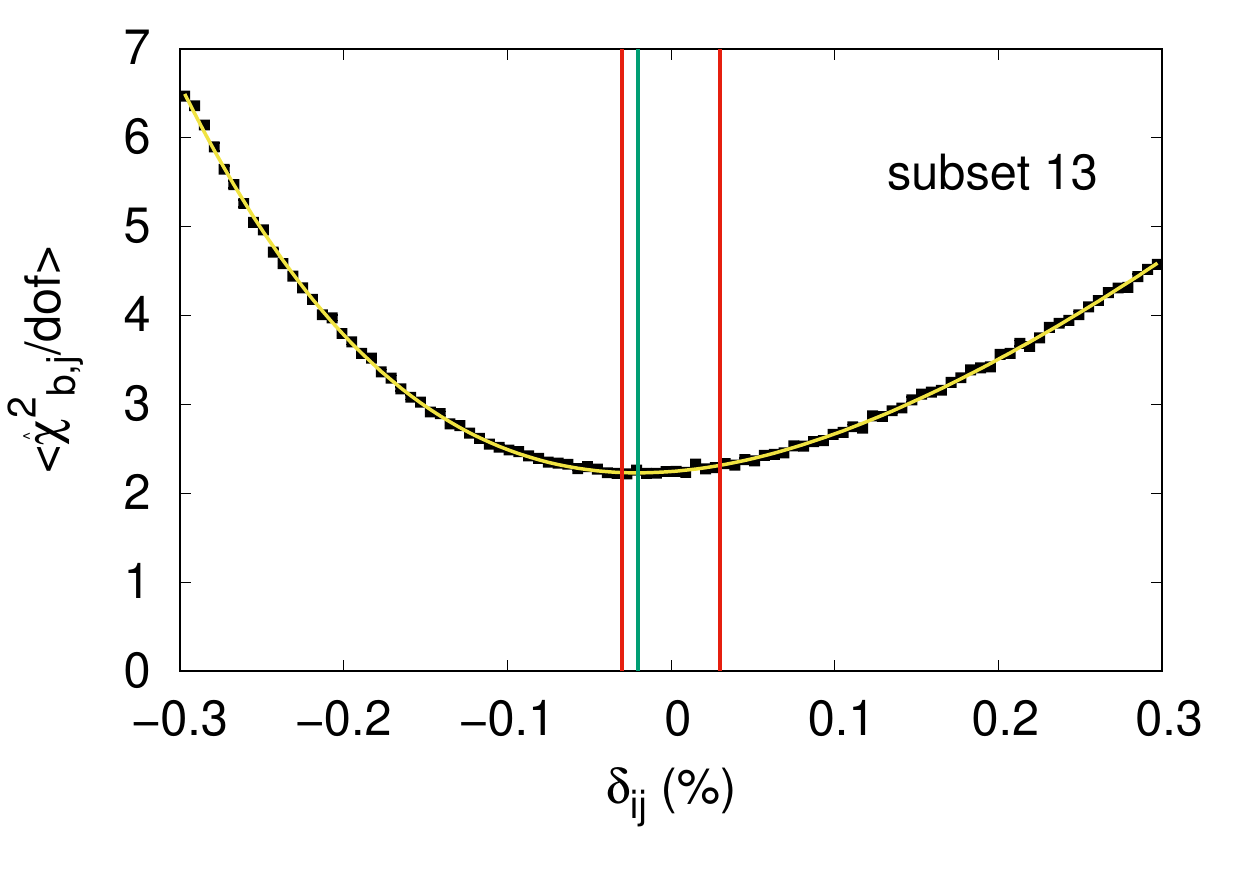}
\label{fig:realsys13}
  \end{subfigure}
  ~ 
  \begin{subfigure}[b]{0.45\textwidth}
  \end{subfigure}  
  \caption{Estimate of $-{\delta}^*$ for data subsets from 7 to 13. In each plot, the black points are the results of the preliminary bootstrap cycle, the yellow curve  gives the result of the quartic polynomial fit, the red curves are set at $\pm \Delta_k$ and the green line is set at $-{\delta}^*$.}\label{fig:realsysall2}
	\end{figure}
Some comments are in order here:
	\beitem
  \item[*] the subsets with very small number of points, or with points lying in kinematical regions not very sensitive to the fit parameter ($\astat$-$\bstat$), basically show a flat distribution. This means that
 the value of the systematic offset does not have a significant impact on the final fit results;
  \item[*] for the majority of the subsets, the estimated systematic offset lies inside the published range\footnote{For more details, see the references quoted in \tref{tab:data}.}, thus cross-checking the validity of the method;
  \item[*] As mentioned before, this technique is model-dependent; for this reason, we do not recommend to use it to automatically discard data from the whole set. Even if the fit model $T$ is correct,
 when the evaluated offset value is outside the estimated
 systematic uncertainty interval, an ad hoc procedure is needed
 to correctly deal with each specific case.
	\enitem

In \tref{tab:normal} the different values of the 
 $\tilde\delta_k$ parameters  determined for all the subsets
are listed and compared to the $(f_k-1)$ parameters evaluated
with the minimization of the modified $\chi^2$ function (see \eqr{eq:chi2sys}).
As done in \secr{sec:profile}, the uncertainty intervals of the $\tilde\delta_k$
  parameters can be 
  estimated from the MINOS method. Also in this case we get
  very similar results as those obtained with the $\chi^2_{mod}$ approach.

The good agreement between these two sets of values gives a further indication
of the correctness and validity of our new method.
As noticed before, the offset estimate is now enough reliable due to the
rather large number of data subsets.

Using the evaluated $\tilde{\delta_{k}}$ values, we can then
rescale all the points of each subset by their estimated systematic offset
and finally
perform the bootstrap sampling taking only into account
their statistical uncertainties, \ie
	\beql{eq:real1}
\mc S_{ij} = (1 + \tilde{\delta_{k}})( E_{i}+\gamma_{ij}\sigma_i),
	\enq
where the only random numbers are the standard Gaussian variables $\gamma_{ij}$.\\

The results thus obtained for $\astat$ and $\bstat$ ( in the usual units of $10^{-4}\text{ fm}^3$, adopted from here on) are:
	\beql{eq:real2}
        \astat = 12.08\pm 0.24, \quad \bstat = 1.69\pm0.24, \quad \hat\chi_r^2 = 0.9, \quad \text{$p$-value } =  20\% , 
	\enq
        which are statistically consistent with the values that can be evaluated from a standard $\chi^2$ fit discarding all the systematic uncertainties, \ie
		\beql{eq:real3}
\astat = 11.99 \pm 0.31, \quad \bstat = 1.81\pm 0.31, \quad \hat\chi_r^2 = 1.25,  \quad \text{$p$-value } =  2\% . 
	\enq
If, on the other hand, we minimize the $\chi_{mod}^2$ function given in \eqr{eq:chi2sys}, we get: 
		\beql{eq:real4}
\astat = 11.94 \pm 0.40, \quad \bstat = 1.86\pm 0.40, \quad \hat\chi_{mod,r}^2 = 1.26,
	\enq
 which are, again, almost identical to the results
 obtained in the two previous cases.
 \\

	\betab
\begin{tabular}{|c|c|c||c|c|c|}
\hline
 k & $f_k -1 $ $(\%)$ & $\tilde\delta_k$ $(\%)$  & k & $f_k -1$ $(\%)$ & $\tilde\delta_k$ $(\%)$\\
\hline
1 & $7.5 \pm 2.3$ & $8.6$ & 8 & $0.0 \pm 1.9$ & $-0.4$\\
2 & $ -0.6\pm 4.8$ & $-2.0$ & 9 & $ -4.7\pm 3.6$ & $-11.6$\\
3 & $ -4.5\pm 2.1$ & $-5.6$ & 10 & $ -5.7\pm 2.5$ & $-9.2$\\
4 & $-2.5 \pm 6.5$ & $-5.6$ & 11 & $ -0.4\pm 2.8$ & $-1.7$\\
5 & $3.0 \pm 4.3$ & $3.7$ & 12 & $ -0.7\pm 2.4$ & $-2.6$\\
6 & $-0.3 \pm 0.9$ & $-2.9$ & 13 & $ -0.1\pm 1.3$ & $-2.1$\\
7 & $ 7.5\pm 2.4$ & $14.8$ &  &  &\\
\hline
\end{tabular}
\caption{Estimate of the experimental offset for each data subset (labeled with $k$) used in the RCS analysis: results from the $\chi_{mod}^2$ method ($f_k -1$) are compared to the results of the bootstrap method ($\tilde\delta_k$).
}\label{tab:normal}
	\entab

        It is noteworthy to observe that  the statistical significance of the results substantially improves after the data rescaling (see \eqrs{eq:real2} and
          \eref{real3}).
          As already discussed in this work, the $p$-value 
          can not be determined for the results given in \eqr{eq:real4},
          where the $\chi^2_{mod}$
          procedure is used.

	\subsection{RCS: goodness-of-fit distribution}
	\label{sec:com_chi2}

  As mentioned before, in the RCS analysis of Ref.~\cite{Pasquini_2019}
  five parameters are sampled from their experimental estimates, while just one, \ie the ($\astat-\bstat$) difference,
  is left as free parameter.\footnote{In Ref.~\cite{Pasquini_2019}
  this fitting condition is labeled as Fit $1$ and Fit $1^\prime$, where the $^\prime$ superscript stands for the inclusion
  of systematic uncertainties in the bootstrap sampling.}
  This  setting is quite similar to  the previously described Fit$_{2p+1s}$ case.
 Thus, after  introducing the notations Fit$_{1p+5s}$ (Fit$_{1p+5s}^\prime$)
 for the exclusion (inclusion) of systematic uncertainties in the bootstrap sampling procedure, we can evaluate
 both the CDFs and the different components
 of ${\hat\chi^2_{th,j}}$ (see \eqr{eq:chi2note}).

		\begin{figure}[h]
\includegraphics[scale=.8]{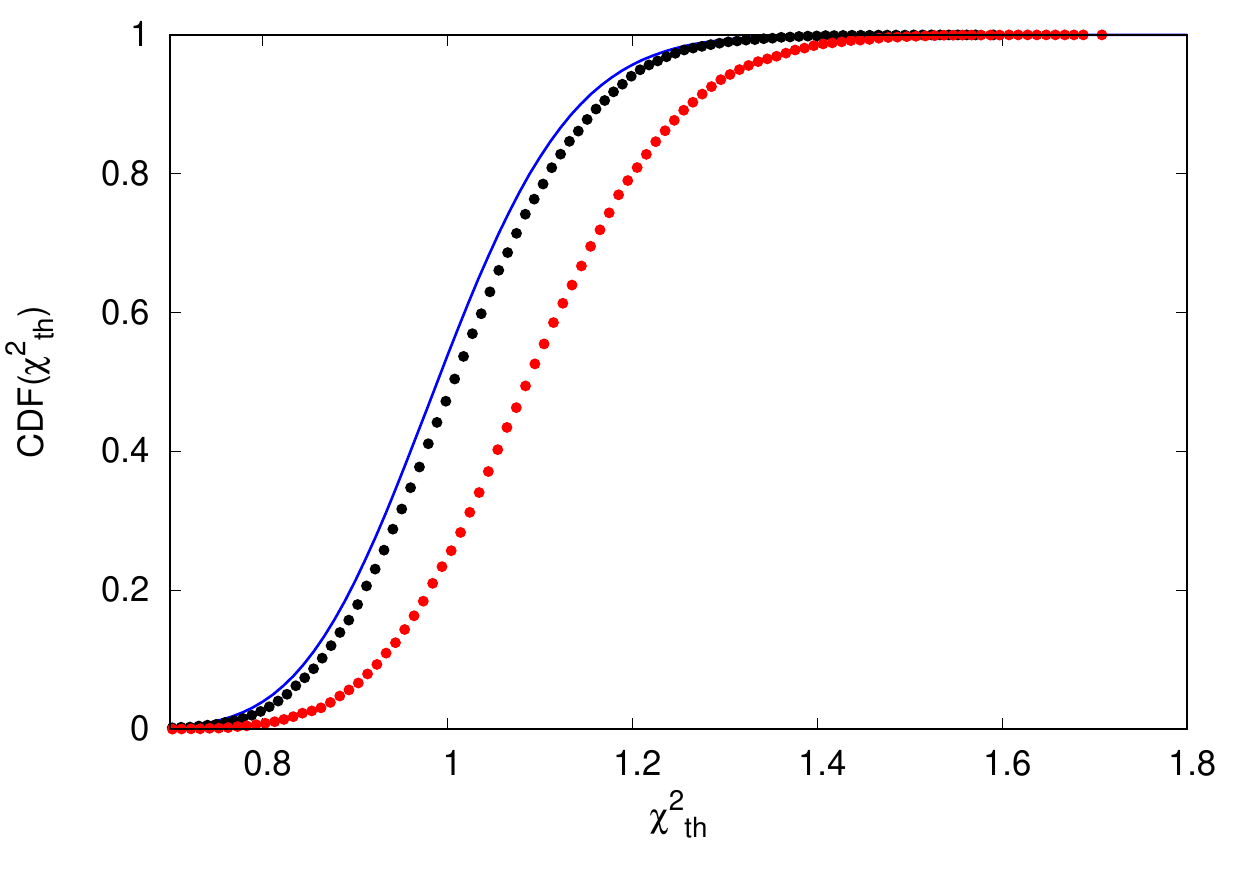}
\caption{CDFs for the $\hat\chi^2_{th,j}$ parameter in the Fit$_{1p+5s}$ (black points) and Fit$_{1p+5s}^\prime$ (red points) conditions, compared to the CDF of  the reduced $\chi^2$-distribution (solid blue curve). See~\tref{tab:chi2_compton} for the meaning of the symbols. 
  \label{fig:chi2cum_compton}}
	\end{figure}

In 
  the cumulative goodness-of-fit probability distribution shown in \fr{fig:chi2cum_compton}, the distortion caused by the
  inclusion of the systematic uncertainties is
clearly visible.
It is also interesting to note that, at odds with
the results previously obtained with the toy model, 
the expected goodness-of-fit distribution 
is not the reduced $\chi^2$-function, even 
when the systematic uncertainties are excluded from the fit.
As already noticed in \secr{sec:profile},
this feature can be related to the non-negligible offsets
present in  the different data subsets. This effect 
can also be quantified with the $\mathbb E\left[{\epsilon^\prime}^2_r\right]$  and $\mathbb E\left[D^\prime_r\right]$ terms,
which are not small enough
to be ignored, even when only the statistical uncertainties are included in the analysis.
   
The expected values 
and the probability distributions of ${\hat\chi^2_{th,j}}$ are then given in
\tref{tab:chi2_compton} and \fr{fig:chi2_compton01} respectively, both
 switching on and off the systematic uncertainties.
\begin{table}[]
\begin{tabular}{|c|c|c|c|c|c|c|}
\hline
\multicolumn{7}{|c|}{ MODEL}  \\
\hline
\hline
 fitting conditions &  $\mathbb E\left[\hat\chi^2_r\right]$ & $\mathbb E\left[\gamma^2_r\right]$  &  $\mathbb E\left[{\epsilon^\prime}^2_r\right]$ & $\mathbb E\left[D^\prime_r\right]$ & $\mathbb E\left[\Phi\right]$  & Symbol \\  
\hline  
 Fit$_{1p+5s}$ & $ 10^{-6} $  & $ 1.01 \pm 0.11 $  & $ (2.09\pm2.97)\cdot 10^{-2}$  & $ (-1.30\pm 2.75)\cdot 10^{-2} $  & $ (3.62 \pm 7.66) \cdot 10^{-2}$  & \onestatth  \\
Fit$_{1p+5s}^\prime$ & $ 10^{-6} $  & $ 1.01 \pm 0.11 $  & $ (10.3\pm 4.8)\cdot 10^{-2}$  & $ (-1.35\pm5.36)\cdot 10^{-2} $  & $ (3.62 \pm 7.86)\cdot 10^{-2} $  & \onesysth  \\
\hline
\end{tabular}
\caption{Decomposition of the $\hat\chi^2_{th,j}$ parameter, with the notation of \eqr{eq:chi2note}, referred to the analysis of RCS proton data ~\cite{Pasquini_2019}. The different symbols refer to the point styles of \fr{fig:chi2cum_compton}.
}\label{tab:chi2_compton}
\end{table}
From the numerical values of \tref{tab:chi2_compton}, we can conclude that:
\beitem
\item[*]
the uncertainties on the fit parameter cannot be small, due to the relatively large values of the $\mathbb E\left[{\epsilon^\prime}^2_r\right]$  and $\mathbb E\left[D^\prime_r\right]$ ;
\item[*]
  the sampling of the additional nuisance parameters is under control,
  being the $\mathbb E\left[\Phi_r\right]$ small;
\item[*]
  the systematic uncertainties have a sizable effect on the fit uncertainties,
  being the $\mathbb E\left[{\epsilon^\prime}^2_r\right]$ term increased by a factor 5 as soon as they are included in the procedure.\\
  \enitem
		\begin{figure}[h]
\includegraphics[scale=.9]{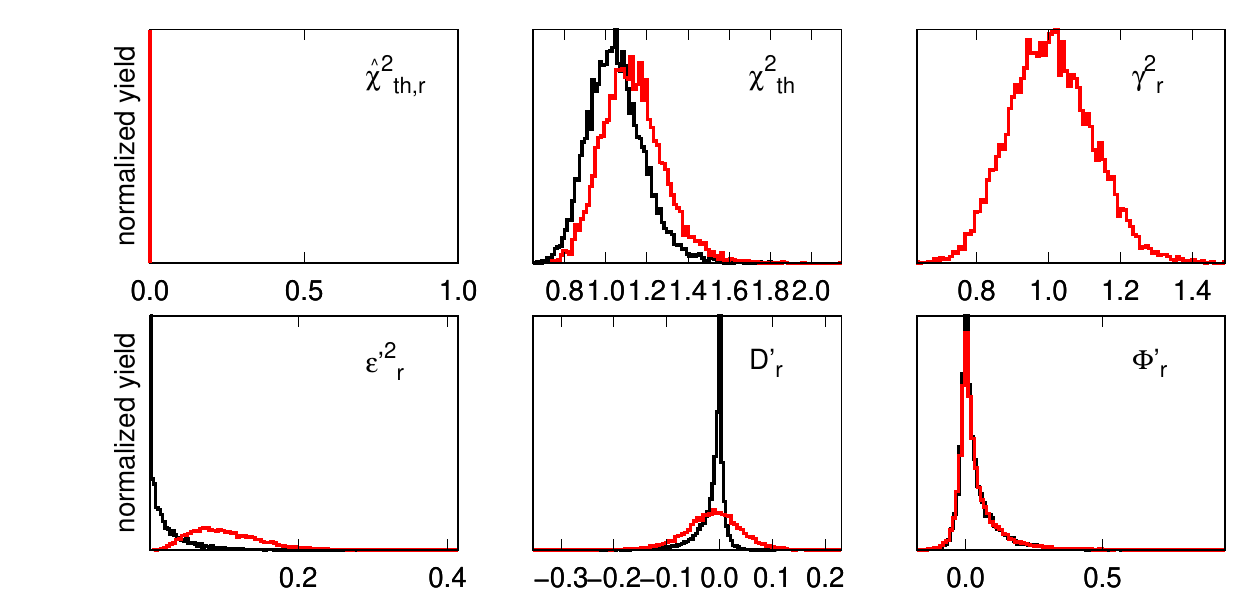}
\caption{Decomposition of the $\hat\chi^2_{th,j}$ parameter
  for the  Fit$_{1p + 5s}$ (black curves) and  Fit$_{1p + 5s}^\prime$ (red curves) configurations.
  Upper panel (from left to right): $\hat\chi^2_r$, $\chi_b^2$ and $\gamma^2_r$ components. Lower panel (from left to right): $\epsilon_r^2$, the $D_r$ and $\Phi_r$ components. See text for the notation.
  \label{fig:chi2_compton01}}
	\end{figure}

	\section{Conclusions}
	\label{sec:concl}

 We presented a new fitting technique based on the parametric bootstrap method and we developed two different toy models
 to completely analyze and cross-check its main features using the results obtained both with the standard $\chi^{2}$ procedures
   and a Hierarchical Bayesian model.
 Furthermore, we applied the fitting technique to an actual physics process, \ie the real Compton scattering off the proton~\cite{Pasquini_2019}, thus confirming the portability of the technique itself.\\
 We showed that this new technique offers several advantages when compared to the other procedures.
 The systematic uncertainties can be taken into account in a straightforward way without the need of additional fit parameters and with a very flexible implementation of any probability distribution. Furthermore, the probability distributions of the fit
 parameters are not assumed to be a priori Gaussian, but are empirically obtained by the procedure itself.
 Another advantage with respect to the standard best-fit methods
 is that the uncertainties on additional nuisance parameters can be easily taken into account, without resorting to the approximated, and often complicated to be implemented, error-propagation formula.\\
The bootstrap framework provides also an estimate of the overall offset of a given data set,
giving results that are in very good agreement with the ones obtained from the standard $\chi_{mod}^2$ method.
This feature can be used as an indication about the quality of the data points, but it should not be used as a fitting strategy by itself.\\
Furthermore, our fitting technique provides the correct $p$-value when systematic uncertainties are present and in all other cases
when the goodness-of-fit distribution is not the reduced $\chi^2$-distribution.\\
All these benefits comw with one drawback: a relevant number of artificial bootstrap  ``measurements'' has to be generated in order to well approximate both the (unknown) true probability distributions of the fit parameters and the fit $p$-value.
Apart from this computational limitation, common to all the Monte Carlo-based methods, the previous considerations lead
us to encourage the use of this technique.
	\section{Acknowledgments}
We are very grateful to Barbara Pasquini, that provided all the theoretical framework for the analysis of the RCS data. We want to thank also Andrea Fontana, Alberto Rotondi, for stimulating discussions and Simone Rodini, for some useful suggestions. We also want to thank Lissa De Souza Campos for a careful reading of the manuscript.

\bibliography{biblio}
\end{document}